%% file: PhysRepOct31-jm.tex
\documentclass{elsart}

\usepackage{amsmath}
\usepackage{amssymb}
\usepackage{amsfonts}
\usepackage{graphicx}
\usepackage{makeidx}
\usepackage{multicol}
\usepackage{bm}

\newcommand{\beq}{\begin{equation}}
\newcommand{\eeq}{\end{equation}}
\newcommand{\bea}{\begin{eqnarray}}
\newcommand{\eea}{\end{eqnarray}}
\newcommand{\fig}[1]{Fig.\ \ref{#1}}
\newcommand{\eq}[1]{Eq.\ (\ref{#1})}
\renewcommand{\vec}{\mathbf}
\newcommand{\showlabel}[1]{} 
\newcommand{\internalcom}[1]{} 

\begin{document}
\begin{frontmatter}
\title{Ultracold Neutral Plasmas}

\author[Rice]{T.\ C.\ Killian},
\author[APS]{T.\ Pattard},
\author[ITAMP]{T.\ Pohl} and
\author[DD]{J.\ M.\ Rost}
\address[Rice]{Rice University, Department of Physics and Astronomy
and Rice Quantum Institute, Houston, Texas, USA}
\address[APS]{APS Editorial Office, 1 Research Road, Ridge, NY 11961}
\address[ITAMP]{ITAMP, Harvard-Smithsonian Center for Astrophysics, 60 Garden
Street, Cambridge, MA 02138, USA}
\address[DD]{Max Planck Institute for the Physics of Complex Systems, Dresden,
Germany}

\date{\today}

\begin{abstract}
Ultracold neutral plasmas, formed by photoionizing laser-cooled atoms
near the ionization threshold, have electron temperatures in the
1-1000\,kelvin range and ion temperatures from tens of millikelvin to
a few kelvin.  They represent a new frontier in the study of neutral
plasmas, which traditionally deals with much hotter systems, but they
also blur the boundaries of plasma, atomic, condensed matter, and low
temperature physics.  Modelling these plasmas challenges computational
techniques and theories of non-equilibrium systems, so the field has
attracted great interest from the theoretical and computational
physics communities.  By varying laser intensities and wavelengths it
is possible to accurately set the initial plasma density and energy,
and charged-particle-detection and optical diagnostics allow precise
measurements for comparison with theoretical predictions.

Recent experiments using optical probes demonstrated that ions in
the plasma equilibrate in a strongly coupled fluid phase. Strongly
coupled plasmas, in which the electrical interaction energy between
charged particles exceeds the average kinetic energy, reverse the
traditional energy hierarchy underlying basic plasma concepts such
as Debye screening and hydrodynamics.  Equilibration in this regime
is of particular interest because it involves the establishment of
spatial correlations between particles, and it connects to the
physics of the interiors of gas-giant planets and inertial
confinement fusion devices.

\end{abstract}
\end{frontmatter}

\setcounter{tocdepth}{100}
\tableofcontents
\clearpage
\section{Introduction}
\input{intro-jm}
\section{Experimental Methods}
\label{sectionexperimentalaspects} \subsection{Creation of ultracold
neutral plasmas}
\input{PlasmaCreation-jm}
\subsection{Detection of ultracold neutral plasmas}
\input{PlasmaDetection-jm}
\section{Theoretical Description}
\label{sectiontheoreticalaspects}
\input{theo_intro-jm}
\subsection{Macroscopic Approaches}
\label{sectionhydrodyn}
\input{hydrodyn2-jm}
\subsection{Microscopic Approaches}
\input{hybrid-jm}
\label{sectionhybrid}
\section{Physical Processes in Ultracold Neutral Plasmas}
\label{sectionphysprocess}
\input{PhysProcOverview-jm}
\subsection{Initial Electron Equilibration}
\label{sectioninidtialelectron}
\input{InitEEqui-jm}
\subsection{Initial Ion Equilibration}
\label{sectionionequilibration}
\input{InitIEqui-jm}
\subsection{Collective Electronic Plasma Modes}
\label{sec:PlasOsc}
\label{sectionplasmaoscillation}
\input{PlasOsc-jm}
\subsection{Plasma Expansion}
\label{sectionexpansion}
\input{PlasExpansion-jm}
\subsection{Electron Heating Mechanisms}
\label{sectionelectronheating}
\input{ElecHeatMech-jm}
\label{sectionelectronheatingtheory}
\label{sectionelectronheatingexperiments}
\subsection{Coulomb Coupling Parameters}
\label{sectioncoulombcoupling}
\input{CoulCoupPar-jm}
\section{Achieving Strong Coupling}
\label{sectionachievingstringcoupling}
\input{future_strong_intro-jm}
\subsection{Ionic coupling}
\label{IonCoupling}
\input{IonCoupling-jm}
\subsection{Electronic coupling}
\label{IceCubes}
\input{icecubes-jm}
\section{Conclusions and Future Directions}
\label{sectionfuture}
\input{future-jm}

\begin{appendix}
\section{Important Quantities}
\input{quantities-jm}
\section{Extracting Plasma Parameters from the Absorption Spectrum}
\input{app-absorptionspectrum-jm}
\label{appendixabsorptionimaging}
\section{The Quasineutral Approximation}
\input{quasineutrality}
\label{appb}

\end{appendix}

\end{document}

%% file: intro-jm.tex
\begin{figure}[]
    \centering
    \includegraphics[width=5in,clip=true]{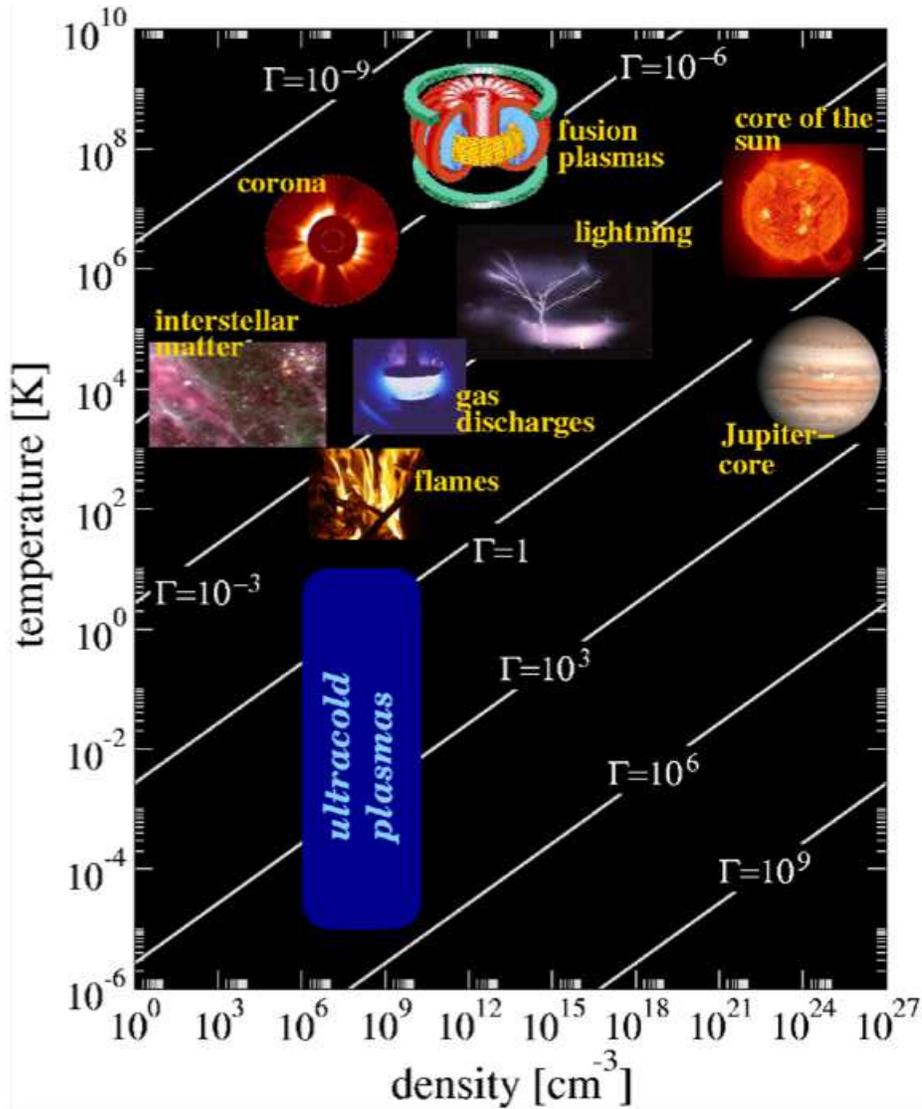}
    \caption{Overview of neutral plasmas in the density-temperature
    parameter plane. Lines of constant Coulomb coupling parameter
    $\Gamma$ (see \eq{coulombcouplingdefinition}) are indicated.
    After \cite{pdiss}.}
    \label{fig:plasmasketch}
\end{figure}

In a conventional neutral plasma, ions and electrons are created
from atoms and molecules by ionizing collisions between particles.
Because a typical ionization potential is on the order of an
electronvolt, most neutral plasmas have temperatures of thousands of
kelvin or more. (See Fig.~\ref{fig:plasmasketch}.)  Plasma physics
in this regime encompasses a wealth of fascinating fundamental
phenomena such as collective modes, instabilities, and transport
mechanisms, and it can contribute to important applications such as
lighting technologies, plasma processing, and the pursuit of fusion
energy.

Ultracold neutral plasmas \cite{kkb99} stretch the boundaries of
traditional neutral plasma physics. They are formed by photoionizing
laser-cooled atoms near the ionization threshold  and have electron
temperatures ranging from 1-1000\,K and ion temperatures of around
1\,K.  While providing access to a new regime, they have also proven
to be clean and simple systems that form an excellent testing ground
for basic plasma theory, and they elucidate physics found across a
much wider spectrum of plasma energy and density.

The clearest distinguishing feature of ultracold neutral plasmas is
that particles can be in or near the strongly coupled regime
\cite{ich82}.  In strongly coupled plasmas the Coulomb interaction
energy between charged particles exceeds the average kinetic energy,
which is quantified by the Coulomb coupling parameter
\begin{equation}\label{coulombcouplingdefinition}
\Gamma= e^2 / [4\pi \varepsilon_0 a k_B T],
\end{equation}
where $a=[3/(4 \pi \rho)]^{1/3}$ is the Wigner-Seitz radius
characterizing the separation between particles in a plasma of
density $\rho$ and temperature $T$. $\Gamma>1$ reverses the
traditional energy hierarchy that underlies our normal understanding
of plasmas based on concepts such as Debye screening and
hydrodynamics. Strongly coupled plasmas exist in dense astrophysical
systems \cite{vho91}, matter irradiated with intense laser fields
\cite{nmg98,sht00,rem05,saa06}, colloidal or dusty plasmas of highly
charged macroscopic particles \cite{mtk99,hol00b}, and non-neutral
trapped ion plasmas  that are laser cooled until they freeze into
Wigner crystals \cite{mbh99}.  Strong coupling is manifested through
the presence of spatial correlations between particles, which can be
important for the equilibration and collective dynamics of the
system.  Optical probes \cite{scg04,cdd05} demonstrated that ions in
an ultracold neutral plasma can be strongly coupled.  Electrons seem
to remain just outside of this regime because the rate of
recombination of electrons and ions to form neutral particles
diverges rapidly as the electron temperature drops.


There is great similarity between the dynamics of ultracold neutral
plasmas and equilibration of plasmas created by  fast-pulse laser
irradiation of solid \cite{ckz00,skh00} and thin film targets
\cite{hbc00,mgf00,bwp01,msp02,hkp02,asa03,kss04,fsk05,rfb05}, rare
gas clusters \cite{dts97}, and gas jets \cite{sbn99,kcn99}. However,
the fast-pulse systems have densities close to or higher than the
condensed
phase.  
This complicates observation of plasma dynamics, which takes place
on the scale of the inverse plasma frequency,
$\omega_{\textrm{p}}^{-1}=[m\varepsilon_{\rm 0}/(e^2\rho)]^{1/2}$,
and is of the order of attoseconds for electron and femtoseconds for
ions. Ultracold plasmas, on the other hand, reach the strong
coupling regime despite the fact that they are extremely dilute
(Fig.~\ref{fig:plasmasketch}). The low density implies that
ultracold plasmas evolve on a time scale that is more easily
accessible experimentally.
 Studies of ion equilibration
have revealed oscillations of the ion kinetic energy
\cite{csl04,ppa05conf,ppr05prl,mur06PRL} that are a fundamental
characteristic of equilibration of strongly coupled systems.
Expansion of a plasma into the surrounding vacuum may be directly
observed with excellent spatial and temporal resolution in ultracold
systems \cite{kkb00,rha03,ppr04,cdd05}.  This phenomenon is
important in fast-pulse experiments, especially in the context of
the generation of high energy particles
\cite{ckz00,skh00,hbc00,mgf00,bwp01,msp02,hkp02,asa03,kss04,fsk05,rfb05,dts97,sbn99,kcn99},
inertial-confinement fusion experiments \cite{lin95}, and x-ray
lasers \cite{dai02}.

 High energy-density and strongly coupled plasmas are often difficult
 to describe theoretically because of the importance of many-body
 interactions, and these problems are often best addressed with
 computational techniques.  Many body interactions are also important
 in ultracold neutral plasmas, and computational physicists have made
 significant contributions to the understanding of these system by
 applying techniques usually used for high energy-density systems
 \cite{kon02,mck02}.  Ultracold plasmas have motivated the development
 and application of specially adapted numerical techniques to
 describe the presence of spatial correlations between the ions
 \cite{ppr04archive} and the evolution of electrons and ions
 on very different timescales \cite{kon02}.  Using these techniques it
 has even been possible to lead the experimental efforts by predicting
 the possibility of Wigner crystallization in an ultracold expanding
 neutral plasma \cite{ppr04} and the conditions that lead to it.
 Again, the excellent experimental control and diagnostics available
 imply that these computational results can be tested with great
 precision in ultracold systems.

Immediately after the creation of an ultracold neutral plasma
through photoionization, the system is far from equilibrium, and
most research in the field has focused on the equilibration process.
Theoretical and computational work
\cite{rha03,kon02,mck02,mur01,kon02prl,rha02,tya00} has  identified
 processes that heat the plasma.
For plasmas in or near the strongly coupled regime, where the
potential energy is comparable to or exceeds the thermal energy,
 excess potential energy arising from a random
distribution of ions \cite{mur01} or electrons \cite{kon02prl} is
converted to thermal energy. This is called \emph{disorder-induced
heating}. Another important heating mechanism is \emph{three-body
recombination} (TBR), in which an ion and two electrons collide to
form an energetic electron and an excited neutral atom.   The TBR
rate varies with temperature as $T^{-9/2}$ \cite{mke69}, and at low
temperatures it can become the fastest process in the plasma.
TBR is a crucial process for the formation of cold antihydrogen
through positron-antiproton recombination \cite{aab02,gbo02}.
Conditions in these anti-matter plasmas are similar to ultracold
neutral plasmas except a large magnetic field is usually present to
contain the particles. Surprisingly, recombination in ultracold
neutral plasmas also has similarities to stellar dynamics in
globular clusters \cite{cvz05}.

The study of collective modes has proven to be a rich area of
research, originally motivated by the possibility of observing
modification of the mode structure due to strong coupling
\cite{kgm93}.  Electron plasma oscillations \cite{kkb00,bsp03} and
Tonks-Dattner modes \cite{fzr06} have served as probes of plasma
dynamics and represent interesting fundamental problems in their own
right.

The review is structured such that Sections
\ref{sectionexperimentalaspects} and \ref{sectiontheoreticalaspects}
provide the experimental and theoretical basis that allows one to
understand and manipulate ultracold neutral plasmas. Section
\ref{sectionphysprocess} of the review describes the physical
processes which take place in an ultracold neutral plasma from the
time it is created until it expands into the surrounding vacuum.
In Section
\ref{sectionachievingstringcoupling}, we discuss proposals and
prospects for achieving very strong coupling of the plasma -- in
some sense the ``holy grail'' of ultracold plasma physics.  We
conclude with Section \ref{sectionfuture} by discussing some
possible future directions of this relatively young research field.

%% file: PlasmaCreation-jm.tex
\begin{figure}
\centering
\includegraphics[width=4.0in,clip=true,trim=55 370 20 160,angle=0 ]{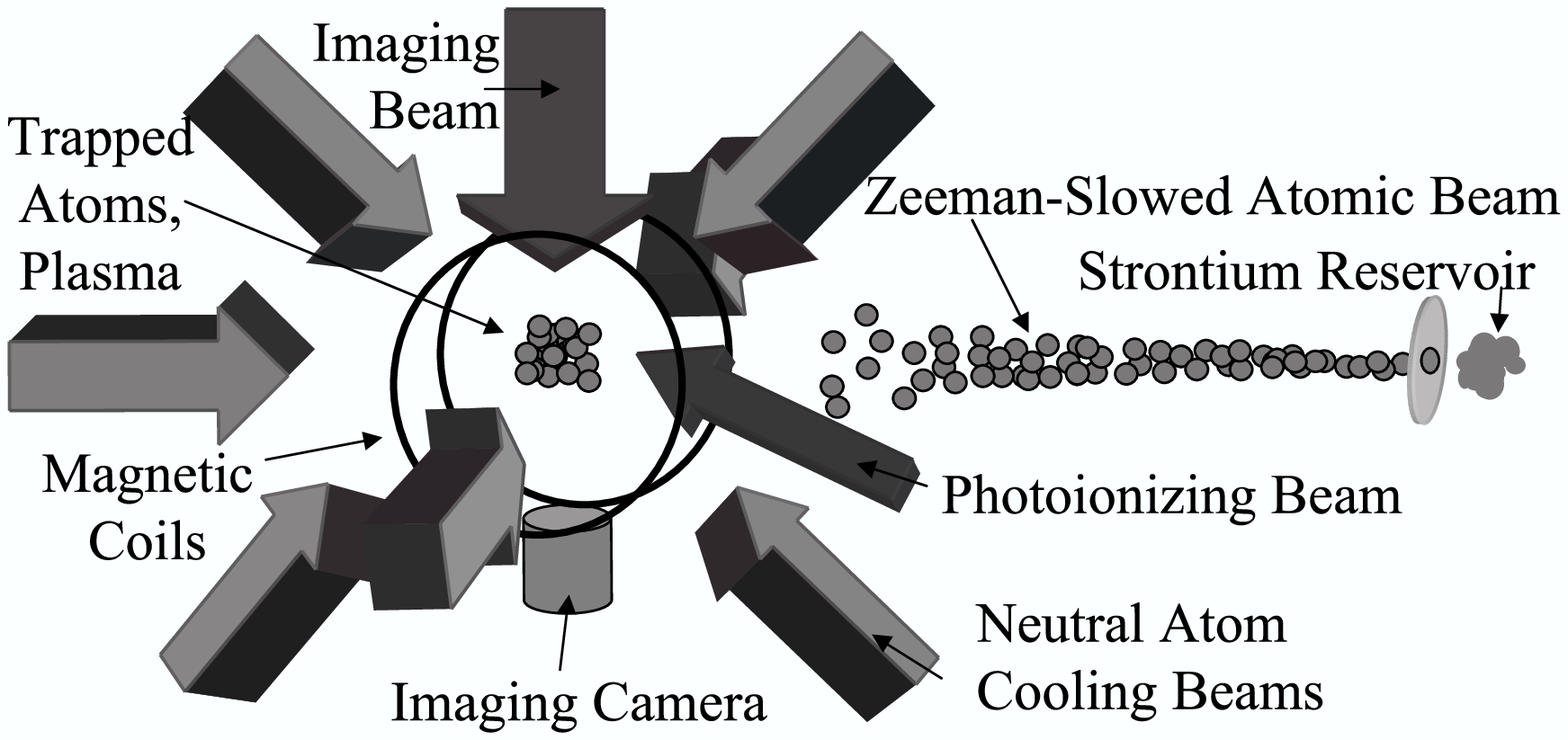}\\
\caption{Schematic setup of a typical ultracold plasma experiment,
exemplified with strontium atoms.  Neutral atoms are laser cooled
and trapped in a magneto-optical trap operating on the
$^1S_0-{^1P_1}$ transition at 461 nm, as described in \cite{nsl03}.
In a second step, $^1P_1$ atoms are ionized by photons from a laser
at $\sim 412$~nm (see Fig.\ \ref{energylevels}A).  Finally, the
ionic plasma component is imaged using the $^2S_{1/2}-{^2P_{1/2}}$
transition in Sr$^+$ at $422$~nm (see Fig.\ \ref{energylevels}B).
Reused with permission from \cite{scg04}. Copyright 2004, American
Physical Society.}\label{apparatus}
\end{figure}

The production of an ultracold neutral plasma  starts with
laser-cooled and trapped neutral atoms. There
are many good resources on this topic, such as the book ``Laser Cooling
and Trapping," by Metcalf and van der Straten \cite{mvs99}. In a
configuration of laser beams and magnetic fields known as a
magneto-optical trap (Fig.\ \ref{apparatus}),
as many as $10^9$ atoms can be cooled to
millikelvin or even microkelvin temperatures, and confined at
densities approaching $10^{11}$\,cm$^{-3}$. The typical spatial
distribution of the cloud is of spherical Gaussian shape,
\beq\label{gaussian-cloud}
\rho({r})=\rho_0{\exp}(-r^2/2\sigma^2)\,,
\eeq
 with a width of
$\sigma= 200-1500\,\mu$m. The first ultracold plasma experiments
were performed at the National Institute of Standards and Technology
in Gaithersburg (NIST)  with metastable xenon \cite{wmw93}, but any
atom that can be laser cooled and photoionized may be used, such as
strontium \cite{nsl03,kcg05}, calcium \cite{cdd05}, rubidium \cite
{wgc04,rtn00}, and cesium \cite{rtn00}. A plasma has also been
produced by photoionizing a Bose-Einstein condensate \cite{cam02}.

%

\begin{figure}
\centering
  \includegraphics[width=4in,clip=true,trim=45 200 20 80,angle=0 ]{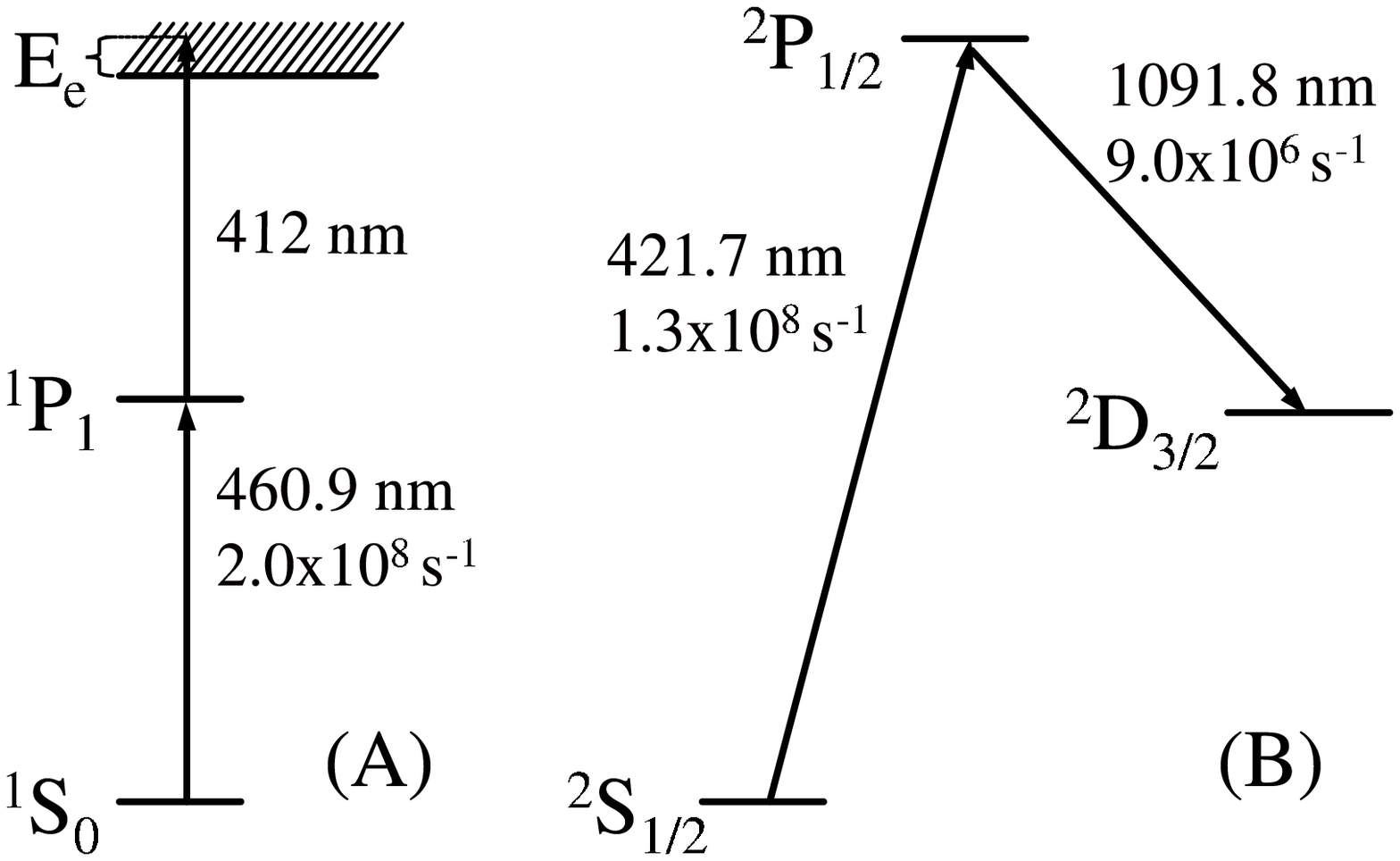}\\
 \caption{Strontium atomic and ionic energy levels with decay rates
 relevant for ultracold plasma experiments.
 (A) Creation of an ultracold plasma starts with laser cooling and
 trapping neutral atoms using the  $^1S_0-{^1P_1}$ transition at 461
 nm. $^1P_1$ atoms are then photoionized with photons from a
 laser at $\sim 412$~nm.  (B)
 Ions are optically imaged using the $^2S_{1/2}-{^2P_{1/2}}$ transition in Sr$^+$  at
 $422$~nm.  $^2P_{1/2}$ ions decay to the $^2D_{3/2}$ state 7\% of the
 time, after which they cease to interact with the imaging beam.  This
 does not affect most experiments because ions typically scatter fewer than
 one photon while the imaging beam is on.  The level scheme
 is similar for calcium \cite{cdd05}.}\label{energylevels}
\end{figure}

A pulsed dye laser excites the laser-cooled atoms above the
ionization threshold (Fig.\ \ref{energylevels}A). Because of their
light mass, the electrons have an initial kinetic energy
($E_{\mathrm{e}}$) approximately equal to the difference between the
photon energy and the ionization potential. $E_{\mathrm{e}}/k_{B}$
can be as low as the bandwidth of the ionizing laser, which is $\sim
100$\,mK with standard pulsed dye lasers, but most studies so far
have dealt with $E_{\mathrm{e}}/k_{B}$ between 1 and $1000$\,K. The
initial kinetic energy for the ions is in the millikelvin range. The
number of atoms ionized ($N_i$), and thus the density of the plasma
($\rho$), is controlled by varying the energy of the photoionizing
laser pulse.

\begin{figure}
\centering
  \includegraphics[width=5in,clip=true,trim=80 50 0 200,angle=0 ]{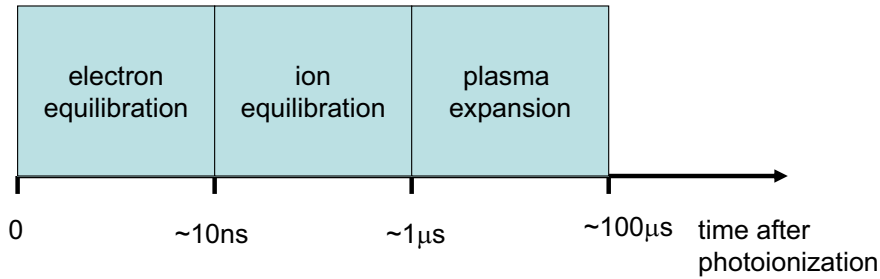}\\
  \caption{Stages and timescales in the dynamics of an ultracold neutral plasma.
  Electron equilibration is a fast process in ultracold neutral plasmas,
  and it occurs in a time equal to the  inverse electron
  plasma frequency,
  $\omega_\mathrm{p,e}^{-1}=\sqrt{m_\mathrm{e}\varepsilon_{0}/(\rho_\mathrm{e}e^{2})}$.
  Analogously, the timescale for ion equilibration in the
  second stage is the inverse  ion plasma
  frequency,
  $\omega_\mathrm{p,i}^{-1}=\sqrt{m_\mathrm{i}\varepsilon_{0}/(\rho_\mathrm{i}e^{2})}$.
  The timescale for plasma expansion is set by the hydrodynamic
  time,
  $\tau_\mathrm{exp}=\sqrt{m_\mathrm{i}\sigma^{2}/k_{B}T_\mathrm{e}}$.
}
  \label{figuretimeline}
\end{figure}

We will discuss the dynamics of the plasma in greater detail in
Section \ref{sectionphysprocess}, but Fig.\ \ref{figuretimeline}
illustrates how the evolution can be divided into three phases. The
fastest process is the equilibration of the electrons because they
are relatively energetic and light. The more sluggish ions then
equilibrate with themselves on a slightly longer timescale. And
finally, the plasma -- which is unconfined in typical experiments --
expands into the surrounding vacuum.

\subsubsection{Photoionization}
The photoionization process is crucial in these experiments, and the
book ``Rydberg Atoms" \cite{gal94} by T.\ Gallagher provides an
excellent introduction to this subject.  Most ultracold plasma
experiments have used two-photon ionization, in which one photon
from the continuous-wave (CW) cooling laser excites atoms to the
upper level of the cooling transition, and a pulsed laser then
excites them to the continuum.  (See Fig.\ \ref{energylevels}.)

In equilibrium
the fraction of atoms in the upper state of the cooling transition
can approach 50\%, and is given by \cite{mvs99}
\begin{equation}\label{excited fraction}
   f={{s_0/2} \over {1 + s_0
   + (2 \delta/\gamma)^2}},
\end{equation}
where $\delta$ is the detuning from resonance of the CW laser,
$\gamma$ is the full linewidth at half maximum for the transition,
and $s_0=I/I_\mathrm{sat}$ is the saturation parameter for the
excitation beam of intensity $I$. $I_\mathrm{sat}$ is typically a
few mW/cm$^{2}$, and it is the intensity required to drive the
transition at a rate equal to $\gamma$.  The excited-state fraction
during normal MOT operation is typically about 10\%, but an extra
excitation beam may also be used to attain $f \approx 0.5$. Highly
saturating the transition with a pulsed-dye-amplified beam yields
even better results \cite{cdd05}.

Photoionization from the intermediate state is best described in
terms of a cross section for absorption of ionizing photons,
$\sigma_\mathrm{PI}$. Cross sections near the ionization threshold
are of the order of the atomic unit of area,
$\sigma_\mathrm{PI}\approx a_{0}^{2} \approx 10^{-18}$ cm$^{2}$.
The probability of ionization is given by
$1-{\rm exp}(-F \sigma_\mathrm{PI})$, where the number of photons per
unit area in the pulse is $F= \int _\mathrm{pulse} dt I/h \nu_\mathrm{laser} $
for instantaneous laser intensity $I$.  A blue (450 nm), 1 mJ, 5
ns pulse has about $2 \times 10^{15}$ photons, and an intensity of
20 MW/cm$^2$ for a pulse area of 1 mm$^2$. This gives a
$2\times 10^{17}$ cm$^{-2}$ integrated photon flux and an ionization
probability of 20\% for atoms in the intermediate level. With more
power it is possible to ionize nearly 100\% of the atoms in the
intermediate state.

Cross sections are normally constant over the
near-threshold-ionization region of interest for ultracold plasmas,
unless there is some physics feature that modifies them, such as an
autoionizing resonance, which is the case in strontium \cite{mbk95}.
In  alkali metals except lithium, cross sections for  single-photon
excitation from the $S$ ground state  to near the ionization
threshold show Cooper minima \cite{gal94}. This is one reason why
most ultracold plasma experiments use two-photon excitation.

The ionization process adds momentum and energy to the electrons and
ions. The particles can be treated as at rest initially, and recoil
during excitation to the intermediate state is negligible. Momentum
and energy conservation for the pulsed laser photons of frequency
$\nu$ gives
\begin{equation}\label{ionizationenergydistribution}
   \Delta E \equiv h\nu -E_{IP}=\frac{p_\mathrm{i}^2}{2 m_\mathrm{i}}+
   \frac{p_\mathrm{e}^2}{2m_\mathrm{e}}\,,
\end{equation}
\begin{equation}
   \hbar \vec k  = {\vec{p}}_\mathrm{i} + {\vec{p}}_\mathrm{e},
\end{equation}
where the ionization potential is $E_{IP}$.
 The photon momentum $k=|\vec k|$
is small. For instance, at 10 K, $\hbar k^{2} \approx \langle
p_\mathrm{e}^{2} \rangle/100$, which implies that $p_\mathrm{i}^2
\approx p_\mathrm{e}^2$. The ion kinetic energy is thus much smaller
than the electron kinetic energy, $E_\mathrm{e}$. Hence, the
electrons take away essentially all the excess photon energy,
$\Delta E$, as kinetic energy:
\begin{equation}
  E_e\equiv \frac{p_\mathrm{e}^2}{2 m_\mathrm{e}}\approx \Delta E.
\end{equation}
The ion kinetic energy is then
\begin{equation}
  \frac{p_\mathrm{i}^2}{2 m_\mathrm{i}}\approx
  \frac{p_\mathrm{e}^2}{2 m_\mathrm{i}}=\Delta E
  \frac{m_\mathrm{e}}{m_\mathrm{i}}\,,
\end{equation}
which is only on the order of millikelvin even for $\Delta E=k_B 1000$ K.
Within a few nanoseconds after photoionization \cite{Spitzer}, the
electrons collisionally thermalize with themselves at
$k_B T_\mathrm{e}\approx  \frac 23 E_\mathrm{e}$. We will describe
below how  processes in the plasma can subsequently heat the
electrons and ions to substantially higher temperatures.

\subsubsection{Spontaneous ionization of a dense Rydberg gas}
An ultracold neutral plasma also forms from spontaneous ionization in
a dense cloud of ultracold Rydberg atoms.  In fact, the first
experiments focused on this phenomenon \cite{rbs98}, and the hope was
to observe a gas-phase analog of the Mott insulator-to-conductor
transition familiar from condensed matter physics \cite{era95}.  The
idea was first explored by Haroche and co-workers \cite{vrg82} when
they saw spontaneous ionization in a beam of Rydberg atoms.

In experiments performed so far, collisions, as opposed to a phase
transition, drive the spontaneous ionization process in beams or
laser-cooled clouds, and after the formation of the plasma, the
dynamics is very similar to the evolution of a system created by
direct photoionization \cite{rha03}.  We will touch upon this work
briefly in Sec.\ \ref{sectionspontaneousdetails}.  However, it will
not be the main focus of this review, and for a deeper discussion we
refer to papers exploring the spontaneous evolution of Rydberg atoms
into a plasma \cite{lnr04} and dynamics of Rydberg atoms embedded in
an ultracold plasma \cite{wgc04}.  Related questions regarding the
stability of a Rydberg gas are important for fundamental interest
and also because this is a proposed architecture for quantum
computing \cite{jcz00}.

%% file: PlasmaDetection-jm.tex
\subsubsection{Charged particle detection}
\label{section_fielddetect}
\begin{figure}
\centering
  \includegraphics[width=3.5in,clip=true,trim=155 140 20 80,angle=0 ]{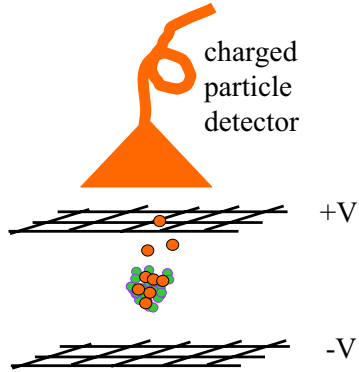}\\
  \caption{Charged particle detection diagnostics.  When electrons
  escape from the Coulomb well formed by the ions, electric fields
  created by potentials on wire-mesh grids direct them to an electron
  multiplier or microchannel plate.  By reversing the electric field,
  ions can be detected.  The grids can also be used to apply
  oscillating radio frequency fields to excite plasma collective
  modes.  }\label{chargedparticle}
\end{figure}

The most common diagnostics for studying ultracold plasmas is
detection of charged particles after they have left the plasma. The
basic idea is shown in Fig.\ \ref{chargedparticle}. The initial
experiments at NIST on plasmas produced by direct photoionization
\cite{kkb99,kkb00,klk01,rfl04}  detected electrons. Experiments
exploring the spontaneous evolution of Rydberg atoms into a plasma
\cite{lnr04} and dynamics of Rydberg atoms embedded in an ultracold
plasma \cite{wgc04,vct05} have used both ion and electron detection.

For monitoring electrons that escape the plasma during normal
evolution, it is essential to not perturb the plasma with large
electric fields.   However, one needs a field of about 5 mV/cm to
guide the escaping charged particles to the detector. This produces
a time-of-flight delay for electrons until they reach the detector
of about 1 $\mu$s.  Since electrons move away from the plasma in all
directions, it also means that the detection efficiency is not
100\%.
easily be ramped to much larger values that can quickly dump all
charged particles of a given sign on the detector.  With careful
design, values well in excess of 100 V/cm can be obtained, which are
also useful for pulsed field ionization of Rydberg atoms
\cite{gal94}.

\subsubsection{Optical probes}
\label{sectionopticalprobes}

Probing an ultracold plasma through charged particles has some
limitations.  For example, remote detection of particles tends to
only provide information on average properties of the plasma, and
cannot resolve variations, such as ion acoustic waves or ion-ion
spatial correlations.  The time resolution is also limited by the
time of flight for electrons to reach the detector.

For detecting electrons, there are currently no options besides
charged particle diagnostics.  For probing ions, however, optical
methods offer a powerful alternative that can provide {\it in situ},
non-destructive measurements, with excellent spatial, temporal, and
spectral resolution.  Absorption imaging \cite{scg04} has been used
to probe ion-ion equilibration and expansion of the plasma during
the first few microseconds after photoionization.  Fluorescence
monitoring has given information on the plasma expansion dynamics
\cite{cdd05} from shortly after photoionization to 50\,$\mu$s later.
Both techniques have great potential to study phenomena such
as ion collective modes \cite{mur00}, 
shock waves \cite{rha03}, recombination, and particle-particle
spatial correlations \cite{mbh99}.

Absorption imaging is an adaptation of one of the most powerful
techniques for studying laser cooled and trapped neutral atoms
\cite{mvs99}.  Implementation of this probe to study ultracold plasmas
requires much higher temporal resolution than typically needed for
neutral atom experiments because the plasma evolves quickly - in as
short as nanoseconds in some cases.  It is also essential to perform
experiments with an ion whose ground state possesses electric-dipole
allowed transitions in the optical regime where cameras and lasers are
available.  Alkaline-earth elements possess this property, which is
why they are used in quantum computing and optical frequency standard
experiments using trapped ions.  Strontium is a good choice for plasma
experiments because its laser cooling is well-developed \cite{nsl03}
and imaging can be performed using the Sr$^+$ ${^2S_{1/2}} \rightarrow
{^2P_{1/2}}$ transition at 422\,nm (Fig.\ \ref{energylevels}B).
Calcium is also a good choice because it has a similar level structure
\cite{cdd05,cdd05physplasmas}.

\begin{figure}
\centering
 \includegraphics[width=3in,clip=true, trim=0 00 0 0]{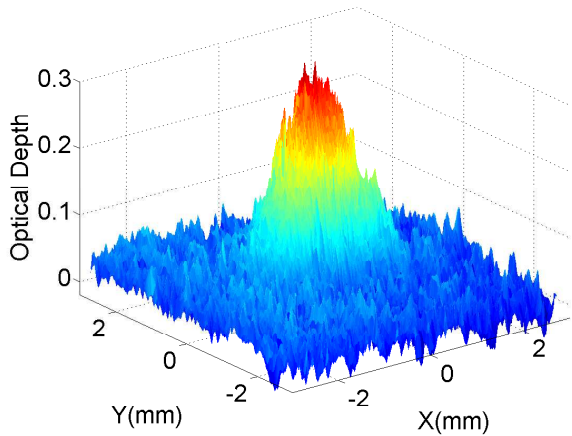}\\
\caption{Optical depth of a strontium ultracold neutral plasma.
  The delay between the formation of the plasma and
image exposure is $85$~ns.  The plasma contains
  $7 \times 10^7$ ions and the initial 
  peak ion density is $\rho_{0\mathrm i}=2 \times 10^{10}$~cm$^{-3}$.
    Reused with permission from \cite{scg04}. Copyright
2004, American Physical Society.  }\label{image}
\end{figure}

To record an absorption image of the plasma, a collimated laser
beam, tuned near resonance with the principal transition in the ions
(Fig.\ \ref{energylevels}B), illuminates the plasma and falls on an
image--intensified CCD camera.  Ions scatter photons out of the
laser beam and create a shadow that is recorded by the camera.  The
optical depth ($OD$) is defined in terms of the image intensity
without the plasma ($I_{0}$) and with the plasma present ($I$) as
\begin{eqnarray}\label{ODexperiment}
OD(x,y)&=&{\rm ln}\left[I_{0}(x,y)/I(x,y)\right]\,.
\end{eqnarray}
 Figure \ref{image} shows a typical absorption
image.
By varying the delay $\Delta t$ between the formation of the plasma
and image exposure, the time-evolution of the plasma can be studied.
The minimum camera exposure gate width for standard intensified CCD
cameras is $\sim 10$ ns.

\label{spectrumsection}
\label{spectrumanalysis}

\begin{figure}
\centering
\includegraphics[width=3in,clip=true]{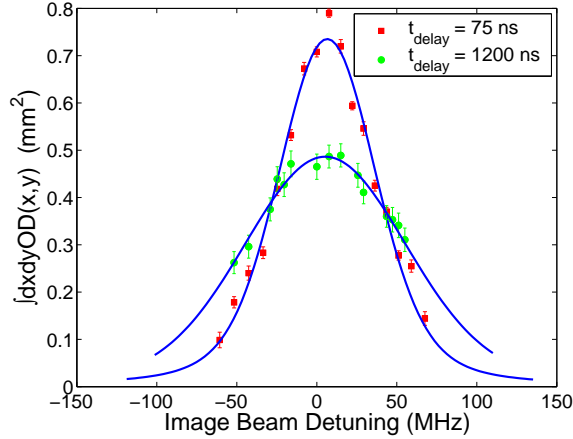}\\
\caption{Absorption spectra of ultracold neutral plasmas (Eq.\
(\ref{ODintegral})).
  The frequency is with respect to a Doppler-free absorption
  feature in a strontium discharge cell.
  Both spectra correspond to $T_e=56$~K and
  an initial peak plasma density of
  $\rho_{0\mathrm i}=2 \times 10^{10}$~cm$^{-3}$.
  Data are fit with Voigt profiles (Eq.\
  (\ref{equationtieffspectrum})). Note the increase in
  linewidth for longer delay $\Delta t$ (see Eq.\ (\ref{equationtieffrelatedtoexpansion})
  in appendix \ref{appendixabsorptionimaging}).
  Reused with permission from \cite{kcg05}. Copyright
2005, Institute of Physics.  }\label{spectrum}
\end{figure}

Valuable information  is provided by the absorption spectrum, which
can be recorded by varying the frequency of the probe beam $\nu$
\cite{scg04}. For an average spectrum of the entire plasma, the
optical depth is integrated over the plane perpendicular to the
laser,
\begin{eqnarray}\label{ODintegral}
S(\nu)\equiv \int dx\,dy\,\, OD(x,y)\,.
\end{eqnarray}
In reality the camera pixels are summed, and typical averaged absorption
spectra are shown in Fig.\ \ref{spectrum}.

In order to quantitatively interpret the spectra, it is important to
account for the inhomogeneous density distribution of the plasma and
other effects that will be discussed in subsequent sections. Among
them are thermal ion motion, plasma expansion, and the lack of
global thermal equilibrium for the ions.  Detailed descriptions of
the spectral analysis were given in \cite{kcg05} and \cite{lcg06}.
Further aspects are also discussed in App.\
\ref{appendixabsorptionimaging}. The $OD$  can be related to
underlying physical parameters according to Beer's law,
\begin{eqnarray}\label{ODtheory}
OD(x,y)&=&\int dz \hspace{.025in}
       \rho_\mathrm{i}(x,y,z) \alpha[\nu, T_\mathrm{i}(r)],
\end{eqnarray}
 where
 $\rho_\mathrm{i}(x,y,z)$ is the ion density, and $\alpha[\nu, T_\mathrm{i}(r)]$ is the
 ion absorption cross section at
 the image beam
frequency, $\nu$. The  absorption cross section is a function of ion
kinetic energy, or temperature, due to Doppler broadening. The ion
temperature varies with density, so for a spherically symmetric but
inhomogenous plasma (e.g., for a gaussian plasma cloud as in
\eq{gaussian-cloud}) $T_\mathrm{i}=T_\mathrm{i}(r)$ and therefore
$\alpha$ varies with position. Equation (\ref{ODtheory}) can be used
to relate the spectrum to the density weighted average of the
absorption cross section
\begin{eqnarray}\label{ODintegral2}
S(\nu)&=&\int d^3r \hspace{.025in}
       \rho_\mathrm{i}(r) \alpha[\nu, T_\mathrm{i}(r)].
\end{eqnarray}

The absorption cross section $\alpha[\nu,T_\mathrm{i}(r)]$ for ions
whose motion can be described by a temperature $T_\mathrm{i}$ is
given by the Voigt profile (see appendix
\ref{appendixabsorptionimaging}, Eq.\
(\ref{absorptioncrossection})).  When this expression, generalized
to include the effect of the plasma expansion (Eq.\
(\ref{absorptionfull})) is substituted into the expression for the
spectrum (Eq.\ (\ref{ODintegral2})), the result is reasonably
complicated.  However, to a good approximation \cite{kcg05,lcg06},
the density-averaged spectrum can be fit by a single Voigt profile
characterized by an effective ion temperature ($T_\mathrm{i,eff}$)
that describes all the Doppler broadening due to ion motion (For the
derivation, see appendix \ref{appendixabsorptionimaging}),
\begin{eqnarray} \label{equationtieffspectrum}
  \hspace{-.25in}S(\nu)&=&N_i {3^* \lambda^2 \over
  {2\pi}}{\gamma_0 \over \gamma_\mathrm{eff}}\int d s {1 \over 1+ 4\left( { \nu-s \over \gamma_\mathrm{eff}/2\pi}\right)^2} {1
  \over \sqrt{2\pi} \sigma_D(T_\mathrm{i,eff})}
  {\rm exp}\left[-{(s-\nu_0)^2 \over
  2\sigma_D(T_\mathrm{i,eff})^2}\right]\,, \nonumber \\
\end{eqnarray}
where $\sigma_D(T)=\sqrt{k_B T/m_\mathrm{i}}/\lambda$ is the Doppler
width, and $\gamma_\mathrm{eff}=\gamma_0+\gamma_\mathrm{ins}$ is the
effective Lorentzian linewidth due to the natural linewidth
$\gamma_{0}$ of the transition and any instrumental linewidth
$\gamma_\mathrm{ins}$. The center frequency of the transition is
$\nu_0=c/\lambda$, and the ``three-star" symbol \cite{sie86} is a
numerical factor that accounts for the polarization state of the
ions and the imaging light.

With reasonable assumptions valid for a wide range of plasma
parameters, $T_\mathrm{i,eff}$ can be related to the average ion
temperature in the plasma, $T_\mathrm{i,ave}=\int d^3r
\hspace{.025in} T_i(r)$, but the relation varies as the plasma
evolves. During the first microsecond after photoionization and
while disorder-induced heating (Sec. \ref{section_DIH}) takes place,
the plasma is not in local thermal equilibrium (LTE). (See Section
 \ref{sectionionequilibration}.)  Strictly speaking,
 ion temperature is not a well defined concept at in this situation.  However,
 $T_\mathrm{i,eff}$ can always be related to the average ion kinetic
 energy and the rms ion velocity along the laser beam
 through $k_BT_\mathrm{i,eff}=\frac{2}{3}\langle E_\mathrm{kin}\rangle=m_i
u_{z,rms}^2$.

Once LTE is established, it is possible to assign an ion temperature
reflecting thermal motion, $T_\mathrm{i}(r)$, but it varies with
position because the plasma lacks global thermal equilibrium
\cite{ppr04PRA,lcg06}. The approach to global thermal equilibrium is
governed by a heat diffusion equation. This leads to a equilibration
time equal to
\begin{equation}\label{hydscale}
\tau_\mathrm{global}\approx \sigma^2/D_{th}\,,
\end{equation}
where the heat diffusion coefficient for a strongly coupled plasma
is predicted to be on the order of \cite{dns98}
\begin{equation}\label{heatdiffusioncoefficient}
D_{\mathrm{th}}\sim a^2 \omega_{\mathrm{p,i}}\,.
\end{equation}
This depends on the plasma density through the Wigner-Seitz radius
and the ion plasma oscillation frequency. $\tau_\mathrm{global}$ is
typically greater than 100\,$\mu$s, which is longer than experiments
or simulations have probed.




The average ion temperature just after establishment of LTE, $T_\mathrm{i,ave}^{LTE}$,
 is an important characteristic scale
for thermal ion energy. For times much less than
\begin{eqnarray}\label{texpansion}
t_\mathrm{exp}= \sqrt{\frac{\sigma^{2}_{0}m_\mathrm{i}}{k_{B}T_\mathrm{e0}}}
 \sqrt{\frac{T_\mathrm{i,ave}^{LTE}}{
T_e}}\equiv \tau_\mathrm{exp} \sqrt{\frac{T_\mathrm{i,ave}^{LTE}}{T_e}},
\end{eqnarray}
 the Doppler shift due to plasma expansion is much less than the
 Doppler width due to thermal ion motion, and the expansion can be
 neglected. Here,
 $\tau_\mathrm{exp}$ is the characteristic time scale
 of plasma expansion which emerges from the kinetic theory of ultracold
 plasmas. (See \eq{exptime} in Section \ref{colless_kin}.)  References \cite{kcg05} and \cite{lcg06} show through
 numerical simulation that under these conditions,
 $T_\mathrm{i,eff}\approx CT_\mathrm{i,ave}$, where $C=0.95\pm0.05$.

At later times, the expansion is not negligible. But for a broad
range of plasma initial conditions, the expansion is quite simple
and can be described by assuming quasi-neutrality and a self-similar
expansion of the plasma (Section \ref{colless_kin}). The effect of
the expansion velocity on the Doppler width can then be incorporated
into the effective temperature of \eq{equationtieffspectrum} as
shown  in appendix \ref{appendixabsorptionimaging}.

Regardless of the exact spatial dependence of $T_\mathrm{i}(\mathbf
r)$, the integral of the spectrum for the entire plasma is
proportional to the number of ions $N_\mathrm{i}$ in the plasma and
the natural linewidth of the transition. This provides an absolute
calibration of the signal through
\begin{eqnarray}\label{spectrumintegral}
\int d\nu \, S(\nu)&=&\int d\nu \, d^3r \,
       \rho_\mathrm{i}(\mathbf r) \alpha[\nu, T_\mathrm{i}(\mathbf r)]=\frac{3^* \lambda^2 N_\mathrm{i}
       \gamma_0}{8\pi}\,.
\end{eqnarray}

If Eq.  (\ref{ODintegral}) is modified so that the \textit{OD} is
integrated over a subsection of the plasma image, then
$T_\mathrm{i,eff}$ describes ion motion in the region bounded by the
integration limits and extending along the propagation direction of
the imaging laser.  The integral over the spectrum is then
proportional to the number of ions in that region.  Analysis of
spectra in annular regions was described in \cite{csl04,lcg06} and
proved very useful for studying kinetic energy oscillations (Section
\ref{sectionionequilibration}).

\begin{figure}
\centering
 \includegraphics[width=4in,clip=true, trim=0 00 0 0]{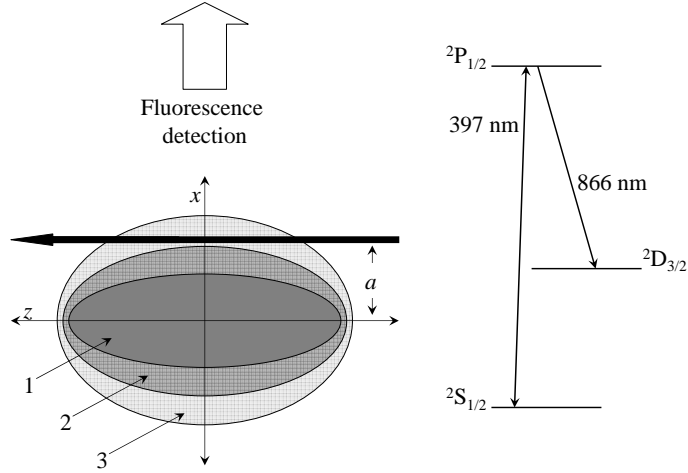}\\
\caption{Fluorescence detection in an expanding  ultracold neutral
calcium plasma. Left - The plasma is elongated along z. Regions 1-3
occur sequentially in time and show how the plasma expands
predominantly in the x-y plane. The laser excitation probe
propagates parallel to the z axis. Right - Electronic levels of the
calcium ion important for the experiment. Reused with permission
from \cite{cdd05physplasmas}. Copyright 2005, American Institute of
Physics.
  }\label{figurefluorescencedetection}
\end{figure}

In fluorescence experiments, which have been performed with calcium
\cite{cdd05,cdd05physplasmas}, fluorescence from a cylindrical region
formed by the intersection of the plasma with a tightly focused laser
beam is collected on a photomultiplier tube.  (See Fig.\
\ref{figurefluorescencedetection}.)  This approach has the advantage
of recording the full time evolution of the system with each
photoionizing pulse, but repeated scans must be taken with the laser
at different positions in order to map the spatial distribution.

When the only Doppler-broadening of the optical transition arises from thermal motion of the ions,
and the ion velocity distribution along the laser beam
can be written as a Maxwellian of the form $P(v_z)=
\exp\left(-v_z^2/2v_\mathrm{th}^2\right)/\sqrt{2\pi}v_\mathrm{th}$,
the amplitude of the fluorescence signal on resonance is related to
the thermal velocity, $v_\mathrm{th}$,
through \cite{cdd05}
\begin{equation}\label{equationfluoressignalrelatedtothermal}
    S\propto \frac{1}{v_\mathrm{th}} \exp\left(-\frac{b^2}{2v_\mathrm{th}^2}\right)
    \textrm{erfc}\left(\frac{b}{\sqrt{2}v_\mathrm{th}}\right),
\end{equation}
where $b=\gamma \lambda /2$ is the velocity that corresponds to a Doppler shift equal to the natural
linewidth of the transition, $\gamma$. If the temperature of the ions is known at some time, the
signal can be calibrated and the evolution of the ion temperature can be extracted \cite{cdd05}.
When Doppler broadening due to the expansion velocity of the ions becomes important,
an analogous expression can be derived that allows one to obtain information on the expansion
dynamics \cite{cdd05physplasmas}.

%% file: theo_intro-jm.tex
In the last decades, a variety of methods have been developed to
describe equilibrium states \cite{han73,dew76,pot99a,vsk04,fgb04},
relaxation processes \cite{zwi99,mbm01}, transport properties
\cite{red97,nfc98}, and atomic processes
\cite{kli81,sgj90,bir91,obb96} of plasmas.  These methods cover a
wide parameter space, ranging from cold, low-density plasmas in
gas-discharges to warm dense matter systems.  A typical ultracold
plasma behaves classically, which facilitates its description.
Indeed, the low particle density in an ultracold plasma implies a
Br\"uckner parameter $r_{\rm s}= (3/4\pi\rho_{\rm e})^{1/3}/a_{0}
\gg1$ and an electronic Fermi energy $E_{\rm F}\ll k_{\rm B} T_{\rm
e}$.  Hence, the values of both of these parameters justify a
classical treatment of the free charges.  Moreover, due to the low
electron temperatures, recombination forms atoms in highly excited
states, such that electron-atom collisions \cite{mke69} and
radiative transitions \cite{fvr03} can be described classically as
well.  One might be tempted to conclude that a macroscopic treatment
based on kinetic theory incorporating some collision phenomena may
suffice to describe ultracold plasmas.

However, as it has turned out this is not true. Even a classical
description of ultracold neutral plasmas is far from trivial and
poses several challenges.  First of all (and this is of course one
reason why ultracold plasmas are attractive) the ionic plasma
component may be strongly coupled, rendering common kinetic theories
for weakly coupled plasmas inapplicable. Strongly coupled problems
are usually tackled numerically with a molecular-dynamics (MD)
approach.  Due to the open boundary conditions and long evolution
times of ultracold plasmas, however, a full MD description of the
plasma dynamics constitutes an intractable computational task
currently. Moreover, the highly non-equilibrium plasma dynamics
exhibits different relaxation processes on very different time
scales, which all have to be appropriately taken into account for a
reliable and realistic understanding of an ultracold plasma.

Hence, one may ask if it is possible at all to describe ultracold
plasmas on a macroscopic level based on kinetic and hydrodynamic
formulations. This is indeed possible to some
extent,  as we will detail in
Section \ref{sectionhydrodyn}.
Starting from a collisionless description of the plasma, additional
effects such as inelastic collisions and strong ion correlations are
successively incorporated into the model.  More sophisticated, but
also technically more demanding, approaches, which describe the
plasma dynamics on a microscopic level, are introduced in Section
\ref{micro}. They can be used and are necessary to assess the
validity and limitations of the macroscopic descriptions.

%% file: hydrodyn2-jm.tex
\subsubsection{Collisionless plasma dynamics}
\label{colless_kin} If we assume the Coulomb coupling parameter of
both the electrons and the ions to be much less than unity and
neglect any collisional processes in the plasma, the dynamical
equations describing the one-particle phase space density
$f_{\alpha}\left({\bf{r}}_{\alpha},{\bf{v}}_{\alpha}\right)$
\cite{lib98} simplify to the well known Vlasov equations
\showlabel{vlasov} \beq \label{vlasov} \frac{\partial
f_{\alpha}}{\partial t}+{\bf{v}}_{\alpha}\frac{\partial
f_{\alpha}}{\partial {\bf{r}}_{{\alpha}}} -
m_{\alpha}^{-1}\frac{\partial f_{\alpha}}{\partial
{\bf{v}}_{{\alpha}}} q_{\alpha}\frac{\partial
\varphi\left({\bf{r}}_{\alpha}\right)}{\partial {\bf{r}}_{\alpha}} =
0 \; .  \eeq Here, $\alpha = {\rm e,i}$ for electrons and ions,
respectively\footnote{Here and in the remainder of this section,
greek indices denote particle species while roman indices label
individual particles of a given species.}, $m_\alpha$ and
$q_{\alpha}$ are the mass and charge of species $\alpha$, and
$\varphi$ is the total mean-field potential determined by the
Poisson equation \showlabel{poisson} \beq \label{poisson} \Delta
\varphi=\frac{e}{\varepsilon_0}\left(\rho_{\rm e}-\rho_{\rm i}\right)\;, \eeq with
$\rho_{\alpha}=\int f_{\alpha}d{\bf v}_{\alpha}$.  As a paradigmatic
problem of plasma physics, the collisionless plasma expansion into
vacuum has been considered for a long time \cite{gpp66,ssc87} and
solutions of Eqs.\ (\ref{vlasov}) have been investigated in a number
of recent publications \cite{dse98,moh03}. Ultracold plasmas for the
first time allow for a combined theoretical and experimental study
of this problem.  Clearly, for the present situation of potentially
strongly coupled plasmas,  the Vlasov equations alone will not
suffice. But they will serve as the basis of our hydrodynamic
approach. A suitable approximate treatment of correlation effects
will subsequently be added step by step, to clarify the influence of
correlations on the expansion dynamics.

Due to its nonlinearity a general, closed-form analytical solution
of Eqs.\ (\ref{vlasov}) cannot be given.  Here we will employ the additional
assumption of quasineutrality \cite{ssc87}, under which a large set of
self-similar solutions exists.  Of particular interest in the
context of ultracold plasmas is the case of a Gaussian phase space
density (see Eq.(\ref{gauansatz})), discussed in \cite{dse98}, since this is the typical
initial shape of the plasma created in a magneto-optical trap.  For
a spherically symmetric plasma it describes a local equilibrium in
velocity space, which holds for arbitrarily long times of the plasma
evolution.  We will use a further approximation, enforcing a
local equilibrium for the electrons (see Eq.(\ref{elquasistat})), in order to treat more general
cases of strongly coupled or non-symmetric plasmas as discussed
below.  Due to the very small mass ratio $m_{\rm e}/m_{\rm i}$, the
timescale for equilibration of the electrons is  much smaller than
the timescale for motion of the ions.  Exploiting this fact, one may
safely use an adiabatic approximation for the electrons, assuming
instant equilibration, such that the electronic phase space density
is given by a quasistationary distribution \showlabel{elquasistat}
\bea \label{elquasistat}
f_{\rm{e}}\left({\bf{r}},{\bf{v}},t\right)&=&f_{\rm{e}}^{\rm{(qs)}}\left({\bf{r}},{\bf{v}},T_{\rm{e}}(t)\right)\nonumber\\
&=&\rho_{\rm
e}\left({\bf{r}},t\right)\phi_{\rm{e}}^{\rm{(qs)}}\left({\bf{v}},T_{\rm{e}}(t)\right)\propto\rho_{\rm
e}\left({\bf{r}},t\right)\exp{\left(-\frac{m_{\rm e}v^2}{2k_{\rm
B}T_{\rm e}(t)}\right)}\;, \eea
where the time dependence of the
velocity distribution function $\phi_{\rm e}^{\rm (qs)}$ is only
implicit in the time dependence of $T_{\rm e}$.

Substituting this expression into the electronic Vlasov equation
(\ref{vlasov}) together with the quasineutrality condition
($\rho_{\rm e}({\bf r})\approx\rho_{\rm i}({\bf r})$) allows one to
express the mean-field potential in terms of the ion density
\showlabel{quasineut} \beq \label{quasineut} e\frac{\partial
\varphi}{\partial
{\bf{r}}}=k_{\rm{B}}T_{\rm{e}}\rho_{\rm{e}}^{-1}\frac{\partial
\rho_{\rm{e}}}{\partial {\bf{r}}}\approx
k_{\rm{B}}T_{\rm{e}}\rho_{\rm{i}}^{-1}\frac{\partial
\rho_{\rm{i}}}{\partial {\bf{r}}} \; , \eeq which together with the
ionic Vlasov equation (\ref{vlasov}) yields a closed set of
equations for the ionic phase space density $f_{\rm i}$.  Note, that
the condition $\rho_{\rm e}({\bf r})\approx\rho_{\rm i}({\bf r})$
does not imply the space charge potential to be zero. Rather, it
replaces the Poisson equation (\ref{poisson}) by Eq.\
(\ref{quasineut}). In appendix \ref{appb} we provide a more detailed
account of this fact, which may seem somewhat contradictorily and is
rarely discussed in the literature.

For a Gaussian spatial plasma density, Eq.\ (\ref{quasineut})
implies
a linear force such that the simple ansatz \showlabel{gauansatz}
\beq \label{gauansatz}
f_{\rm{i}}\propto\exp{\left(-\sum_k\frac{r_k^2}{2\sigma_k^2}\right)}\exp{
\left(-\sum_k\frac{m_{\rm{i}}(v_k-\gamma_kr_k)^2}{2k_{B}T_{{\rm{i}},k}}\right)}
\; \eeq provides a selfsimilar solution of the ionic kinetic
equation. Here the index $k = x,y,z$ labels the (cartesian)
components, $\sigma$ is the rms-radius of the Gaussian spatial
distribution, and $\gamma$ is a parameter determining the local mean
of the velocity distribution, i.e. the hydrodynamic ion velocity.
Coincidentally, the Gaussian form (\ref{gauansatz}) corresponds to
the initial distribution of ions, which reflects the Gaussian
density profile and zero mean velocity (i.e.\ $\gamma = 0$) of atoms
in the magneto-optical trap. Hence, within the collisionless
quasineutral approximation, the expanding plasma retains its shape
for all times, and its evolution can be parameterized by the
macroscopic parameters $\sigma_k$, $\gamma_k$, $T_{{\rm i},k}$ and
$T_{\rm e}$ only.

Considering a spherically symmetric plasma of Gaussian form as in
\eq{gaussian-cloud} and substituting the ansatz Eq.\
(\ref{gauansatz}) into Eq.\ (\ref{vlasov}) yields the following set
of equations for the plasma parameters \showlabel{hydro1}
\begin{subequations}\label{hydro1}
\begin{eqnarray}
\label{hydro1a}
\frac{\partial}{\partial t}\sigma^2&=&2\gamma\sigma^2 \;,  \\
\label{hydro1b} \frac{\partial}{\partial
t}\gamma&=&\frac{\left(k_{{\rm B}}T_{{\rm e}}
+k_{{\rm B}}T_{{\rm i}}\right)}{m_{{\rm i}}\sigma^2}-\gamma^2\;,  \\
\label{hydro1c} \frac{\partial}{\partial t} (k_{{\rm B}}T_{{\rm
i}})&=&-2\gamma
k_{{\rm B}}T_{{\rm i}}\;,  \\
\label{hydro1d} \frac{\partial}{\partial t} (k_{{\rm B}}T_{{\rm
e}})&=&-2\gamma k_{{\rm B}}T_{{\rm e}}\;.
\end{eqnarray}
\end{subequations}
In addition to the conserved total energy $E_{\rm
tot}=\frac{3}{2}N_{\rm i}k_{\rm B}(T_{\rm e}+T_{\rm
i})+\frac{3}{2}N_{\rm i}m_{\rm i}\gamma^2\sigma^2$ Eqs.\
\ref{hydro1} possess two integrals of motion $\sigma^2T_{\rm e}={\rm
const.}$ and $\sigma^2T_{\rm i}={\rm const.}$, reflecting the
adiabatic cooling of both the electrons and the ions during the
plasma expansion. Using these constants of motion, a simple
analytical solution of Eqs.\ (\ref{hydro1}) describing the
collisionless plasma expansion can be found \showlabel{expansion}
\begin{subequations}\label{expansion}
\bea
\label{expansiona}
\sigma^2(t)&=&\sigma^2(0)\left(1+t^2/\tau_{\rm exp}^2\right)\;, \\
\label{expansionb}
\gamma(t)&=&\frac{t/\tau_{\rm exp}^2}{1+t^2/\tau_{\rm exp}^2}\;, \\
\label{expansionc}
T_{\rm i}(t)&=&\frac{T_{\rm i}(0)}{1+t^2/\tau_{\rm exp}^2}\;, \\
\label{expansiond}
T_{\rm e}(t)&=&\frac{T_{\rm e}(0)}{1+t^2/\tau_{\rm exp}^2}\;,
\eea
\end{subequations}
where the characteristic plasma expansion time is given by
\showlabel{exptime} \beq \label{exptime} \tau_{\rm
exp}=\sqrt{\frac{m_{\rm i}\sigma^2(0)}{k_{\rm B}T_{\rm e}(0)}}\;,
\eeq and the expansion velocity is \beq \label{expvelocity}
\mathbf{u}(\mathbf{r},t)=\gamma(t) \mathbf{r}\;. \eeq Measurements
of the average density dynamics for ultracold plasmas are in very
good agreement with these simple relations for high initial electron
temperature ($T_{\rm e}(0)> 70$\,K) \cite{kkb00}. However, the same
experiments have revealed significant deviations from this average
density dynamics for lower electron temperature.  As discussed in
\cite{rha02}, they mainly arise from inelastic collisions, i.e.,
from the formation of Rydberg atoms during the plasma expansion.
This is one consequence of collisions in the plasma which we will
discuss next.

\subsubsection{Collisional processes} \label{section_coll}
There are two types of collisional processes in plasmas: elastic
collisions, which do not affect the total system temperature $T_{\rm
e}+T_{\rm i}$, and inelastic collisions, which may drastically
change the temperature of the electrons.  On the shortest timescale
of several ns, elastic electron-electron collisions lead to an
equilibration of the electron velocities, giving rise to the
electronic temperature of the plasma.  Implicitly, we have already
accounted for this process in the previous section within the
adiabatic approximation Eq.\ (\ref{elquasistat}), which assumes an
instantaneous relaxation towards a Maxwellian electron velocity
distribution.  This assumption is well justified since the timescale
for electron-electron collisions is by far the shortest one for a
typical ultracold plasma.  The ion velocity distribution, on the other hand,
has the initial Maxwellian form of the atomic velocity distribution. Hence, the
ion-ion collision integral vanishes and does not change
the functional velocity dependence of the ion distribution as described by Eq.\ (\ref{gauansatz}).

While electron-electron and ion-ion collisions lead to an individual
electron and ion temperature, respectively,
 electron-ion collisions tend to
equilibrate the ion and the electron temperature.  However, elastic
electron-ion collisions may safely be neglected for the description
of a freely expanding plasma\footnote{Note that the situation is
drastically different in laser-cooled plasmas considered in Section
\ref{section_lcool}}, since the corresponding ion heating rate
\cite{Spitzer} is much less than the rate of the initial disorder
induced ion heating to be discussed in Section \ref{section_DIH}.
During later times, the expansion leads to an adiabatic cooling of
the ions, but also to a reduction of the collisional heating rate. Hence,  {\sl
elastic} electron-ion collisions remain negligible at all times
\cite{ppr05jpb}.

 This is not the case, however, for {\sl inelastic} electron-ion
 collisions due to the low temperature of the electrons, which leads to
 efficient recombination into bound Rydberg states.  The formation of
 bound states during the evolution of the plasma can be described most
 accurately within a classical molecular dynamics treatment taking
 into account all electron-ion interactions \cite{kon02,mck02}.
 However, as discussed above, such an approach is limited to short
 times due to the fact that the small timescale on which the
 electronic motion takes place has to be resolved.  Within the
 present macroscopic approach the problem is simplified significantly
 by adopting a {\sl chemical picture}, i.e., by introducing
 neutral atoms as an additional particle species.  Naturally,
 this transition is accomplished by splitting the electron-ion two-particle
 distribution function $f_{\rm ei}$ into a free and a bound part
 \cite{kli81}, where the bound part is identified with the
 distribution function $f_{\rm a}$ of the atoms.  The evolution
 equation for the atom distribution function then involves an integral
 over the three-particle distribution function describing correlations between pairs of electrons and an ion, which accounts for inelastic collisions, i.e. recombination, ionization and electron impact (de)-excitation.
 Without going into detail, the corresponding inelastic processes can
 be identified as three-body recombination, electron-impact
 ionization, and electron impact excitation or deexcitation.
 Consequently, the
 electron-ion kinetic equation (\ref{vlasov}) splits into
 separate equations for free electrons, ions, and atoms.
 Neglecting atom-atom interactions and still retaining the Vlasov
 approximation, the  kinetic equations for ions and atoms  read
 \showlabel{atoms} \bea
\label{atoms}
\frac{\partial f_{\rm{i}}}{\partial t}+{\bf{v}}_{\rm{i}}\frac{\partial
f_{\rm{i}}}{\partial {\bf{r}}_{\rm{i}}}-em_{\rm i}\frac{\partial f_{\rm i}}{\partial {\bf v}_{\rm i}}\frac{\partial \varphi}{\partial {\bf r}_{\rm i}} & = &
J_{\rm{ei}}^{\rm{(3)}}, \nonumber \\
\frac{\partial f_{\rm{a}}}{\partial
t}+{\bf{v}}_{\rm{a}}\frac{\partial f_{\rm{a}}}{\partial
{\bf{r}}_{\rm{a}}} & = & J_{\rm{ae}} \; . \eea
The collision terms
on the right-hand sides of \eq{atoms} are given by
\showlabel{Jei}
\beq \label{Jei} J_{\rm{ei}}^{\rm{(3)}}=\sum_n
K_{\rm{ion}}(n,\rho_{\rm{e}}({\bf{r}}_{\rm{i}}),T_{\rm{e}})f_{\rm{a}}(n,{\bf{p}
}_{\rm{i}},{\bf{r}}_{\rm{i}})-\sum_n
K_{\rm{tbr}}(n,\rho_{\rm{e}}({\bf{r}}_{
\rm{i}}),T_{\rm{e}})f_{\rm{i}}({\bf{p}}_{\rm{i}},{\bf{r}}_{\rm{i}})
\eeq and \showlabel{Jae} \bea \label{Jae}
J_{\rm{ae}}&=&\sum_p\left[K_{\rm{bb}}(p,n,\rho_{\rm{e}}({\bf{r}}_{\rm{a}}),
T_{\rm{e}})f_{\rm{a}}(p,{\bf{p}}_{\rm{a}},{\bf{r}}_{\rm{a}})-K_{\rm{bb}}(n,p
,\rho_{\rm{e}}({\bf{r}}_{\rm{a}}),T_{\rm{e}})f_{\rm{a}}(n,{\bf{p}}_{\rm{a}},{
\bf{r}}_{\rm{a}})\right]\nonumber
\\
&&+K_{\rm{tbr}}(n,\rho_{\rm{e}}({\bf{r}}_{\rm{a}}),T_{\rm{e}})f_{\rm{i}}({\bf{p}
}_{\rm{a}},{\bf{r}}_{\rm{a}})-K_{\rm{ion}}(n,\rho_{\rm{e}}({\bf{r}}_{\rm{a}}),
T_{\rm{e}})f_{\rm{a}}(n,{\bf{p}}_{\rm{a}},{\bf{r}}_{\rm{a}}) \; ,
\eea
where the rate coefficients $K_{\rm tbr}$, $K_{\rm ion}$ and
$K_{\rm bb}$ for three-body recombination, electron impact
ionization and electron impact induced bound-bound transitions
depend on the atomic principal quantum numbers $n$ and $p$, the
electronic density $\rho_{\rm e}$, and the temperature $T_{\rm e}$.
Since these coefficients describe three-body processes, no exact
analytical expressions for the rates are known.  For the highly
excited Rydberg states in ultracold plasmas an accurate description
is expected from the rate coefficients of \cite{mke69}, which have
been obtained by fitting analytical functions to the results of
classical trajectory Monte Carlo calculations.  Since the collision
integrals Eqs.\ (\ref{Jei}) and (\ref{Jae}) introduce a nonlinearity
in the spatial particle densities, an exact selfsimilar solution of
the corresponding kinetic equations can no longer be found.
Nevertheless, we assume the phase space distribution to be the same
for atoms and for ions and to still be given by Eq.\
(\ref{gauansatz}), which was shown to yield a good description of
freely expanding ultracold plasmas by comparison with more
sophisticated calculations \cite{ppr04archive}.  Based on this
approximation and from the  moments of the distribution function
\showlabel{moments}
\begin{subequations}
\label{moments}
\bea
\label{momentsa}
\left<r^2\right>&=&3\sigma^2=N_{\rm{i}}^{-1}\int
r^2f_{\rm{i}}({\bf{r}}_{\rm{i}},{\bf{v}}_{\rm{i}},t)\:d{\bf{r}}_{\rm{i}}d{\bf{v}
}_{\rm{i}} \\
\left<{\bf{v}}{\bf{r}}\right>&=&3\gamma\sigma^2=N_{\rm{i}}^{-1}\int
{\bf{v}}_{\rm{i}}{\bf{r}}_{\rm{i}}f_{\rm{i}}({\bf{r}}_{\rm{i}},{\bf{v}}_{\rm{i}}
,t)\:d{\bf{r}}_{\rm{i}}d{\bf{v}}_{\rm{i}} \\
\left<v^2\right>&=&3\left(\frac{k_{B}T_{\rm i}}{m_{\rm i}}+\gamma^2\sigma^2\right)=N_{\rm{i}}^{-1}\int
v_{\rm{i}}^2f_{\rm{i}}({\bf{r}}_{\rm{i}},{\bf{v}}_{\rm{i}},t)\:d{\bf{r}}_{\rm{i}
}d{\bf{v}}_{\rm{i}} \; .
\eea
\end{subequations}
we finally obtain simple evolution equations for the parameters of the plasma
\showlabel{hydro2}
\begin{subequations}\label{hydro2}
\begin{eqnarray}
\label{hydro2a}
\frac{\partial}{\partial t}\sigma^2&=&2\gamma\sigma^2  \\
\label{hydro2b}
\frac{\partial}{\partial t}\gamma&=&\frac{N_{\rm i}}{N_{\rm i}+
N_{\rm a}}\frac{k_{{\rm B}}T_{{\rm e}}
+k_{{\rm B}}T_{{\rm i}}}{m_{{\rm i}}\sigma^2}-\gamma^2 \\
\label{hydro2c}
\frac{\partial}{\partial t} (k_{{\rm B}}T_{{\rm i}})&=&-2\gamma k_{{\rm B}}T_{{\rm i}}\\
\label{hydro2d}
\frac{\partial}{\partial t} {\mathcal N}_{\rm a}(n)& =& \sum_{p\neq n}\left[
R_{\rm bb}(p,n){\mathcal N}_{\rm a}(p)- R_{\rm bb}(n,p) {\mathcal N}_{\rm a}(n) \right] \nonumber \\
&&+R_{\rm tbr}(n)N_{\rm i}-R_{\rm ion}(n) {\mathcal N}_{\rm a}(n) \\
\label{hydro2e}
E_{{\rm tot}} & = &\frac{3}{2}N_{\rm i}\left[k_{\rm B}T_{\rm e}+k_{\rm B}
T_{\rm i} +m_{\rm i} \gamma^2\sigma^2\right]-\sum_{n}{{\mathcal N}_{\rm a}(n)\frac{{\mathcal R}}{n^{2}}} = \mbox{const.}\; ,
\end{eqnarray}
\end{subequations}
where $\mathcal R$ is the Rydberg constant which has the value
${\mathcal R}=13.6$\,eV in electron Volts and ${\mathcal R}=2.18
\times 10^{-18}$\,J in SI-units. The number of atoms with principal
quantum number $n$ is denoted by ${\mathcal N}_{\rm a} (n)$, where
$\sum_n {\mathcal N}_{\rm a} (n)= N_{\rm a}$, and the density
averaged collision rates are defined as \showlabel{avrate} \bea
\label{avrate} R_{\rm{bb}}(n,p)&=&N_{\rm{i}}^{-1}\int
\rho_{\rm{i}}({\bf{r}})K_{\rm{bb}}\left(n,p,\rho_{\rm{i}}({\bf{r}},T_{\rm{e}})
\right)d{\bf{r}}\nonumber\\
R_{\rm{ion}}(n)&=&N_{\rm{i}}^{-1}\int
\rho_{\rm{i}}({\bf{r}})K_{\rm{ion}}\left(n,\rho_{\rm{i}}({\bf{r}},T_{\rm{e}})
\right)d{\bf{r}}\nonumber\\
R_{\rm{tbr}}(n)&=&N_{\rm{i}}^{-1}\int
\rho_{\rm{i}}({\bf{r}})K_{\rm{tbr}}\left(n,\rho_{\rm{i}}({\bf{r}},T_{\rm{e}})
\right)d{\bf{r}} \; .
\eea
As can be seen from the energy relation \eq{hydro2e}, inelastic
collisions not only change the total number of atoms and ions, i.e.,
the degree of ionization of the plasma, but also affect the
temperature of the electronic component.  \emph{Three-body
recombination} heats the electron plasma, since one electron carries
away the excess energy gained in the recombination of another
electron with an ion.  Subsequent de-exciting
\emph{electron-Rydberg atom collisions} further heat up the
plasma, \emph{electron
impact ionization and excitation} tend to cool the electrons.

The total three-body recombination rate has a strong temperature
dependence and scales as $\sum_n R_{\rm tbr}(n)\propto T_{\rm
e}^{-9/2}$.  Hence, for very high initial electron temperatures
inelastic collisions are of minor importance and Eqs.\
(\ref{hydro2}) reduce to Eqs.\ (\ref{hydro1}), such that the plasma
expansion is well described by the simple relations Eqs.\
(\ref{expansion}).  On the other hand, at low initial $T_{\rm e}$,
when the electronic Coulomb coupling parameter approaches unity,
three-body recombination and subsequent electron-atom collisions
quickly heat up the electron gas, strongly decreasing $\Gamma_{\rm
e}$ \cite{Rob02}.  This justifies our assumption of an ideal
electron gas, inherent in the kinetic description.  The ionic
temperature, on the other hand, remains practically unaffected by
inelastic collisions.  Hence, ionic correlation effects may not
necessarily be negligible, and will be accounted for in the next section.

\subsubsection{Strongly coupled ions} \label{section_hydcoupl}
Kinetic theories of non-ideal plasmas are commonly based on a
small-$\Gamma$ expansion.  For ultracold plasmas such approaches are
clearly not applicable, since the ion component can exhibit Coulomb
coupling parameters on the order of or even much larger than unity.
Alternatively, one can combine a hydrodynamical treatment of the
expansion dynamics with numerical results for homogeneous strongly
coupled plasmas \cite{ppr04archive}.  The starting point is the
kinetic equation for the ions, which accounts exactly for ion-ion
correlations,
\showlabel{Vlapluscorr}
\beq
\label{Vlapluscorr}
\left(\frac{\partial}{\partial
t}+{\bf{v}}_{\rm{i}}\frac{\partial}{\partial
{\bf{r}}_{\rm{i}}}-\frac{e}{m_{\rm{i}}}\frac{\partial
\varphi}{\partial {\bf{r}}_{\rm{i}}}\frac{\partial}{\partial
{\bf{v}}_{\rm{i}}}\right)f_{\rm{i}}({\bf{r}}_{\rm{i}},{\bf{v}}_{\rm{i}},t)=
\frac{\partial}{\partial {\bf p}_{\rm i}}\int{\left(\frac{\partial
V_{\rm{ii}}}{\partial {\bf{r}}_{\rm{i}}}\right)c_{\rm{ii}}({
\bf{r}}_{\rm{i}},{\bf{v}}_{\rm{i}},{\bf{r}}_{\rm{i^{\prime}}},{\bf{v}}_{\rm{i^{
\prime}}},t)\:d{\bf{r}}_{\rm{i^{\prime}}}d{\bf{v}}_{\rm{i^{\prime}}}}
\; , \eeq
where $c_{\rm ii}$ is the ion-ion correlation function.  To
simplify notation we neglect inelastic collisions whose contributions
can straightforwardly be added, as described in the previous section.

As before, from Eq.\ (\ref{Vlapluscorr}) we derive the time
evolution of the moments of $f_{\rm i}$ up to second order, which
together with the ansatz \eq{gauansatz} yields \showlabel{momgau}
\begin{subequations}
\label{momgau}
\bea
\frac{\partial}{\partial t}\sigma^2&=&2\gamma \sigma^2\\
\frac{3}{2}m_{\rm{i}}\frac{\partial}{\partial
t}\gamma \sigma^2&=&\frac{3}{2}m_{\rm{i}}\gamma^2\sigma^2+\frac{3}{2}k_{\rm{B}}
T_{\rm{i}}+\frac{3}{2}k_{\rm{B}}T_{\rm{e}} \nonumber \\ &&
+\frac{1}{2N_{\rm{i}}}\int{\rho_{\rm{i}}({\bf{r}}_{\rm{i}}){\bf{r}}_{\rm{i
}}{\bf{F}}_{\rm{ii}}({\bf{r}}_{\rm{i}})\:d{\bf{r}}_{\rm{i}}}\\
\frac{3}{2}\frac{\partial}{\partial
t}\left(k_{\rm{B}}T_{\rm{i}}+m_{\rm{i}}\gamma^2\sigma^2\right)&=&\gamma k_{\rm{B
}}T_{\rm{e}}-\frac{\partial U_{\rm{ii}}}{\partial t} \; .
\eea
\end{subequations}
The correlation pressure force ${\bf F}_{\rm ii}$ and the correlation energy $U_{\rm ii}$ are obtained from
\showlabel{corrforce}
\beq
\label{corrforce}
{\bf{F}}_{\rm{ii}}=-\frac{1}{3}\left(\frac{u_{\rm{ii}}}{\rho_{\rm{i}}}+\frac{
\partial u_{\rm{ii}}}{\partial \rho_{\rm{i}}}\right)\frac{\partial \rho_{\rm{i
}}}{\partial {\bf{r}}_{\rm{i}}}\;,
\eeq
and
\showlabel{corren}
\beq
\label{corren}
U_{\rm{ii}}=\frac{1}{N_{\rm{i}}}\int{\rho_{\rm{i}}({\bf{r}}_{\rm{i}})u_{\rm{ii}}
({\bf{r}}_{\rm{i}})}\:d{\bf{r}}_{\rm{i}}\;,
\eeq
respectively, where
\showlabel{correnpp}
\beq
\label{correnpp}
u_{\rm{ii}}=\frac{e^2}{2}\rho_{\rm{i}}({\bf{r}}_{\rm{i}})\int{\frac{g_{\rm{ii}}
\left(y,\rho_{\rm{i}}({\bf{r}}_{\rm{i}})\right)}{y}\:d{\bf{y}}}
\eeq
is the mean correlation energy per particle and $g_{\rm ii}(r,\rho({\bf r}_{\rm i}))$ is the spatial pair correlation function of a homogeneous plasma of density
$\rho(\mathbf{r_i})$. Using
\showlabel{fequ}
\beq
\label{fequ}
\frac{1}{N_{\rm{i}}}\int{\rho_{\rm{i}}({\bf{r}}){\bf{r}}{\bf{F}}_{\rm{ii}}({\bf{
r}})\:d{\bf{r}}} = -\frac{1}{N_{\rm{i}}}\int{u_{\rm{ii}}
\rho_{\rm{i}}\:d{\bf{r}}}=-U_{\rm{ii}} \; ,
\eeq
Eqs.\ (\ref{momgau}) are rewritten as
\showlabel{momgau2}
\begin{subequations}
\label{momgau2}
\bea
\label{momgau2a}
\frac{\partial \sigma^2}{\partial t}&=&2\gamma\sigma^2\\
\label{momgau2b}
\frac{\partial \gamma}{\partial t}&=&\frac{k_{\rm{B}}T_{\rm{e}}+k_{\rm{B}}T_{
\rm{i}}+\frac{1}{3}U_{\rm{ii}}}{m_{\rm{i}}\sigma^2}-\gamma^2\\
\label{momgau2c}
\frac{\partial k_{\rm{B}}T_{\rm{i}}}{\partial t}&=&-2\gamma k_{\rm{B}}T_{\rm{i}}
-\frac{2}{3}\gamma U_{\rm{ii}}-\frac{2}{3}\frac{\partial U_{\rm{ii}}}{\partial
t}\\
\label{momgau2d}
\frac{\partial k_{\rm{B}}T_{\rm{e}}}{\partial t}&=&-2\gamma k_{\rm{B}}T_{\rm{e}}
\; ,
\eea
\end{subequations}
where the last equation follows from energy conservation,
\showlabel{encons} \beq \label{encons}
E_{\rm{tot}}=N_{\rm{i}}\frac{m_{\rm{i}}}{2}\left<v^2\right>+\frac{3}{2}N_{\rm{i}
}k_{\rm{B}}T_{\rm{e}}+N_{\rm{i}}U_{\rm{ii}} \; . \eeq While the
description of the influence of ionic correlations on the plasma
dynamics via the single macroscopic quantity $U_{\rm ii}$
constitutes a significant simplification of the problem, the system
of eqs.\ (\ref{momgau2}) is still not complete since it does not
contain an equation for the time evolution of $U_{\rm ii}$. An
accurate description of the evolution of $U_{\rm ii}$ on a kinetic
level is rather complicated, and beyond the capabilities of the
simple approach pursued here. A useful and reasonable approximation
for the evolution of $U_{\rm ii}$ is the so-called correlation-time
approximation \cite{bon96} \showlabel{ctapp} \beq \label{ctapp}
\frac{\partial U_{\rm{ii}}}{\partial t}\approx
-\frac{U_{\rm{ii}}-U_{\rm{ii}}^{ \rm{(eq)}}}{\tau_{\rm{corr}}} \; ,
\eeq where $\tau_{\rm corr} = \sqrt{m_{\rm i}\varepsilon_{0}/(e^2
\rho_{\rm i})} = \omega_{{\rm p, i}}^{-1}$ is the timescale on which
pair correlations relax towards their equilibrium value,
\showlabel{Ueq} \beq \label{Ueq} U_{\rm{ii}}^{\rm{(eq)}}=\int
\rho_{\rm{i}}({\bf{r}})u_{\rm{ii}}^{\rm{(eq)}}({ \bf{r}})\:d{\bf{r}}
\; , \eeq and $u_{\rm ii}^{\rm (eq)}$ is the correlation energy of a
homogeneous one-component plasma (OCP) of density $\rho_{\rm
i}(\mathbf{r})$ in thermodynamical equilibrium \cite{don99}.  It is
a special property of the Coulomb potential that the temperature
scaled correlation energy is uniquely determined by the Coulomb
coupling parameter $\Gamma_{\rm i}$, such that numerical values of
$U_{\rm ii}$ can be tabulated over a broad range of $\Gamma_{\rm
i}$'s \cite{sdd80,ng74} and accurate formulae exist.  The following
formula interpolates between the low-$\Gamma$ Abe limit and the
high-$\Gamma$ behavior
 \cite{cpo98} \showlabel{ueqanalyt}
\begin{equation} \label{ueqanalyt}
u_{\rm{ii}}^{\rm{eq}}({\mathbf{r}})=k_{\rm B} T_{\rm i} \Gamma^{3/2}\left(
\frac{A_1}{\sqrt{A_2+ \Gamma}}+ \frac{A_3}{1+\Gamma}\right) \;,
\end{equation}
with $A_1=-0.9052$, $A_2=0.6322$ and $A_3=-\sqrt{3}/2-A_1/\sqrt{A_2}$.

This expression is valid only for an OCP, i.e., a system of ions
embedded in a neutralizing negative electron background. Therefore,
ultracold plasma experiments provide an ideal testing ground for the
validity of the widely used OCP model in real two-component systems.
While the OCP model should yield a good description for $\Gamma_{\rm
e}\rightarrow 0$, we would certainly expect deviations from Eq.\
(\ref{ueqanalyt}) due to electron screening of the ion-ion
interaction for low initial electron temperatures.  For correlated
but still weakly coupled electron components  this effect is well
described within the Debye-H\"uckel theory by an exponentially
screened Yukawa-type interaction potential \showlabel{debye} \beq
\label{debye} V_{\rm ii}=\frac{e^2}{4\pi
\varepsilon_0}\frac{e^{-r/\lambda_{\rm D}}}{r}\,, \eeq where \beq
\label{screening-length} \lambda_{\rm
D}=\sqrt{\varepsilon_{0}k_{B}T_{\mathrm{e}}/(e^{2}\rho_{\mathrm{e}})}
\eeq is the Debye screening length, characterizing the mean distance
on which local charge imbalances are screened inside the plasma.  As
this screening effect introduces an additional length scale in the
plasma the equilibrium correlation energy now depends on two
parameters, $\Gamma_{\rm i}$ and $\kappa=a/\lambda_{\rm D}$,
\showlabel{Uscreened} \beq \label{Uscreened} U_{\rm ii}^{\rm
(eq)}=k_{\rm B}T_{\rm i} \Gamma_{\rm
i}\left(\tilde{U}+\frac{\kappa}{2}\right)\;. \eeq The quantity
$\tilde{U}(\Gamma_{\rm i},\kappa)$ has been obtained from molecular
dynamics simulations and tabulated over a wide range of $\Gamma_{\rm
i}$ and $\kappa$ values \cite{hfd96,hfd97}.

The range of applicability of the Debye H\"uckel is not strictly defined. The theory is
rigorously valid only for $a\ll\lambda_{\rm D}$. Ultracold plasma experiments, however, showed
that Eq.(\ref{Uscreened}) yields a surprisingly good description even for $a\approx\lambda_{\rm D}$ ($\Gamma_{\rm e}\approx1/3$) -- i.e. with only one electron per Debye sphere \cite{scg04}.

Eq.\ (\ref{ctapp}) and Eqs.\ (\ref{momgau2})  constitute a closed
set of equations, describing the time evolution of a neutral plasma
of electrons and ions. Compared to more involved approaches, such
as, e.g., Molecular Dynamics methods, the macroscopic evolution
equations are simple enough to allow for significant physical
insight into effects of ion correlation  on the plasma dynamics.
While Eq.\ (\ref{momgau2d}) shows no direct influence on the
electron temperature, the time evolution of the ion temperature is
significantly modified by the interionic correlations. In addition
to the adiabatic cooling, which is given by the first term on the
right-hand side (rhs) of (\ref{momgau2c}) in complete analogy to the
electronic component, the ionic temperature is increased (note that
$U_{\rm ii} < 0$) by the development of correlations (last term on
the rhs of (\ref{momgau2c})). This effect, which has been named
``disorder-induced heating'' or ``correlation-induced heating,'' has
been widely studied for homogeneous systems
\cite{zwi99,kon02,kon02prl,mur01,gmu03,gms03,ppr04PRA,ppr05prl}.
Ultracold plasmas not only allow for experimental tests of present
theories but also for studies of the ionic correlation dynamics in
expanding systems with a steadily changing local equilibrium state,
as described by the hydrodynamical evolution equations.

The modified time evolution of the ionic temperature leads to a
modified expansion dynamics of the plasma through the change in the
ionic pressure, which contains a correlation part in addition to the
thermal part. The development of ionic correlations also
reduces of the total electrostatic interaction energy compared to
a fully uncorrelated plasma. Together, these two effects result in the
term $U_{\rm ii}/3$ on the rhs of (\ref{momgau2b}), which
constitutes an effective negative acceleration slowing down the
plasma expansion. This term can be regarded as a time dependent
``external'' force, giving rise to an additional effective potential
in which the ions move. This potential, however, is not static, but
changes during the expansion of the plasma, and leads (due to
conservation of total energy) to an additional heating of the ions
(second term on the rhs of (\ref{momgau2c})).

Eqs.\ (\ref{momgau2a})-(\ref{momgau2c}), (\ref{hydro2d}) and
(\ref{hydro2e}) form the final set of equations for the macroscopic
parameters of the plasma state. They are quickly and easily solved
numerically.  Since the numerical effort is independent of the
number of particles involved, the hydrodynamic approach provides a
convenient method for simulations of large plasma clouds over long,
i.e., millisecond, timescales.  It can be used to quickly gain
insight into the plasma dynamics by efficiently scanning a broad
range of initial-state parameters, and it is able to simulate
plasmas that are too large to be treated effectively with molecular
dynamics methods. Moreover, and maybe even more importantly, the
kinetic model provides physical insight by reducing the plasma
dynamics to a few macroscopic parameters. However, the accuracy of
these approximate macroscopic treatments, particularly when they
describe  dynamics influenced by ion correlation, must be assessed
carefully. This can be done to some extent by comparison with
experiment, but more rigorously by developing numerically accurate
microscopic approaches relaxing most of the approximations invoked
in the macroscopic approach.

%% file: hybrid-jm.tex
\label{micro} Microscopic approaches are by definition more
accurate.  They are of great value for justifying macroscopic
descriptions, and of course are the only alternatives for plasma
conditions that preclude macroscopic descriptions.  This refers to
detailed information about the very early system evolution which is
not accessible to hydrodynamics, and the effects of strong particle
correlations, which are of particular interest in the context of
ultracold plasmas.

\subsubsection{Molecular Dynamics Simulations} \label{moldyn}
Molecular dynamics simulations of Coulomb-type interacting particles
have been employed for investigating a variety of different problems
in plasma physics and are a corner stone of astrophysical studies of
galaxy and globular clusters \cite{aar99}.  Conceptually, the
problem seems rather simple as it only involves the solution of a set
of coupled differential equations describing the classical motion of
charged particles.  However, there are several technical challenges,
which in the case of ultracold neutral plasmas arise from the open
boundary of the expanding system, the attractive nature of the
electron-ion interaction and the small mass ratio of the electrons
and the ions.

A general difficulty  for simulations of neutral plasmas  arises
from the bound state dynamics of atomic electrons, whose orbital timescale is many orders of magnitude smaller than the plasma evolution time. In particular for highly eccentric
orbits conventional integrators, such as higher order Runge-Kutta
methods, do not provide a stable solution, as they lead to a
significant drift of the total energy.  More appropriate propagation
schemes such as symplectic integrators without energy drift and
originally designed for the integration of gravitational systems,
have clear advantages, but also several draw-backs and the method of
choice depends on the particular problem to be studied.  For
simulating ultracold plasmas higher order symplectic integrators
\cite{hvm06}, predictor-corrector schemes \cite{kon02prl} and
timescaling methods combined with regularization procedures
\cite{psg06}  can yield stable and reliable results.

Although a single electron and ion pair can be propagated very
stably, many-body simulations have to fight additional
complications. Electrons are bound to different ions with different
binding energies and hence their orbital times
 can be very different. Moreover, these orbital times are
generally much smaller than the  timescale on which the  heavier
ions move.  With a typical ion timescale on the order of
microseconds, one has to cover timescales from femtoseconds to
microseconds in order to describe the initial relaxation of the
ionic component. Mazevet et al.\ \cite{mck02} and Kuzmin and O'Neil
\cite{kon02prl} have used a small ion mass and scaled the obtained
timescales to realistic experimental situations to obtain a
quantitative estimate of the experimental results.  Such a scaling,
however, does not apply to all aspects of the plasma dynamics, since
e.g., the timescales for electron-ion equilibration and ion-ion
equilibration scale with a different power of the ion mass.

Several methods to numerically tackle  very different  timescales
have been discussed in the literature (see, e.g., \cite{tbm90} and
\cite{aar99,hea94} for an overview and further references).  The
common idea of such approaches is to introduce individual timesteps
for the different particles.  The criteria determining the timestep
and the procedures for the synchronization of the different
particles can however be quite different and have to be adapted to
the specific problem under consideration.  Kuzmin and O'Neil
\cite{kon02prl,kon02} used a modified version of the method
originally developed by Aarseth \cite{aar85} to perform a full scale
simulation of ultracold neutral plasmas with a realistic xenon ion
mass and a total number of 8192 particles.

The number  of particles $N$ is generally limited by the number of
force evaluations that have to be carried out for every timestep.
Since for every particle one has to sum up $N-1$ forces of the
surrounding charges the numerical effort quickly increases as
$N\cdot (N-1)\sim N^2$ with increasing number of particles.
Simulations of large, extended plasmas typically consider a smaller
part of the system and employ periodic boundary conditions to mimic
the properties of the larger system \cite{hea94}.  For ultracold
plasmas such an approach is not very appropriate as it neglects an
important part of the dynamics -- the expansion of the plasma.  The
$N^2$-scaling of the numerical effort can, however, be reduced by
so-called hierarchical tree-methods \cite{bar86} or fast multipole
methods \cite{gro87}, originally developed to simulate globular
clusters.  Such approaches still account for the exact forces of
nearby charges but approximate the force of distant particles by the
 average force up to a certain multipole order of the corresponding
charge distribution. This  reduces the scaling of the numerical
effort with the number of particles to $\sim N\ln N$ or even $\sim
N$.

\subsubsection{Particle-In-Cell Approaches} \label{pic}
 An alternative and conceptually different method to reduce the
 numerical effort connected with the summation of the inter-particle
 forces is to neglect correlations between the charges but retain
 their meanfield interaction due to the long-range part of the Coulomb
 potential.  The so-called particle-in-cell simulations (PIC)
 \cite{bla95} have become a standard method to treat various plasma
 physics problems, such as the collective dynamics of fusion devices
 \cite{bsb99,zc02} or of cold plasmas in gas discharges  \cite{gc04,ss04,s04},
 i.e., weakly coupled plasmas.  Clearly, for ultracold neutral plasmas
 this method seems questionable, as we are interested in the regime of
 strong coupling.  However, the approach has proven to yield a
 reliable and efficient description of certain aspects of the plasma
 dynamics as long as the plasma expansion is mainly driven by the
 thermal pressure of the electrons \cite{ppr04archive}.

The striking advantage of the PIC method is the possibility to treat
very large systems, by replacing a collection of real particles by a
single super-particle, with an increased charge and mass, retaining
the
charge-to-mass ratio of the physical particle.  A single propagation
timestep of such an ensemble of super-particles consists of the
following sub-steps:
\begin{enumerate}
\item At the beginning of each propagation timestep the positions of
these super-particles are used to determine the charge density of the
system, represented on an appropriately chosen grid.  To reduce
numerical noise due the finite particle number the charges are
represented by a shape function.  Most common shape functions are
rectangles (particle in cell), tent functions (cloud in cell) or
Gaussians centered around the actual particle position.  Further
problems arise for particle weighting in curved coordinates, as
discussed in \cite{ver01}.
\item Having obtained the charge density
one evaluates the corresponding static electric field from the
Poisson equation (Eq.\ \ref{poisson}).  If the system exhibits
additional spatial symmetries this is easily done by employing
Gauss' theorem. In the general case the electric mean-field is most
efficiently evaluated within an FFT procedure.
\item The calculated electric field is
subsequently transformed to the actual particle positions taking
into account the chosen shape function, which finally allows
advancement of all particle positions and velocities according to
the classical equations of motion.
\end{enumerate}
The positions and velocities of the test-particles have no physical
meaning.  The relevant quantities are the phase space distributions
of the charges determined by the ensemble of test-particles.  It is
easy to see that the resulting dynamics provides a numerical
solution of the Vlasov equation (\ref{vlasov}).

Hence, the PIC treatment as described so far does not account for
any collisional processes, such as formation of Rydberg atoms and
subsequent electron-atom scattering.  Such collisions can however be
incorporated into the PIC-treatment on the basis of the collision
integrals introduced in Eqs.\ (\ref{atoms})-(\ref{Jae}).  Within
such so-called particle-in-cell-Monte-Carlo-collision simulations
(PICMCC) \label{PICMCC} \cite{bir91} the collision terms in Eq.\
(\ref{vlasov}) are evaluated by random realizations of collision
events, with a probability distribution determined by the respective
collision integral.

For the collision integrals given in Eqs.\ (\ref{Jei}) and
(\ref{Jae}) describing inelastic collisions and Rydberg atom
formation, this stochastic evaluation procedure is particularly
simple.  First, one introduces an additional species by providing
each ion with an additional label $n$, which characterizes its
internal state.  If the particle is an ion, then $n=0$, and $n>0$
for atoms, where $n$ denotes the principle quantum number of the
atom. During each timestep of length $\Delta t$, an ion at position
$\mathbf{r}$ is transformed into an atom with probability $P_{\rm
tbr} = \Delta t \sum_n K_{\rm tbr}(n,\rho_{\rm e} (\mathbf{r}_{\rm
i}), T_{\rm e})$. In this case, the corresponding principal quantum
number is sampled from the probability distribution $P_{\rm tbr}(n)
= \Delta t K_{\rm tbr}(n, \rho_{\rm e}(\mathbf{r}_{\rm i}), T_{\rm
e}) / P_{\rm tbr}$. At the same time the electron number is
decreased by unity. Analogously, for bound states, an electron-atom
collision occurs with the probability $P_{\rm ea} = \Delta t (
K_{\rm ion}(n,\rho_{\rm e} (\mathbf{r}_{\rm a}), T_{\rm e}) +
\sum_{p \ne n} K_{\rm bb}(n,p,\rho_{\rm e} (\mathbf{r}_{\rm a}),
T_{\rm e}))$. With a probability $P_{\rm ion} = \Delta t K_{\rm
ion}(n,\rho_{\rm e} (\mathbf{r}_{\rm a}), T_{\rm e}) / P_{\rm ea}$,
this leads to ionization of the atom. Otherwise, the collision
results in a bound-bound transition to a state with principal
quantum number $p$ with probability $P_{\rm bb}(n,p) = \Delta t
K_{\rm bb}(n,p,\rho_{\rm e} (\mathbf{r}_{\rm a}), T_{\rm e}) /
P_{\rm ea}$.

We finally note, that the tractable collision integrals are not
limited to such a simple case, but can also be nonlinear in the
distribution functions.  In particular, elastic electron-ion and
ion-ion collisions can be described in a similar way
\cite{rha03,ppr05jpb}.

\subsubsection{Hybrid Methods} \label{hyb}
So far we have discussed two complementary approaches which either provide the full
microscopic information about the early plasma evolution or allow to follow the
long-time dynamics on the cost of neglecting possible strong correlations between
the plasma particles.

While the ionic component may indeed form a strongly correlated
state, it turns out that the electrons remain weakly coupled during
the entire plasma evolution (see Section
\ref{sectionelectronheating} and \ref{sectioncoulombcoupling}).  In
\cite{ppr04,ppr04archive} this different behavior of the two
particle species has been used to develop a
hybrid-molecular-dynamics approach. It treats the electrons and ions
on different levels of approximations, combining the advantages of a
mean field treatment (electron dynamics) and molecular dynamics
simulations (ion motion).  Collisions between electrons and the
heavy particles (atoms and ions) are accounted for within the
Monte-Carlo procedure described in the previous subsection. By
eliminating the short dynamical timescale of the bound electron
motion this method significantly reduces the numerical workload,
while simultaneously enabling a  virtually exact numerical
simulation of the ion dynamics fully incorporating ion-ion
correlations.

The numerical effort may be further reduced within an adiabatic
approximation for the electrons.  As discussed in Section
\ref{colless_kin}, the  small electron-to-ion mass ratio permits an
adiabatic approximation, where instantaneous (on the timescale of
the ion dynamics) equilibration of the electrons is assumed. Keeping
this approximation in a (hybrid) MD approach means that much larger
timesteps may be chosen, since the electronic dynamics does not need
to be resolved.  In this way, an MD simulation of the ion motion
over the whole timescale of the experiments becomes feasible.
According to Eq.\ (\ref{elquasistat}), the electronic distribution
function is then given by a quasistatic distribution $f_{\rm e}^{\rm
(qs)}$ at all times.

The exact form of $f_{\rm e}^{\rm (qs)}$, however, merits some
discussion.  The assumption of an ideal equilibrium, i.e.\ a
Maxwell-Boltzmann distribution \showlabel{MaxBo} \beq \label{MaxBo}
\rho_{\rm{e}}\propto
\exp{\left(\frac{e{\varphi}({\mathbf{r}}_{\rm{e}})}{
k_{\rm{B}}T_{\rm{e}}}\right)} \; , \eeq leads to fundamental
problems. Since the mean-field potential $\varphi$ has a finite
depth, the corresponding Maxwell-Boltzmann distribution approaches a
finite non-zero value even for $r \to \infty$ and hence is not
normalizable. In \cite{rha03}, this problem was circumvented by
introducing a spatially dependent electron temperature which is
constant within a finite region around the plasma center, but
decreases quadratically for large distances. An analogous problem
was discussed long ago in an astrophysical context, in connection
with the formation of star clusters, where the assumption of a
Maxwellian velocity distribution implies an infinite mass of the
star cluster in complete analogy to the present case. As discussed
by Chandrasekhar \cite{cha43}, such a velocity distribution contains
particles (stars) with arbitrarily high velocities which are not
bound to the cluster potential. The rate with which these stars
evaporate from the cluster has also been considered in \cite{cha43},
and later in \cite{sha58}. If the evaporation is much slower than
the relaxation of the stars remaining in the cluster, a
quasistationary distribution forms. Comparison with numerical
simulations shows that this distribution is well described by a
truncated Maxwell-Boltzmann distribution \showlabel{MiKi1} \beq
\label{MiKi1}
f_{\rm{e}}(\mathbf{r}={\bf{0}},{\bf{v}}_{\rm{e}})\propto
\exp{\left(-\frac{m_{\rm{e}}v_{
\rm{e}}^2}{2k_{\rm{B}}T_{\rm{e}}}\right)}-\exp{\left(-\frac{m_{\rm{e}}w_{\rm{e}
}^2}{2k_{\rm{B}}T_{\rm{e}}}\right)} \eeq in the center of the
cluster. The escape velocity $w_{\rm e} = \sqrt{2 e \delta\varphi /
m_{\rm e}}$ is determined by the depth of the potential well,
$\delta\varphi = \varphi(r \to \infty) - \varphi(\bf{0})$. Following
\cite{kin66}, this may be generalized to arbitrary positions,
\showlabel{MiKi2} \beq \label{MiKi2}
f_{\rm{e}}({\bf{r}}_{\rm{e}},{\bf{v}}_{\rm{e}})\propto
\exp{\left(\frac{e{ \varphi (\mathbf{r}_{\rm
e})}}{k_{\rm{B}}T_{\rm{e}}}\right)}\left[\exp{\left(
-\frac{m_{\rm{e}}v_{\rm{e}}^2}{2k_{\rm{B}}T_{\rm{e}}}\right)}-\exp{\left(-
\frac{m_{\rm{e}}w_{\rm{e}}^2}{2k_{\rm{B}}T_{\rm{e}}}\right)}\right]
\; . \eeq If the potential is a monotonic function of the distance
from the cluster center, as in the case of gravitational forces, the
space-dependent escape velocity is given by  $w_{\rm e} = \sqrt{2 e
[\varphi(\infty) - \varphi(r_{\rm e})] / m_{\rm e}}$ for a spherical
symmetric system, and one obtains the so-called Michie-King
distribution \cite{mic63,kin66}. In the present case of a neutral
plasma, the potential may be non-monotonic due to the presence of
charges with different sign. Hence, the escape velocity has to be
defined more generally as \showlabel{escvel} \beq \label{escvel}
\frac{m_{\rm{e}}}{2}w_{\rm{e}}^2=\max_{r\ge
r_{\rm{e}}}\left[e{\varphi}(r_{ \rm{e}})-e{\varphi}(r)\right] \; .
\eeq With $W_{\rm e} \equiv m_{\rm e} w_{\rm e}^2 / (2 k_{\rm B}
T_{\rm e})$, the electronic density is then given by
\showlabel{MiKi3} \beq \label{MiKi3} \rho_{\rm{e}}(\mathbf{r}_{\rm
e}) = \int_{0}^{w_{\rm{e}}}4\pi v_{\rm{e}}^2 f_{\rm{e}}dv_{\rm{e}}
\propto \exp{\left(\frac{e{\varphi(\mathbf{r}_{\rm e})}
}{k_{\rm{B}}T_{\rm{e}}}\right)}\int_{0}^{W_{\rm{e}}}e^{-x}x^{3/2}dx
\;, \eeq which approaches a Maxwell-Boltzmann distribution in the
center of the plasma for sufficiently low temperatures (i.e.\
$W_{\rm e} \gg 1$) but decays significantly faster towards the edge
of the plasma.

The Poisson equation, particle number conservation, energy
conservation, and distribution \eq{MiKi3} constitute a closed set of
equations \showlabel{eldensHMD}
\begin{subequations}
\label{eldensHMD}
\bea
\label{eldensHMDa}
\Delta\varphi&=&4\pi e\left(\rho_{\rm{e}}-\rho_{\rm{i}}\right)\\
\label{eldensHMDb}
\rho_{\rm{e}}&\propto&\exp{\left(\frac{e{\varphi}}{k_{\rm{B}}T_{\rm{e}}}\right)}
\int_{0}^{W_{\rm{e}}}e^{-x}x^{3/2}dx\\
\label{eldensHMDc}
N_{\rm{e}}&=&\int{\rho_{\rm{e}}d{\bf{r}}_{\rm{e}}}\\
\label{eldensHMDd}
E_{\rm{ges}}&=&\frac{m_{\rm{e}}}{2}\int{v_{\rm{e}}^2f_{\rm{e}}}d{\bf{r}}_{\rm{e}
}d{\bf{v}}_{\rm{e}}-\int{\rho_{\rm{e}}\varphi}d{\bf{r}}_{\rm{e}}+E_{\rm{i}
}+E_{\rm{a}}={\rm{const.}} \; ,
\eea
\end{subequations}
that uniquely determines the electronic density, temperature, and
mean-field potential $\varphi$ at all times. (In \eq{eldensHMDd},
$E_{\rm i}$ is the sum of potential and kinetic energies of all
ions, and $E_{\rm a}$ is the sum of the electronic binding energies
and kinetic energies of all atoms.) Having calculated the electron
density profile at time $t$, the knowledge of the corresponding
electron mean field force is used to advance the ion positions and
velocities under the influence of the electronic mean field and the
full ion-ion interaction according to \showlabel{ionfor} \beq
\label{ionfor}
m_{\rm{i}}\ddot{{\bf{r}}}_{{\rm{i}},j}={\bf{F}}_{\rm{ei}}+\sum_{k
\neq j}{\bf{F}
}_{\rm{ii}}\left({\bf{r}}_{{\rm{i}},j},{\bf{r}}_{{\rm{i}},k}\right)
\;, \eeq where $\mathbf{r}_{{\rm{i}},j}$ is the position of the
$j$th ion, $\mathbf{F}_{\rm ei}$ is the force due to the electronic
mean-field, \showlabel{elmf} \beq \label{elmf} \mathbf{F}_{\rm ei} =
-e \frac{\partial \varphi_{\rm e}}{\partial \mathbf{r}_{\rm i}}
\qquad ; \qquad \Delta \varphi_{\rm e} = 4 \pi e \rho_{\rm e} \; ,
\eeq and
\begin{equation}
\mathbf{F}_{\rm ii}(\mathbf{r}_{{\rm i},j},\mathbf{r}_{{\rm i},k}) =
\frac{\mathbf{r}_{{\rm i},j}-\mathbf{r}_{{\rm i},k}}{\left|
\mathbf{r}_{{\rm i},j} - \mathbf{r}_{{\rm i},k} \right|^3}\, .
\end{equation}
In this propagation scheme it is still numerically quite costly to
calculate the inter-ionic forces. However, as demonstrated in
\cite{ppr05prl}, with usage of hierarchical tree-methods for the
direct propagation of the ions, this hybrid-MD approach allows the
description of large systems, with $\sim10^6$ ions over simulation
times of several tens of microseconds.

%% file: PhysProcOverview-jm.tex

The previous section presented the experimental and theoretical
tools to generate, detect, and understand ultracold neutral plasmas.
In this section we will use these tools to discuss physical
processes.  We proceed along the timeline of the dynamics
(\fig{figuretimeline}), i.e., we start in Section
\ref{sectioninidtialelectron} with electron equilibration after
photoionization. Then we will cover the establishment of local
thermal equilibrium for the ions in Section
\ref{sectionionequilibration}.  Before we come to the plasma
expansion (Section \ref{sectionexpansion}) into the surrounding
vacuum, we introduce collective excitations of the plasma electrons,
which may be used to probe the plasma expansion dynamics (Section
\ref{sec:PlasOsc}).

Two more subsections are devoted to important topics. First, a
detailed analysis of the initial heating of the electrons, which
affects the expansion dynamics is given in Section
\ref{sectionelectronheating}. Secondly, Section
\ref{sectioncoulombcoupling} will discuss in some detail the
evolution of the electronic and ionic Coulomb coupling parameters,
which are important since they indicate the degree of correlations
that may be realized in a plasma.



%% file: InitEEqui-jm.tex
\subsubsection{Trapping of Electrons}
\label{sectiontrapping}

\begin{figure}
\centering
 \includegraphics[width=2.5in]{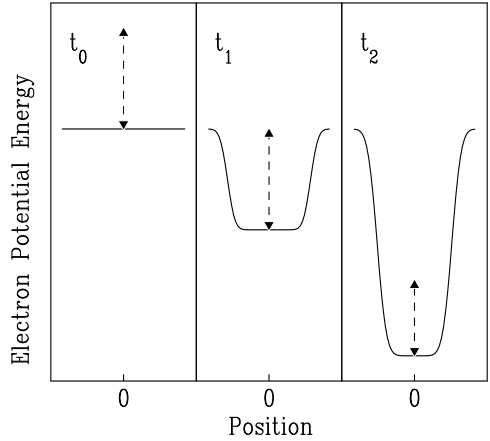}
\caption {Schematic of the potential energy seen by a test
electron when enough atoms are photoionized to result in trapping
of electrons. Photoionization occurs at $t_0=0$ and the sample is
neutral everywhere.
 Because of the  kinetic energy imparted by the laser, some electrons leave
and a charge imbalance develops. At $t_{1}\approx 10\,$ns the
resulting potential  well equals the initial kinetic energy,
trapping the remaining electrons.
As electrons in the well  thermalize, evaporation occurs. The well
depth increases  and  the electrons cool slightly. By $t_{2}\approx
1\,\mu$s evaporation essentially stops. The dashed line  indicates
the average  kinetic energy of the electrons. Reused with permission
from \cite{kkb99}. Copyright 1999, American Physical Society.}
\label{screenmodel}
\end{figure}

The first experimental study of ultracold neutral plasmas
\cite{kkb99} explored the creation of the plasma and trapping of the
electrons by the ionic background charge. (See Fig.\
\ref{screenmodel}.)  Immediately after photoionization, the charge
distribution is neutral everywhere. Due to the kinetic energy of the
electrons, the electron cloud expands on the timescale of the
inverse electron plasma frequency
$\tau_\mathrm{e}=\omega_\mathrm{p,e}^{-1}=\sqrt{m_\mathrm{e}
\varepsilon_0/ \rho_\mathrm{e} e^2} \sim$\,1\,ns, where
$\rho_\mathrm{e}$
is the electron density, and $\omega_\mathrm{p,e}$ is the electron
plasma oscillation frequency.  On this timescale the ions are
essentially immobile.  The resulting local charge imbalance creates
a Coulomb potential energy well that traps all but a small fraction
($<5$\%) of the electrons.  Smaller clouds, more ions and electrons,
and lower initial electron kinetic energy all lead to a larger
fraction of electrons trapped by the ions.  Simple PICC simulations
\cite{kkb99} show that electrons escape mostly from the edges of the
spatial distribution.

\begin{figure}
\centering
 \includegraphics[width=3in]{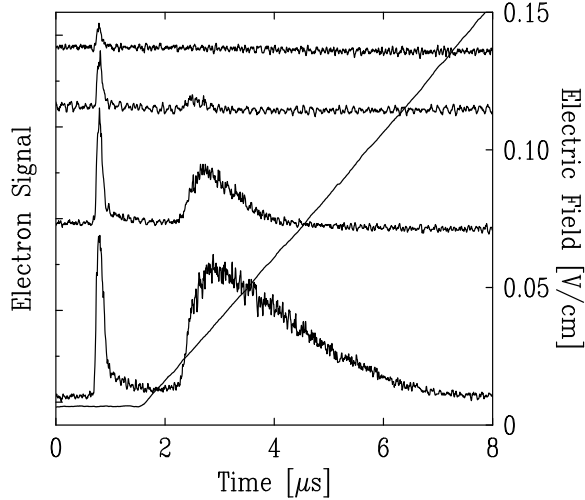}
\caption {Electron signals recorded for four different  pulse
energies of the photoionizing laser, i.e., different densities of
charged particles ($10^{5}-10^{7}\,$cm$^{-3}$). The uppermost curve
corresponds to the lowest energy. The photoionization occurs at
$t=0$. The initial kinetic energy of the electrons is $\Delta
E/k_{B}=0.6$\,K. The data shown is an average over 20 cycles of the
experiment. Also shown is the magnitude of the applied electric
field.  Reused with permission from \cite{kkb99}. Copyright 1999,
American Physical Society. } \label{electrons}
\end{figure}

 This description implies that there is a threshold for trapping
 electrons.  If the well never becomes deeper than the initial kinetic
 energy, all the electrons escape.  This is demonstrated in Fig.\
 \ref{electrons}, and the diagnostic used to obtain this data was
 charged particle detection of electrons after they have left the
 plasma, as shown in Fig.\ \ref{chargedparticle}.  In each cycle of
 the experiment, the atoms are first laser cooled and an electric field
 of approximately $5\,$mV/cm is applied.  Subsequently, the atoms are
 photoionized, and after about $500\,$ns of time of flight a pulse of
 electrons arrives at the detector (see the first peak in the curves
 of Fig.\ \ref{electrons}).  A few microseconds later,
 the electric field is linearly
 increased. If electrons have been trapped by the ionic background
 potential, they will be liberated now  to produce the second peak in
 the curves of Fig.\ \ref{electrons}.
If the laser intensity is so low that the well never becomes
 deeper than the initial kinetic energy, no electrons are trapped
 and the
 second peak is missing (top trace).   In addition, when electrons are trapped,
 the first peak develops a
 tail.  Trapped electrons thermalize within $10 - 100 \,$ns
 \cite{Spitzer}, and  as charges are promoted to energies above the trap
 depth, they leave the well.
On the time scale of this experiment, the ions are
 essentially stationary.


 One can quantify the threshold for the trapping effect.  The last
 electron to leave the plasma must climb out of the potential formed
 by the ions, with no neutralizing electrons.  At threshold, the depth
 of this well equals $E_\mathrm{e}$.  The density profile of the ions
 follows that of the neutral atoms \eq{gaussian-cloud},
 where the maximum density in the center is given by $\rho_{\rm i}=N_\mathrm{i}/(2 \pi
 \sigma^2)^{3/2}$  with the total number of ions, $N_\mathrm{i}$.  The
 electrostatic potential produced by this charge distribution is
\begin{equation}
U(r,N_{\mathrm{i}})=-\int \frac{e^2}{4\pi\varepsilon_0 |{\bf r}-{\bf
r}^{\prime}|} \rho_\mathrm{i}(r^{\prime})d{\bf r}^{\prime}=
-N_\mathrm{i}\frac{e^2}{4\pi\varepsilon_0 r}{\rm
erf}\left(\frac{r}{\sqrt{2\sigma}}\right)\;
\end{equation}
where ${\rm erf}(x)$ is the error function.
The depth of the potential well is given by its value at the center
$r=0$,
\begin{equation}\label{gaussianionpotential2}
    U(0,N_{\mathrm{i}})=-\frac{N_\mathrm{i} e^2}{4 \pi \varepsilon_0 \sigma}
    \sqrt{\frac{2}{\pi}}\,.
\end{equation}

\begin{figure}
\centering
 \includegraphics[width=3in] {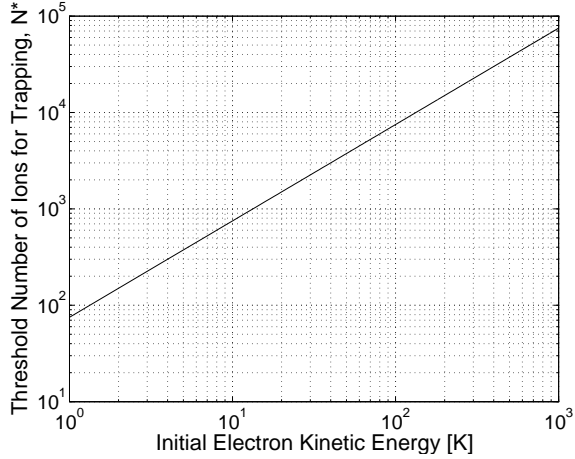}\\
  \caption{$N^{*}$, threshold number of ions required for
  trapping electrons for a Gaussian plasma with $\sigma = 1$ mm. }\label{nstar}
\end{figure}

When $-U(0,N_{\mathrm{i}})$ equals the electron kinetic energy,
$E_\mathrm{e}$, then the last electron to leave the well is trapped.
In other words, the minimum number of ions required to trap any
electrons, defined as $N^*$, is found from
$E_\mathrm{e}=-U(0,N^{*})$.
Figure \ref{nstar} shows typical
values of $N^{*}$.

The threshold condition $N_{\mathrm{i}} = N^{*}$ is  mathematically
equivalent to $\lambda_{D}=\sigma $, where $\lambda_D$ is the Debye
screening length as defined in \eq{screening-length} with
 $T_\mathrm{e}=\frac 23E_\mathrm{e}/k_\mathrm{B}$.
Note, that the relation $\sigma>\lambda_\mathrm{D}$ is the standard
condition for a system to be considered a plasma \cite{gru95},
namely that the system size ($\sigma$) has to be larger than the
characteristic length scale ($\lambda_\mathrm{D}$) on which local
charge imbalances are screened.  In the present context, we can also
interpret $\lambda_{D}$ as the displacement of electrons from their
equilibrium positions when their energy in the local internal
electric field in the plasma equals their kinetic energy
\cite{chen}.  If $\lambda_{D} > \sigma $, the electrons are free to
escape to infinity.  If $\lambda_{D} < \sigma $, electrons are
trapped by the ion cloud.  In the center of a typical ultracold
neutral plasma, $\lambda_{D} = 1-10$ $\mu$m.  Typical plasma sizes
are on the order of some hundred $\mu$m, such that the requirement
for the existence of a plasma state is well fulfilled under
experimental conditions.

\begin{figure}
\centering
\includegraphics[width=4.0in]{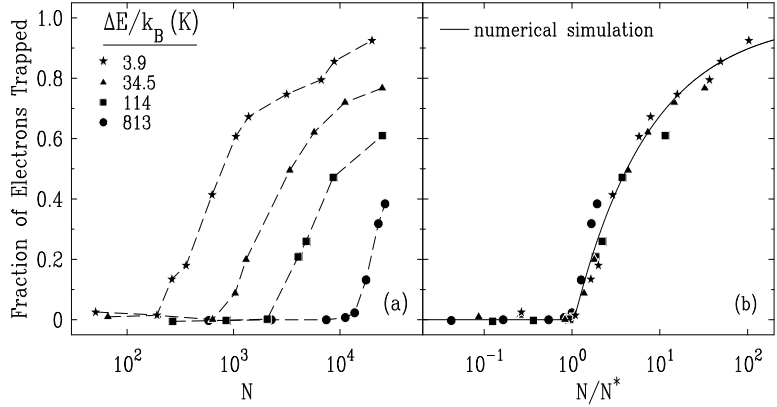}
\caption {(left) The fraction of electrons trapped is plotted versus
the number of photoions created. Each curve corresponds to a
different (green) laser frequency. The corresponding initial
energies of the electrons are displayed in the legend.
(right) Same as (left) but  the number of ions is scaled
by $N^{*}$, see text.
There is a scale uncertainty of about 10\%  in determining
 the fraction of electrons trapped. Reused with permission from \cite{kkb99}. Copyright
 1999, American Physical Society. }
\label{scurve}
\end{figure}

Fig.\ \ref{scurve} \cite{kkb99} illustrates the trapping effect and
the role of $N^{*}$. We see the number of ions required to begin
trapping electrons grows with increasing $E_\mathrm{e}$ (left
panel), i.e., the threshold $N^{*}$ increases. When the number of
ions is scaled by $N^{*}$, the fraction of electrons trapped follows
a universal curve, clearly exhibiting the threshold
$N_{\textrm{i}}/N^{*}=1$ (right panel). For $N_\mathrm{i}<N^{*}$ no
electrons are trapped. This figure also shows that for  $N/N^{*} \gg
1$, the non-neutrality is small  with the fraction of trapped
electrons almost reaching unity.

The fraction of trapped electrons is surprisingly well described by
the simple formula \cite{cvz05,pdiss}
\begin{equation} \label{nn_fit}
\frac{N_\mathrm{e}}{N_\mathrm{i}}=
\frac{\sqrt{N_\mathrm{i}/N^{\star}}-1}{\sqrt{N_\mathrm{i}/N^{\star}}}\,.
\end{equation}
An analytic derivation of this dependence has not yet been found.

A theoretical discussion assuming a quasi-neutral plasma gives a
better understanding of the degree and nature of non-neutrality in
the plasma.
%
%
First note that the quasi-neutrality condition does not imply an
exact equality of the electron and ion densities and hence a
vanishing space charge potential according to the Poisson equation
(\ref{poisson}). Instead, the Poisson equation is replaced by the
electron Vlasov equation which in adiabatic approximation yields for
the space charge potential (\eq{quasineut})
\begin{equation}
e\varphi(r)=-k_\mathrm{B}T_\mathrm{e}\ln \frac{\rho_\mathrm{i}(r)}{\rho_\mathrm{i}(r=0)}=
\frac{1}{2}k_\mathrm{B}T_\mathrm{e}\frac{r^2}{\sigma^2}\;.
\end{equation}
Now we can use this expression for a backward calculation of the charge separation
from the Possion equation, which gives a spatially constant excess charge of
\begin{equation} \label{noneutr}
\rho_\mathrm{i}-\rho_\mathrm{e}=3 \frac{k_\mathrm{B} T_\mathrm{e} \varepsilon_0}{ e^2\sigma^2}
\end{equation}
Introducing the local Debye length, which is a function of density
and thus radius, we find from Eq.\ (\ref{noneutr}), that the
relative charge difference, i.e., the local degree of non-neutrality
\begin{equation} \label{plasma_cond}
\frac{\rho_\mathrm{i}-\rho_\mathrm{e}}{\rho_\mathrm{e}}=\frac{\lambda_\mathrm{D}^2(r)}{\sigma^2}\;,
\end{equation}
is solely determined by the ratio of the Debye length and the system
size $-$ just like the global non-neutrality (Eq.\ (\ref{nn_fit})).
The fractional non-neutrality is small as long as $\lambda_D <
\sigma$, typically less than 1\% in the center of the plasma, where
the density is high.  It becomes large near the edge, however,
violating the quasi-neutrality assumption, which may lead to
deviations from the Gaussian density profile (see Section
\ref{section_ionexpansion} for a more detailed discussion).  For the
major parts of the plasma volume $\lambda_\mathrm{D}^2(r)<\sigma^2$,
justifying the statement that the plasma is essentially in a
quasineutral state.

\begin{figure}[t]
\centering
\includegraphics[width=2.7in]{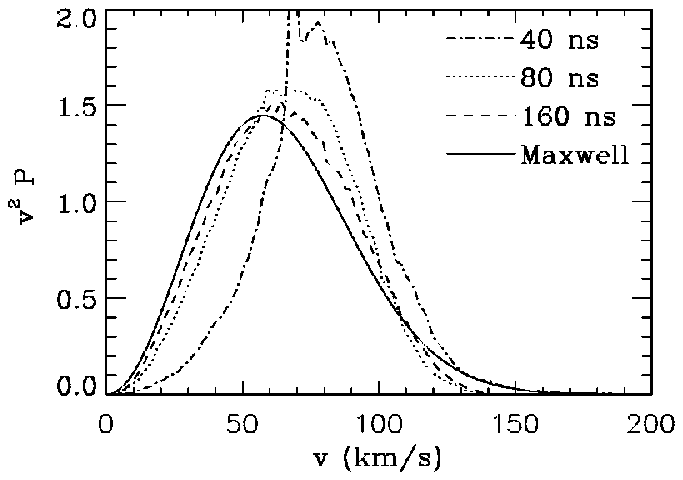}
\includegraphics[width=2.7in]{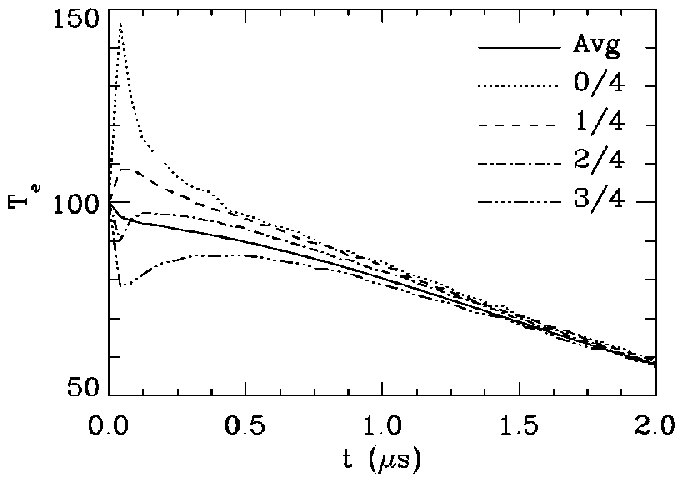}
\caption {(a) Electron velocity distribution at $t=40$\,ns
(dash-dotted line), $t=80$\,ns (dotted line) and $t=160$\,ns (dashed
line) compared to a Maxwell distribution (solid line). The
distribution has been sampled from a shell $1/4$ out from the center
of the plasma. (b) Time evolution of the electron temperature at the
plasma center (dotted line), $1/4$ out from the center (dashed
line), $2/4$ out from the center (dash-dotted line) and $3/4$ out
from the center (dash-dot-dot-dot line). The solid line shows the
average temperature of the innermost $80\%$ of the electrons. The
initial state parameters are
$\bar{\rho}_\mathrm{e}=10^9$\,cm$^{-3}$, $T_\mathrm{e}=100$\,K and
$\sigma=196$\,$\mu$m. Reused with permission from \cite{rha03}.
Copyright 2003, American Institute of Physics.} \label{etemp_rob}
\end{figure}

Non-neutrality is also related to a spatially dependent electron
temperature. When using the definition of the Debye length
$\lambda_\mathrm{D}$ in the above argumentat we have implicitly
assumed the existence of an electronic temperature.  While this is
certainly a good approximation, a global thermal electron velocity
distribution is just developing during the initial phase of the
plasma relaxation.  PICMCC simulations (Section \ref{pic}), which
yield very good agreement with the measured evaporating electron
flux \cite{rha03}, show that the velocity distribution in the
central plasma region quickly reaches its Maxwellian form within a
few hundred $100$\,ns (see Fig.\ \ref{etemp_rob}a).  As shown in
Fig.\ \ref{etemp_rob}b it takes considerably longer for a global
equilibrium to develop.  The center of the plasma is seen to heat
up, while the outer region cools down.  The total electron kinetic
energy is, however, almost unaffected by these local temperature
imbalances.  In reality, the electron gas heats up due to the
formation of Rydberg atoms \cite{klk01,rha02,rha03,ppr04archive} and
an initial build-up of spatial electron correlations.  The latter
process takes place simultaneously with the initial electron
evaporation, but is not observable in Fig.\ \ref{etemp_rob}, due to
the inability of the PIC treatment to describe the  correlated
dynamics of the electrons. A more detailed discussion of the various
electron heating mechanisms will be given in Section
\ref{sectionelectronheating}.

\subsubsection{Spontaneous Ionization of a Dense Rydberg Gas}
\label{sectionspontaneousdetails}

Compared to photoionization of a trapped cloud of ground state
atoms, the dynamics of plasma formation is more complicated when it
is created by spontaneous ionization in a dense cloud of ultracold
Rydberg atoms.  Reference \cite{rtn00} showed that some Rydberg
atoms are initially ionized by black-body radiation, collisions with
hot background atoms, or Penning collisions with other Rydberg
atoms. Attractive dipole-dipole and van der Waals interactions that
pull Rydberg atoms together and enhance the Penning ionization rate
are particularly important for high principal quantum number
($n>50$) \cite{ltg05}.  Resulting electrons escape the cloud in the
first few microseconds until a few thousand ions build up to form a
Coulomb well that traps subsequently produced electrons. This is
reminiscent of the model presented in Fig.\ \ref{screenmodel}. At
this point, the electrons seed an ionization avalanche.  As much as
2/3 of the initial Rydberg population is ionized, and the remaining
fraction is collisionally de-excited to lower Rydberg levels,
satisfying energy conservation \cite{wgc04}. During the avalanche,
\textit{l}-changing electron-Rydberg collisions, $n$-changing
electron-Rydberg collisions, electron-Rydberg ionizing collisions,
and Penning ionizing Rydberg-Rydberg collisions all play a role
\cite{wgc04,ppr03}. By exciting Rydberg atoms with a  relatively low
density in a pre-existing ultracold plasma, it was shown in
\cite{vct05}  that very tightly bound Rydberg atoms ($n\approx 20$)
could be ionized by electron-Rydberg collisions in these systems.
The energetics of electrons in plasmas created in this way, such as
the equilibrium temperature after the first few tens of nanoseconds,
is a question that has not been well-explored.  Theory \cite{rha03}
predicts no drastic change as the excitation laser crosses the
ionization threshold and the experiment changes from direct
photoionization to initial excitation of Rydberg atoms.  Recent
experimental work supports this prediction \cite{cdd05physplasmas}.


%% file: InitIEqui-jm.tex

The establishment of LTE for the ions occurs on a longer timescale
than it does for the electrons, and it is conveniently studied
experimentally.  In addition, several inelastic processes (most
notably three-body recombination) that heat the electrons do not
affect the ions. In this unusual regime, interesting phenomena are
expected to be observable.

Theoretically, the ion dynamics can be studied within the Hybrid
Molecular Dynamics approach outlined in Section \ref{hyb}. In this
approach, the ions are propagated individually, so that full
information about their positions, velocities, as well as spatial
correlations is available.  Experimentally, on the other hand, the
optical techniques described in Section \ref{sectionopticalprobes}
provide an excellent tool for following the ion dynamics.

\subsubsection{Disorder-induced heating}
\label{section_DIH}

\label{sectionDIH}
\begin{figure}[tb]
\centerline{\includegraphics[width=4in,clip=true]{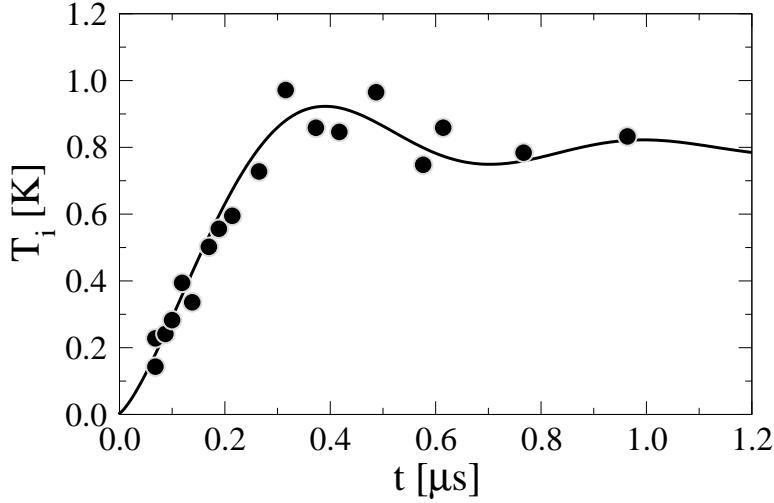}}
\caption{\label{DIHfig1} Time evolution of the effective ion
temperature for a Sr plasma with initial peak density $\rho_0(0) = 2
\cdot 10^9$cm\,$^{-3}$ and electron temperature
$T_{\rm{e}}(0)=38$\,K. The results of a hybrid-MD calculation (solid
line) are compared to experimental data extracted from an absorption
imaging measurement (dots). Reused with permission from
\cite{ppr05prl}. Copyright 2005, American Physical Society.}
\end{figure}

Figure \ref{DIHfig1} shows that the experimental evolution of the
ion temperature of an ultracold plasma and the theoretical curve
obtained from a hybrid-MD simulation are in good quantitative
agreement over a time relevant for ion thermalization
\cite{ppr05prl}.
The most striking observation from Fig.\ \ref{DIHfig1} is the rapid
heating of the plasma ions by several orders of magnitude, which is
the so-called ``disorder-induced heating'' or ``correlation
heating.''  It originates from the fact that the system is created
in an undercorrelated state far from thermodynamic equilibrium.
Immediately after photoionization, the ions in an ultracold neutral
plasma have very little kinetic energy. But they are spatially
uncorrelated, and there is significant excess potential energy in
the system compared to the equilibrium state in which spatial
correlations keep ions from being close to each other
\cite{mbd98,mbh99}.  During early evolution of the plasma,
correlations develop and potential energy is converted into
(thermal) kinetic energy \cite{mur01,gmu03,gms03} leading to
equilibration. However, during the disorder-induced heating phase,
the plasma is not in LTE, and the notion of an ion ``temperature"
should be interpreted as a measure of the average ion kinetic
energy.

Within the framework of the hydrodynamical model of Section
\ref{section_hydcoupl}, the above interpretation of disorder-induced
heating is immediately apparent in Eq.\ (\ref{momgau2c}): As long as
expansion of the plasma is still negligible ($\gamma = 0$), the
change in ion temperature is directly determined by the build-up of
ion-ion correlations (last term on the right-hand side of
(\ref{momgau2c})). This process has been discussed in theoretical
papers in many contexts.  Early interest was generated by
non-equilibrium plasmas created by fast-pulse laser irradiation of
solid targets \cite{hpm74,gma75,bsk97,zwi99,mbm01,mno03}.
Experimental results were lacking, however, because of the fast
timescales involved and limited possibilities for diagnostics.
Ultracold neutral plasmas -- much better suited for a detailed study
of the process --  reinvigorated interest in the mechanism of
disorder-induced heating
\cite{kon02,mck02,mur01,gmu03,gms03,ppr04jphysb}.

Qualitatively, one expects the timescale for the ion heating to be
given by the inverse of the ionic plasma frequency,
$\omega_\mathrm{p,i}^{-1}$, which is the typical timescale on which
spatial correlations develop in a plasma.  Physically,
$\omega_\mathrm{p,i}^{-1}$ is the time for an ion to move one
interparticle spacing when accelerated by a typical Coulomb force of
$ {e^2 / 4\pi \varepsilon_0 a^2}$, where $a=(4\pi
\rho_\mathrm{i}/3)^{-1/3}$ is the Wigner-Seitz radius.  For the
typical plasma of Fig.\ \ref{DIHfig1}, $\omega_\mathrm{p,i}^{-1}
\approx 270\,$ns in good agreement with the rise-time of the
temperature observed in the figure. This timescale of $\sim
10^{2}$\,ns is several orders of magnitude larger than in
laser-generated high-density plasmas as mentioned above, and it is
easily resolved experimentally.   One would guess the ion
temperature after equilibration would be on the order of the Coulomb
interaction energy between neighboring ions,
\begin{equation}\label{ionTC}
k_{B}T_C\approx \frac{e^2}{4\pi \varepsilon_0}\frac 1 a \equiv
U_{a}\, ,
\end{equation}
which is a few kelvin. A quantitative analysis \cite{mur01},
assuming complete initial disorder and incorporating the screening
effects of the electrons, predicts from a simple energy conservation
argument an equilibrium ion temperature of
\begin{eqnarray}\label{iontemp}
  T_\mathrm{i}&=&\frac 2{3k_{B}}U_{a}\left| \frac{\tilde{U}}{U_{a}} +{\kappa \over
 2}\right| .
\end{eqnarray}
 Here, $\kappa =a/\lambda_D$ is the Debye screening constant, and
$\tilde{U}$ is the excess potential energy per ion. $\tilde{U}$ has
been studied with molecular dynamics simulations \cite{fha94} for a
homogeneous system of particles interacting through a Yukawa
potential, $\phi(r)= U_{a} (a/r)\,{\exp}(-r/\lambda_D)$, which
describes ions in the background of weakly coupled electrons
\cite{dresdennotesdebyenote}. To obtain $T_\mathrm{i}$ for given
 $\rho_\mathrm{i}$ and $T_\mathrm{e}$, one needs to selfconsistently
solve Eq.\ (\ref{iontemp}) with an analytic expression for $\tilde{U}$
\cite{hfd97} since $\tilde{U}$ depends itself on $T_\mathrm{i}$.

\begin{figure}
\centering
  \includegraphics[width=5in,clip=true]{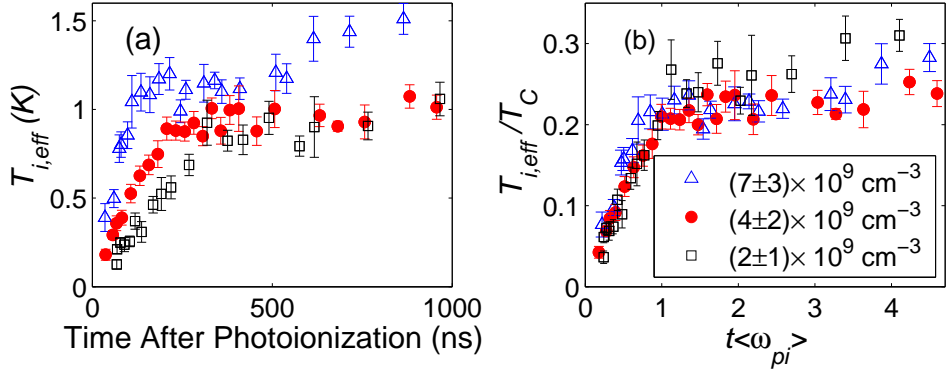}\\
  \caption{The effective ion temperature, $T_\mathrm{i,eff}$ (Eq.\ (\ref{equationtieffspectrum})), versus
  time after photoionization for initial electron temperature of
  $T_e=2 E_e/3k_B=38\pm 6$\,K and various plasma densities.  (a) The
  data is plotted on absolute temperature and time scales.  (b) The
  time is scaled by the inverse of the average plasma period, and
  $T_\mathrm{i,eff}$ is scaled by $T_\mathrm{C}$ (Eq. \ref{ionTC}).
  Reused with permission from \cite{csl04}. Copyright 2004, American Physical Society.
  }
 \label{fig1}
\end{figure}

\begin{figure}
\centering
  \includegraphics[width=4.25in,clip=true]{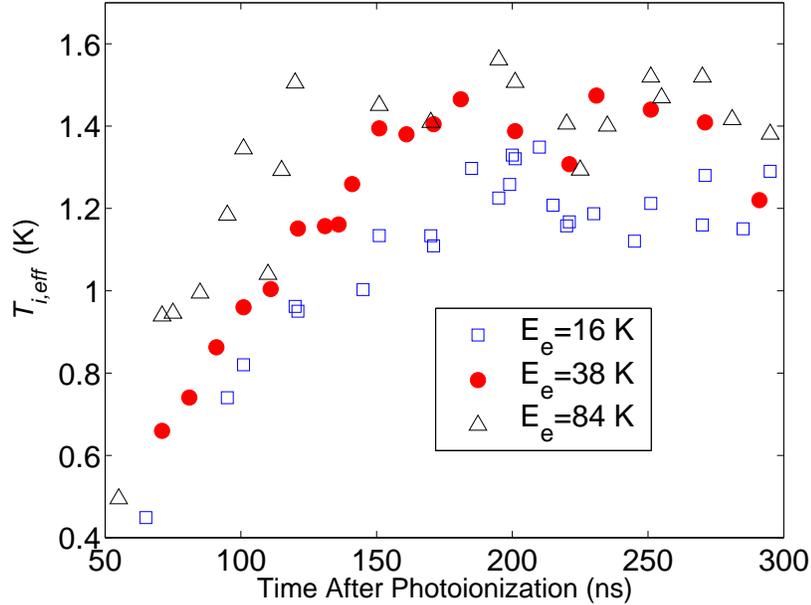}\\
  \caption{Effective ion temperature versus time after
  photoionization for various initial electron kinetic energies.
  The peak ion density for all data is $\rho_\mathrm{0i}=(1.4 \pm 0.5)\times
  10^{10}$\,cm$^{-3}$. Reused with permission  from \cite{csl04}. Copyright 2004, American Physical Society.}
  \label{fig2}
\end{figure}

The arguments given above imply that both the timescale and the
energy scale of the disorder-induced heating depend on the plasma
density only (apart from a small dependence of the final temperature
on the screening $\kappa$, which will be discussed below).  Fig.\
\ref{fig1} confirms this expectation.  If the measured effective ion
temperature $T_\mathrm{i,eff}$ is scaled with $T_{C}$ (\eq{ionTC})
and time with the inverse plasma frequency $\omega_{\mathrm{p,i}}$,
the measured curves for different densities coincide quite well
(Fig.\ \ref{fig1}(b)).  Small differences in the curves are due to
electron screening of the ion-ion interaction.  The effect of
electron is demonstrated in Fig.\ \ref{fig2}, where
$T_\mathrm{i,eff}$ obtained from absorption spectra (Eq.\
(\ref{equationtieffspectrum})) is shown for different initial
electron energies but the same ion density distribution: Smaller
initial electron energy (temperature) implies more screening and a
lower  equilibrium ion temperature. The latter is directly related
to the initial excess potential energy (\eq{iontemp}), which is
lowered if screening through electrons is increased.   It was found
that the measured ion temperature just after the plasma reaches LTE
agrees well with the theoretical value $T_\mathrm{i,ave}$ according
to Eq.\ (\ref{iontemp}) and averaged over the ion distribution
\cite{csl04}.

As emphasized in \cite{ppr04jphysb}, it is not surprising that that
Eq.\ (\ref{iontemp}) accurately predicts the ion temperature for
equilibrating ultracold neutral plasmas, because it basically
expresses energy conservation.
However, it is worth pointing out that for data in Ref.\
\cite{csl04} with the lowest fit $T_\mathrm{e}$ and highest
$\rho_\mathrm{0i}$, the peak value of $\kappa$ in the plasma is 0.7.
This corresponds to three electrons per Debye sphere
($\kappa^{-3}=\rho_\mathrm{e}4\pi\lambda_{D}^{3}/3$). One might not
necessarily expect Eq.\ (\ref{iontemp}) to be accurate in this
regime because it assumes a Yukawa potential for ion-ion
interactions, which is based on Debye screening and is normally
derived for $\kappa^{-3}\gg 1$.

\subsubsection{Kinetic energy oscillations}

\begin{figure}[bt]
\centering
  \includegraphics[width=3.4in,clip=true]{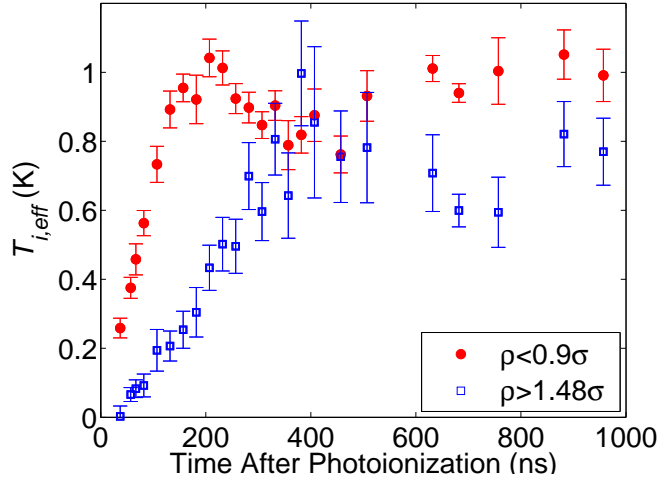}\\
  \caption{Effective ion temperature obtained from different
  selected regions of a plasma cloud with
  $\rho_\mathrm{0i}=(4\pm 2)\times 10^{9}$\,cm$^{-3}$
  and initial $T_\mathrm{e}=2 E_\mathrm{e}/3k_B=38\pm 6$\,K.
  Oscillations are clearly visible at the end of the
  disorder-induced heating process.
  Reused with permission from \cite{csl04}. Copyright
2004, American Physical Society.  }\label{fig4}
\end{figure}

\begin{figure}[bt]
\centering
  \includegraphics[width=4.0in,clip=true]{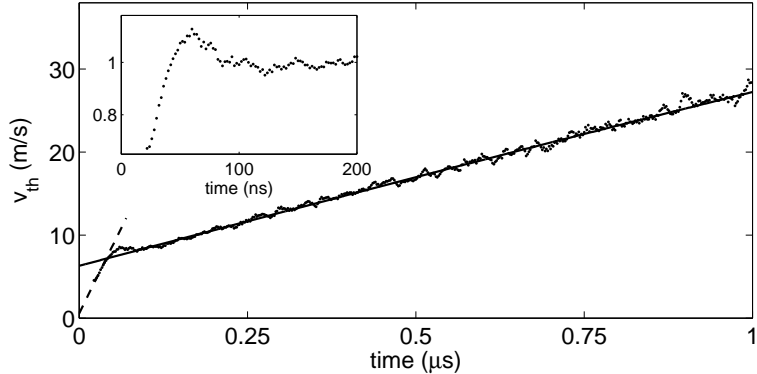}\\
  \caption{The mean $z$-component of the ion velocity in a calcium
  plasma as a function of time after photoionization in a region
  defined by the excitation laser beam.  The velocity was extracted
  using Eq.\ (\ref{equationfluoressignalrelatedtothermal}), and the
  absolute calibration was obtained by assuming the ions had an
  initial velocity distribution equal to that of the laser-cooled
  atoms.  The rapid increase of the velocity  at early times is due to disorder-induced
heating.  The subsequent slower increase reflects the expansion of
the plasma, which will be discussed in Section
\ref{sectionexpansion}. The solid line is a fit of the velocity to a
model for the plasma
  expansion  plus an offset representing the random thermal
  ion velocity. The inset shows the ion velocity divided by the fit,
  which highlights the oscillation
  of the kinetic energy.   Reused with permission from
  \cite{cdd05}. Copyright 2005, American Physical Society.}\label{bergesonvelocity.eps}
\end{figure}

\begin{figure}[tb]
\centerline{\includegraphics[width=3.0in,clip=true]{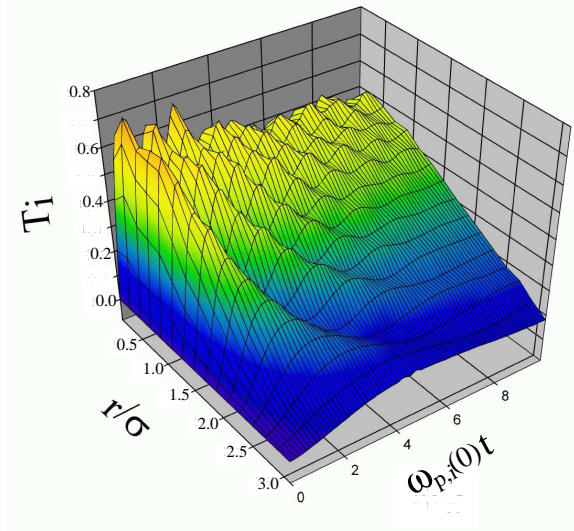}}
\caption{\label{spateosc} Spatio-temporal evolution of the ion
temperature, obtained from HMD-simulations for $40000$ ions and
electrons with an initial Coulomb coupling parameter of $\Gamma_{\rm
e}(0)=0.07$. Reused with permission from \cite{ppa05conf}. Copyright
2005, Institute of Physics.}
\end{figure}

A close inspection of Figs.\ \ref{fig1} and \ref{fig2} reveals that
at the end of the disorder-induced heating phase, the ion
temperature overshoots its equilibrium value before settling to it.
This phenomenon is even more evident in Fig.\ \ref{fig4}, where
$T_\mathrm{i,eff}$ is shown for an inner and outer region of the
plasma image ($\rho=\sqrt{x^2+y^2}< 0.9\, \sigma$ and $\rho>1.48\,
\sigma$, respectively).  Each selected annular region contains 1/3
of the ions and probes a region with significantly less variation in
density than in the entire plasma.  The region with lower density
has lower ion temperature, as expected from Eq.\ (\ref{iontemp}),
but the oscillation is the most striking observation.  The
oscillation was also observed in a calcium plasma with fluorescence
measurement \cite{cdd05} (Fig.\ \ref{bergesonvelocity.eps}).  The
contrast of the oscillations is strong in this case because the
small volume of the plasma excited by the fluorescence excitation
laser significantly reduced the variation of the sampled density.
Numerical simulations clearly show the oscillations and also provide
their temperature and spatial dependence (Fig.\ \ref{spateosc}).

These oscillations display universal relaxation dynamics of a
strongly coupled Coulomb system.   More precisely, they reflect the
fact that the two-particle distribution function (i.e.\ spatial
correlations) and the one-particle distribution function (i.e.\
temperature) relax on the same timescale ($1/\omega_\mathrm{p,i}$)
due to the strong long-range interactions. Normally, correlations
relax much more quickly, and the separation of these time scales is
known as the Bogoliubov assumption \cite{bog46}. Studies with
ultracold neutral plasmas represent the first experimental
observation of kinetic energy oscillations, but it has been the
subject of intense theoretical study through analytic calculations
\cite{gma75} and simulations
\cite{hpm74,bsk97,zwi99,mbm01,mno03,ppr04jphysb,ppr05prl} of
equilibrating strongly coupled plasmas.  Calculations
\cite{hpm74,hmp75} also show oscillations in the velocity
autocorrelation function in equilibrium systems.

Intuitively, one may relate this phenomenon to the motion of each
ion in its local potential energy well.  The discussion of
disorder-induced heating in the previous section qualitatively
ascribed the increase in ion kinetic energy to each ion moving
towards the minimum of its well.  The time scale for this motion is
$\omega_\mathrm{p,i}^{-1}$.  However, the ion will not suddenly stop
at the bottom of its potential; it will overshoot and climb up the
hill again.  It is reasonable to expect harmonic motion to persist
for some time, as kinetic and potential energy are interchanged.
This suggests a kinetic energy oscillation at $2
\omega_\mathrm{p,i}$, which is in line with the observations.  Also
consistent with this description, the oscillation period observed
experimentally in Fig.\ \ref{fig4} is longer in the outer region
where the average density is lower and hence the plasma frequency is
smaller.  This also explains why averaging over the entire cloud
such as in Fig.\ \ref{DIHfig1} obscures the oscillation: the motion
dephases because of the variation in $\omega_\mathrm{p,i}$.  In a
disordered system corresponding to the ultracold plasma scenario,
there is probably no collective or long-range coherence to the
motion, so in spite of the fact that $\omega_\mathrm{p,i}$ sets the
timescale of the oscillation, one must be cautious about describing
the motion as an ion plasma oscillation.

The numerical hybrid-MD approach  permits the definition of a local
temperature.  Hence, it provides better temporal and spatial
resolution of $T_\mathrm{i}$ than the experimental optical probes.
As was shown in \cite{ppr04PRA,ppa05conf}, the ion temperature shows
clear oscillations in space (i.e.\ as a function of the distance
from the center of the plasma) and, in fact, a wavelike oscillation
pattern emanating from the plasma center (Fig. \ref{spateosc}).

From the experimental data, it
 is not straightforward to comment on the damping of the observed
 kinetic energy oscillations because
even in Fig.\ \ref{fig4}, which provides  an annularly-resolved,
more-local probe of the system, the analysis still averages over the
density variation along the z-axis of the plasma. This introduces
dephasing because the
 ion temperature oscillates with the local plasma frequency.
A phenomenological model was developed in \cite{lcg06} 
 to describe the ion equilibration and kinetic energy oscillations in
 a manner that could extract information on damping.  The main
 assumptions of this model are that ions execute damped harmonic
 motion at their local plasma oscillation frequency and Eq.\
 (\ref{iontemp}) determines the equilibrium ion kinetic energy and the
 amplitude of oscillation around this value.  Fits to the data using
 this model showed that the damping time is close to
 $1/\omega_\mathrm{p,i}$, and higher particle density and lower
 electron temperature give slightly stronger damping.
Numerical studies of {\em homogeneous} strongly coupled plasmas
\cite{zwi99,ppa05conf} suggest a similar damping time for
$\Gamma_\mathrm{i} \ge 5$, with lower $\Gamma_\mathrm{i}$ leading to
faster damping.

%% file: PlasOsc-jm.tex
\label{plasmaoscillationssection}
Oscillation of the electron plasma, or more precisely of the
electron density, is a fundamental  collective plasma mode  first
described by Tonks and Langmuir \cite{tla29}.  Excitation of this
mode in an early experiment \cite{kkb00} firmly established that the
photoionized ultracold gas forms a plasma.

\begin{figure}
\centering
  \includegraphics[width=3.5in]{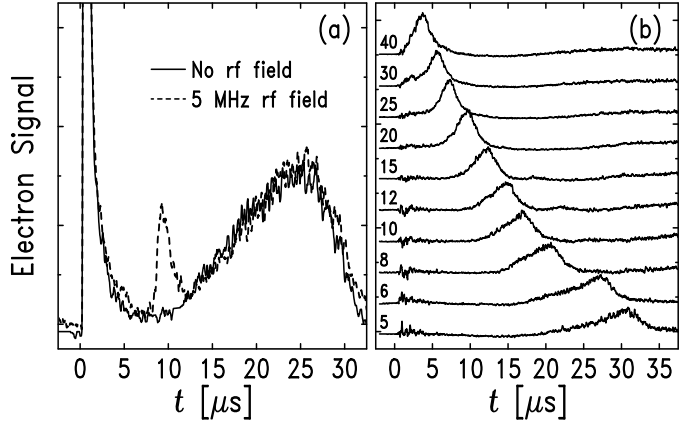}\\
\caption{Electron signals from ultracold xenon plasmas created by
photoionization at $t=0$.  (a) $3 \times 10^{4}$ atoms are
photoionized and $E_\mathrm{e}/k_\mathrm{B}=540\,$K. Signals with and
without rf field are shown.  The rf field is applied continuously.
(b) $8 \times 10^{4}$ atoms are photoionized and
$E_\mathrm{e}/k_\mathrm{B}=26\,$K. For each trace, the rf frequency in
MHz is indicated, and the nonresonant response has been subtracted.
The signals have been offset for clarity.  The resonant response
arrives later for lower frequency, reflecting expansion of the plasma;
Reused with permission from \cite{kkb00}. Copyright 2000, American
Physical Society.} \label{compositepaper}
\end{figure}

A straightforward way to resonantly excite this mode is with
 a radio-frequency (rf) electric field (Fig.\ \ref{chargedparticle}).
 A small DC bias directs electrons that escape the plasma to a charged
 particle detector in order to measure the response.
Typical electron signals from such an experiment \cite{kkb00}
are shown in Fig. \ref{compositepaper}a.
The plasma is created at $t=0$ and some electrons promptly leave the
sample and arrive at the detector at about $1$\,$\mu$s, producing the
first peak in the signal.  The resulting excess positive charge in the
plasma creates the Coulomb potential well that traps the remaining
electrons \cite{kkb99}.
As the plasma expands, the depth of the Coulomb well decreases,
allowing the remaining electrons to leave the trap.  This produces the
broad peak at $\sim 25$\,$\mu$s.

In the presence of an rf field an additional peak appears in the
electron signal, indicating that resonant excitation of plasma
oscillations has pumped energy into the plasma and raised the electron
temperature.  The elevated electron temperature causes more electrons
to reach the detector because it increases the evaporation rate of
electrons out of the Coulomb well.

This data \cite{kkb00} was initially analyzed assuming an expression
for the angular oscillation frequency that is valid in a homogeneous
gas ($\omega_\mathrm{p,e}=\sqrt{e^{2}
\rho_\mathrm{e}/\varepsilon_\mathrm{0}m_\mathrm{e}}$ \cite{tla29}),
and assuming that the excited mode was localized in regions of near
resonant density.  By assuming the peak response arrived when the
average density in the plasma was in resonance, the changing resonant
frequency showed the decrease in plasma density with time as the
plasma expanded into the surrounding vacuum
(Fig.~\ref{compositepaper}b).  Section \ref{sectionexpansion} will
discuss the expansion dynamics in detail.

Reference \cite{bsp03} pointed out that even in the zero-temperature
limit, the electron plasma resonance frequency should be modified
because of the inhomogeneous density distribution.
This effect also
causes the well known result that for a spherically symmetric
flat-top density distribution, the frequency becomes
$\omega=\omega_\mathrm{p,e}/\sqrt{3}$.  For a Gaussian density
distribution, the plasma has a continuous spectrum of
electron-density oscillation frequencies, but spectral weight is
concentrated in a damped quasimode that is concentrated in lower
density regions of the cloud and has a frequency between 1/4 and 1/3
of $\omega_\mathrm{p,e}$ for the peak density in the distribution.
This implies that the approach of \cite{kkb00} underestimated the
density by about a factor of three, but does not change the general
conclusion that the plasma oscillation can be used to follow the
evolution of the plasma density.

\begin{figure}
\centering
  \includegraphics[width=3in]{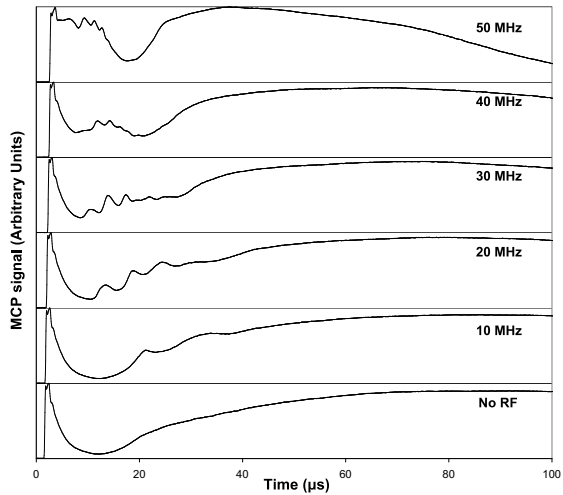}\\
  \caption{Electron emission from an expanding ultracold plasma.
  Data is taken as in Fig.\ \ref{compositepaper}, but the plasma
  is more dense and the
  applied rf field is more intense. For a given rf frequency, the earliest
   rf response peak is the cold plasma mode at the frequency predicted by the theory of
\cite{bsp03}. The later modes are thought to be Tonks-Dattner
resonances \cite{png64}. Reused with permission from \cite{fzr06}.
Copyright 2006, American Physical Society.}
\label{figuretonksdattnerrawdata}
\end{figure}

In very recent work, this idea has been extended by
Fletcher \textit{et al.} \cite{fzr06} who observed
what appear to be Tonks-Dattner modes, which are electron
density-waves characteristic of plasmas with inhomogeneous density
distributions in which thermal effects are important.  (See Fig.\
\ref{figuretonksdattnerrawdata}.)
If one goes beyond cold plasma theory, and includes the effect of electron temperature,
electron plasma oscillations follow the Bohm-Gross dispersion relation \cite{bgr49}
\begin{equation}\label{equationbohmgross}
    \omega^2=\omega_\mathrm{p,e}^2(r)+
    \frac{3 k_\mathrm{B} T_\mathrm{e}}{m_\mathrm{e}} k^2(r),
\end{equation}
where $k(r)$ is the local wavenumber.  When the temperature or
wavenumber becomes high, the thermal term becomes dominant and the
electron density oscillation becomes a sound-wave.  Tonks-Dattner
modes are resonant electron-density standing waves resulting from
this thermal effect, and they were originally seen in cylindrical
plasmas designed to simulate meteor tails \cite{her51,dat63}.  In
these experiments, the plasma density increased towards the central
axis, and the modes resonated between the outer edge of the plasma
and the region where $\omega=\omega_\mathrm{p,e}$.  For higher
frequency modes, more wavelengths fit in the resonant region.  It is
at first surprising that thermal effects are important in ultracold
plasmas. But because of the small plasma size, the wavelength for
density waves must be small and the wavevector must be large.  This
makes the thermal term in the Bohm-Gross dispersion relation (Eq.
(\ref{equationbohmgross})) comparable or larger than
$\omega_\mathrm{p,e}$.

An accurate quantitative description of Tonks-Dattner modes was
given by \cite{png64}, and it was adapted to explain the
observations in ultracold plasmas.  The agreement is striking in
spite of the fact that the authors had to make a somewhat \textit{ad
hoc} assumption about the outer boundary for the standing wave since
there is not a well-defined edge to a Gaussian-shaped ultracold
plasma.  With additional theory input to understand this point,
Tonks-Dattner modes should provide an accurate diagnostic of the
electron temperature with good temporal resolution.

Hence, both types of collective electron excitations can provide
information on the time evolution of the plasma, its changing
density and electron temperature, and may be used to monitor
dynamical effects in the plasma expansion.

%% file: PlasExpansion-jm.tex

\subsubsection{Physical Description}
\begin{figure}
\centering
  \includegraphics[width=3in]{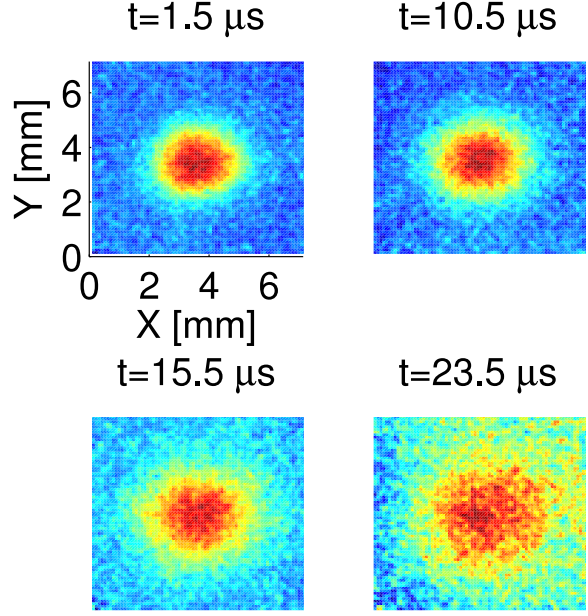}\\
  \caption{False color images of expanding ultracold neutral strontium plasmas.
  The time after photoionization is indicated.} \label{figureexpansionimages}
\end{figure}

In all experiments performed so far, the plasma is unconfined and
expands into the surrounding vacuum.  On a time scale of about 10
$\mu$s, the plasma changes in size noticeably, as shown with
absorption images of strontium plasmas in Fig.\
\ref{figureexpansionimages}.

As discussed in Section \ref{sectiontrapping} in conjunction with
the initial electron trapping, the local charge imbalance due to the
finite electron temperature causes a confining space charge
potential. This traps the electrons such that a quasineutral plasma
state develops, but the
 potential experienced by the ions has
opposite sign, which drives the expansion of the plasma at later
times. In an alternative, but equivalent description, simple
hydrodynamic arguments \cite{kkb00} show that the expansion can also
be related to the thermal pressure of the electrons, since the
hydrodynamic force per ion due to the electron thermal motion and
plasma spatial gradient is given by (see eq.(\ref{quasineut}))
\begin{equation}
{\bf F}= -\frac{k_{\rm B} T_{\rm e}}{\rho_{\rm i}}{\bf
\nabla}\rho_{\rm i} =\frac{k_{\rm B}T_{\rm e}}{\sigma^2}{\bf r}\,.
\end {equation}
This yields the same results as found from collisionless plasma
dynamics using the Vlasov equation as described in Section
\ref{colless_kin}.\footnote{Strong coupling  of the ions
\cite{ppr04archive} causes negligible modifications of the expansion
dynamics (Section \ref{hyb}).}

This model predicts that the plasma phase space density retains its
Gaussian shape (\eq{gauansatz}) during the expansion, with a
quadratically increasing rms-size (Eq.\ \ref{expansiona})
\begin{equation}\label{eq_expansion}
    \sigma(t)^2=\sigma(0)^2+v_{0}^2t^2,
\end{equation}
where the expansion velocity $v_0=\sqrt{k_{\rm B} T_{\rm e}(0)/m_i}$
is solely determined by the initial electron temperature. The
mechanical effect of the electrons on the ions is to induce a
radially outward directed ion velocity (Eq.\ \ref{expvelocity})
\begin{equation}\label{expansionvelocity}
    {\bf u}({\bf r},t)=\frac{k_{\rm B} T_{\rm e}(0)}{m_{\rm
    i}}\frac{t}{\sigma(0)^2+v_{0}^2t^2}{\bf r}\;
\end{equation}
that is distinct from random thermal velocity, such that the ion
kinetic energy due to expansion increases as $E_{\rm
i}=\frac{3}{2}k_{\rm B}T_{\rm e}(0)v_0^2t^2/\sigma^2(t)$. This
energy increase has to be compensated by a decreasing temperature of
the electrons which cool according to $T_{\rm e}(t)=T_{\rm
e}(0)\sigma^2(0)/\sigma^2(t)$ (Eq.\ \ref {expansiond}). Thus,
asymptotically all the initial electron energy is transfered to the
ions. Note that this energy transfer is not due to elastic
collisions between the electrons and the ions, which would require
milliseconds to equilibrate the energy of two plasma components.

The expansion of neutral plasmas in different geometries has been
intensively studied for decades, and  much theoretical work has
recently been done on the dynamics of quasineutral plasmas expanding
into a vacuum \cite{dse98,kby03,ccb05,mor03}.  This problem is
important for understanding plasmas created by irradiation of solid
and thin film targets \cite{ckd00,mgf00,hbc00,msp02,hkp02}, clusters
\cite{dts97}, and gas jets \cite{kcn99} by intense laser pulses, and
dusty plasmas \cite{thz04}, which are important in astrophysical
systems and plasma processing.

\subsubsection{Studying the Expansion with Electron Plasma Oscillations}

\begin{figure}
\centering
  \includegraphics[width=3in]{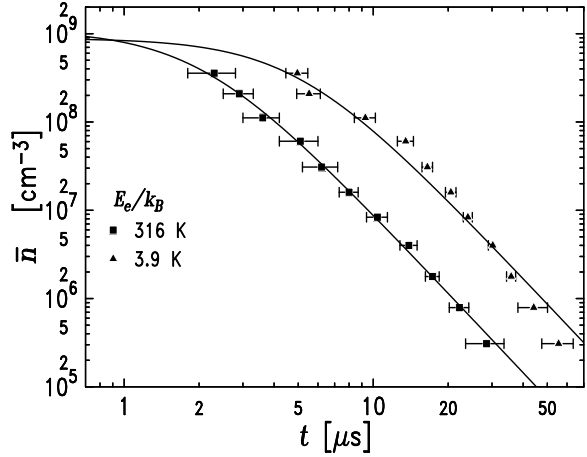}\\
  \caption{Expansion of the plasma for $N= 5 \times 10^{5}$
photoionized atoms. The  expansion  is  described by $\bar n=
N/[4\pi (\sigma_0^{2}+v_{0}^{2}t^{2})]^{3/2}$, where $\sigma_{0}$ is
the initial  rms radius, $v_{0}$ is the rms radial velocity at long
times, and $\bar n$ is the density in resonance with the rf field,
which is assumed to equal the average density in the plasma.
Horizontal  error bars arise from uncertainty in  peak arrival times
in data such as Fig.\ \ref{compositepaper}b.  The fits were
consistently poor at low $E_{\mathrm{e}}$, as  in the $3.9\,$K data.
Reused with permission from \cite{kkb00}. Copyright 2000, American
Physical Society.} \label{paperalltemp}
\end{figure}

\begin{figure}
\centering
  \includegraphics[width=3in]{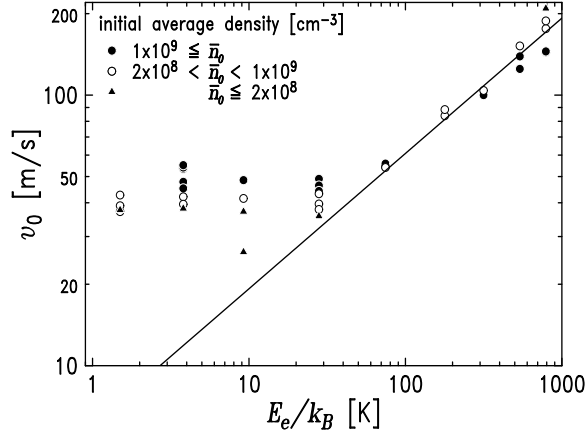}\\
  \caption{Expansion velocities, $v_{0}$, found from fits to data such as in
Fig.\ \ref{paperalltemp}. The initial average density,
$\bar{n}_{0}$, varies from $6 \times 10^{6}$ to $2.5 \times
10^{9}\,{\rm cm}^{-3}$. The solid line,
$v_{0}=\sqrt{E_{\mathrm{e}}/\alpha m_{\mathrm{i}}}$, with
$\alpha=1.7$, is a fit to data with $E_{\mathrm{e}}/k_{B}\ge 70\,$K.
The behavior of low $E_{\mathrm{e}}$ data is discussed in the text.
Uncertainty in  $v_{0}$ is typically equal to the size of the
symbols. There is a $0.5\,$K uncertainty in $E_{\mathrm{e}}/k_{B}$
reflecting uncertainty in the  dye laser wavelength. Note that for
$E_{\mathrm{e}}/k_{B}< 70\,$K, $v_{0}$ shows a  systematic
dependence on $\bar{n}_{0}$. Reused with permission from
\cite{kkb00}. Copyright 2000, American Physical
Society.}\label{expansionpaper}
\end{figure}

The first quantitative experimental study of the expansion was
performed using excitation of electron plasma oscillations as a
probe of the density as described in Section
\ref{sectionplasmaoscillation}. Data such as in Fig.\
\ref{paperalltemp} was fitted assuming
$\sigma(t)^2=\sigma(0)^2+(v_{0}t)^2$ for the expansion.  The
oscillation probe is only sensitive to times when the plasma has
already reached a final velocity.  Figure \ref{expansionpaper} shows
the extracted values of $v_{0}$.  For data with $E_\mathrm{e}\ge
70\,$K, the expansion velocities approximately follow
$v_{0}=[E_\mathrm{e}/(\alpha m_\mathrm{i})]^{1/2}$, where
$m_\mathrm{i}$ is the ion mass and $\alpha=1.7$ is a fit parameter.
This equates to  $v_{0}=[0.9 k_B T_\mathrm{e}(0)/
m_\mathrm{i}]^{1/2}$ if $k_B T_\mathrm{e}(0)= \frac 23E_\mathrm{e}$,
in reasonable agreement with the theory.


For very low $E_\mathrm{e}$, the plasma expands as if the electrons
had much more energy than $E_\mathrm{e}$ (\fig{expansionpaper}).
Different authors addressed this issue in a series of theory papers
pointing to the importance of continuum lowering \cite{hah01},
disorder-induced heating \cite{kon02,mck02}, and recombination
\cite{rha02} as sources of electron heating.  A more detailed
discussion of these issues is given in Section
\ref{sectionelectronheating}.  The heating becomes important just as
$\Gamma_\mathrm{e}$ is approaching one, which is interesting, as it
might indicate the onset of strong coupling effects of the
electrons.  On the other hand this heating also implies that it is
difficult to create strongly coupled electrons.  There is still more
work to be done to study the dependence of the heating on various
parameters in order to work out the details of the dominant heating
sources.

\subsubsection{Studying the Expansion with Optical Probes}
\label{sectionexpansionopticalprobes}

The expansion can also be studied with optical probes.  Because of the
better time resolution of these techniques, it is possible to follow
the entire expansion from plasma formation, through ion acceleration,
to terminal velocity.  One can thus make an accurate comparison
between experiment and theory.

References \cite{cdd05physplasmas} and \cite{cdd05} used the ion
fluorescence to follow the spatial expansion of a calcium plasma. By
placing the focused excitation laser at various distances from the
center of the plasma, they measured fluorescence profiles  up to 50
$\mu$s after photoionization (Fig.\ 1 of \cite{cdd05}) and found
reasonable agreement with  a modified version of the theory
\cite{rha03,ppr04PRA} that addressed the lack of spherical symmetry.
 They also found that for
$E_e<100$\,K, the plasma expanded with more energy than expected
from $E_e$ \cite{cdd05physplasmas}. The ion velocity was probed
spectrally using fluorescence from a resonant excitation beam
passing through the plasma center \cite{cdd05}. Changes in the
fluorescence level could be ascribed to changes in Doppler
broadening and ion velocity, resulting in the data in Fig.\
\ref{bergesonvelocity.eps}.

\begin{figure}
  \includegraphics[width=3in, angle = -90 ,clip=true]{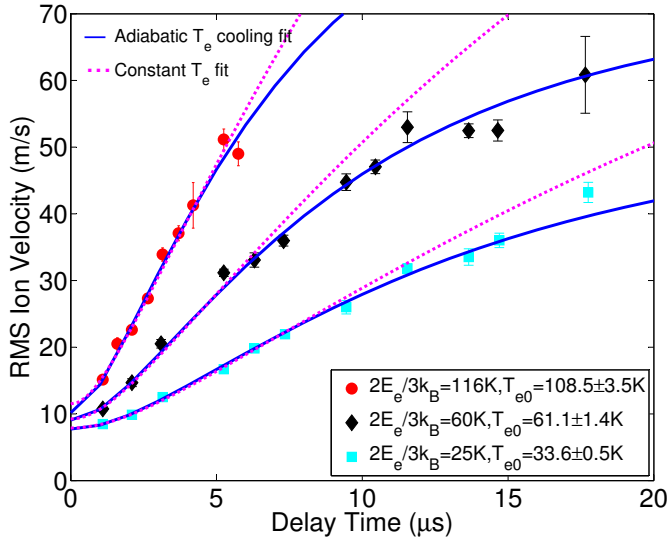}\\
  \caption{Studying plasma expansion with optical absorption imaging.
  The rms velocity along
  the laser beam is extracted from the width of the absorption image
  (\ref{sectionopticalprobes}). The dominant contribution is from
  the expansion velocity  but
  there is a contribution from the ion thermal velocity that
  produces the 10\,m/s offset at short times.
  The initial peak density of these plasmas was
  $n_0 = 3.5 \times 10^{15}$\,m$^{-3}$ and initial
size $\sigma_0 = 1$\, mm. The initial electron kinetic energies are
indicated in the legend. The solid-line fits to the data assume a
self-similar expansion of the plasma as described in Eq.\
\ref{expansionvelocity}. This implies that the 1-dimensional rms ion
velocity is given by $\sqrt{k_B T_{\textrm{i,eff}}/m_i}$, where
$T_{i,eff}$ is given by Eq.\
(\ref{equationtieffrelatedtoexpansion}). The initial electron
temperature can be extracted from the fits, and the agreement with
expected values for the two higher temperature plasmas indicates
that three-body recombination and disorder-induced heating of
electrons is not significant for this data. If the electron
temperature did not drop due to adiabatic expansion, the ion
velocity would follow the dashed-line curves.}
\label{tempevolutiondata}
\end{figure}

In \cite{csl04}, the ion kinetic energy was found with the
absorption spectra, and this was related to the expansion velocity
and the average ion temperature. Initially, the technique was only
applied to early times before adiabatic cooling of the electrons
became important, but recent improvements have allowed study of
longer times (Fig.\ \ref{tempevolutiondata}). For a broad range of
conditions, agreement is excellent with  Eq.
(\ref{equationtieffrelatedtoexpansion}), which assumes a
self-similar expansion (Eq.\ \ref{expansion}) and that after the
first few microseconds the only mechanism effecting the electron
temperature is adiabatic cooling.  At lower $E_e$ and higher
density, similar measurements show that electron heating effects
contribute more significantly and for longer times (Section
\ref{sectionelectronheatingexperiments}). The data shows strong
indications of the approach to a terminal velocity, connecting the
acceleration phase to the measurements of Figs.\ \ref{paperalltemp}
and \ref{expansionpaper}.

\subsubsection{Plasma expansion and ion dynamics}
\label{section_ionexpansion}

The effect of the plasma expansion on the ionic temperature and, in
turn, the influence of the ion relaxation on the expansion, can be
understood qualitatively from the hydrodynamical description of
Section \ref{sectionhydrodyn}.  First, neglecting ionic correlations
($U_\mathrm{ii} = 0$) the expansion leads to adiabatic cooling of
the ions, analogous to the electron dynamics (see Eq.\
(\ref{expansion})). In addition, however, the ion temperature also
changes due to the development of ionic correlations mainly through
disorder-induced heating discussed in Section \ref{sectionDIH} and
described by eqs.\ (\ref{momgau2}).  Moreover, these correlations
reduce the ion-ion interaction (compared to the spatially
uncorrelated case, where pairs of closely neighboring atoms carry a
large potential energy), and therefore lead to an effective negative
acceleration, expressed by the $U_{\rm{ii}}/3$-term in eq.\
(\ref{momgau2b}), in addition to the ideal thermal pressure.  This
contribution, which corresponds to the average nonideal pressure
known from homogeneous systems \cite{ich82,don99}, also leads to an
effective potential in which the ions move.  As they expand in this
potential, the thermal energy increases due to energy conservation.
Finally, (\ref{momgau2b}) shows that the expansion is to the largest
extent driven by the electronic pressure, since $T_\mathrm{e} \gg
T_\mathrm{i}$ even after the disorder-induced heating, since
$U_\mathrm{ii}$ is of the same order as $T_\mathrm{i}$.

\begin{figure}[tb]
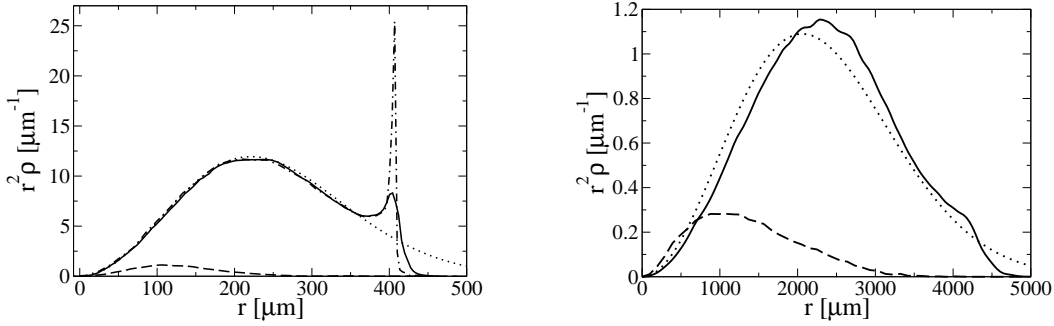

\centerline{\includegraphics[width=2.5in]{PRAfig4a.eps} \hfill
\includegraphics[width=2.5in]{PRAfig4b.eps}}
\caption{\label{PRAfig4} Spatial densities $\rho_{\rm i}$ (solid)
and $\rho_{\rm a}$ (dashed) of the ions and recombined atoms,
respectively, at $t=3\:\mu$s (a) and $t=31.3\:\mu$s (b), compared to
the Gaussian profile assumed for the hydrodynamical model (dotted).
Additionally, $\rho_{\rm i}$ obtained from the particle simulation
using the mean-field interaction only is shown as the dot-dashed
line in (a). Reused with permission from \cite{ppr04PRA}. Copyright
2004, American Physical Society.}
\end{figure}
As the plasma expands, the spatial profile of the ions must deviate
from its original Gaussian shape \cite{rha03}.  This is mainly due
to deviations from quasineutrality, e.g., deviations from the linear
space dependence of the outward directed acceleration, at the plasma
edge.  The influence of the nonlinear correlation pressure on the
density profile is of minor importance, as can be seen by comparing
the solid and dot-dashed line of Fig.\ \ref{PRAfig4}a in the inner
plasma region.  As known from earlier studies of expanding plasmas
based on a mean-field treatment of the particle interactions
\cite{gur66,sac85b,rha03}, a sharp spike develops at the plasma
edge, shown by the dot-dashed line in Fig.\ \ref{PRAfig4}a.  At
later times, this spike decays again when the maximum of the
hydrodynamic ion velocity passes the position of the density peak,
so that the region of the peak is depleted.  Ultimately, at long
times, the entire plasma approaches the profile of a quasineutral
selfsimilar expansion \cite{sac85b}.  From Fig.\ \ref{PRAfig4}a it
becomes apparent that with ionic correlations, the peak structure is
less pronounced than in mean field approximation.  This is due to
dissipation caused by ion-ion collisions which are fully taken into
account in the hybrid-MD simulation.  As shown in \cite{sac85b}, by
adding an ion viscosity term to the hydrodynamic equations of
motion, dissipation tends to stabilize the ion density and prevents
the occurrence of wavebreaking which was found to be responsible for
the diverging ion density at the plasma edge in the case of a
dissipationless plasma expansion. Furthermore, the initial
correlation heating of the ions largely increases the thermal ion
velocities leading to a broadening of the peak structure compared to
the zero-temperature case.

%% file: ElecHeatMech-jm.tex

\begin{figure}[t]
  \centering
  \includegraphics[width=3.0in]{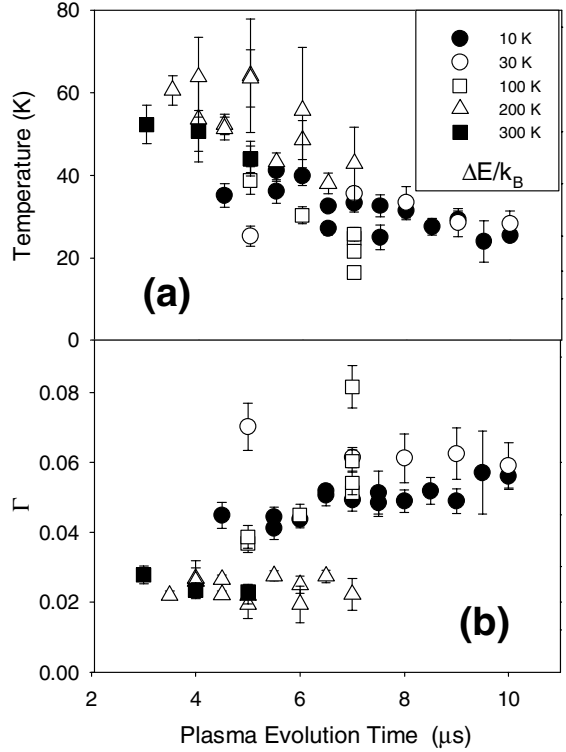}
     \caption {Measured time evolution of the electron temperature (a)
     and electronic Coulomb coupling parameter (b) for initial ion
     numbers of $\sim 10^6$, initial plasma sizes of
     $0.4$\,mm$\le\sigma\le0.7$\,mm and different initial electron
     temperatures.  Reused with permission from \cite{rfl04}. Copyright 2004, American Physical Society.}
 \label{etemp_exp}
\end{figure}
Figures \ref{paperalltemp} and \ref{expansionpaper} suggest that
plasmas strongly deviate from the the simple selfsimilar expansion
(\eq{expansion}) when the plasma becomes dense and the initial
kinetic energy of the electrons, $E_\mathrm{e}$, becomes small,
suggesting that additional physical processes heat the electrons.
This question has been addressed in a series of theory papers
\cite{tya00,hah01,mck02,ppr03,kon02prl,kon02,rha02,rha03,tya01},
proposing different heating mechanisms. The general conclusion is
that the electron heating processes restrict the temperature of the
electrons such that the electron Coulomb Coupling parameter is
limited to
$\Gamma_{\mathrm{e}}$\raisebox{-.6ex}{$\stackrel{<}{\sim}$}$ 0.2$.

The NIST group \cite{rfl04} determined the electron temperature
directly as a function of time after photoionization by tipping the
confining Coulomb potential and measuring the fraction of electrons
that escaped over the lowered barrier.  The potential was tipped by
applying a small electric field pulse.  The result in
\fig{etemp_exp} shows that over a wide range of initial electron
kinetic energies, $E_\mathrm{e}$, the electrons cooled or heated
within a couple of microseconds to $T_\mathrm{e}\approx 60$\, K.
$T_\mathrm{e}$ subsequently dropped over the next 10\,$\mu$s along a
similar curve for all conditions.  While the high-temperature data
are consistent with the collisionless expansion (Eq.\
(\ref{expansion})), the low-temperature results clearly indicate
electron heating as the final temperature is well above the initial
kinetic energy given to the electrons.


\subsubsection{Disorder-induced electron heating}

\begin{figure}[t]
\centering
\includegraphics[width=4.0in]{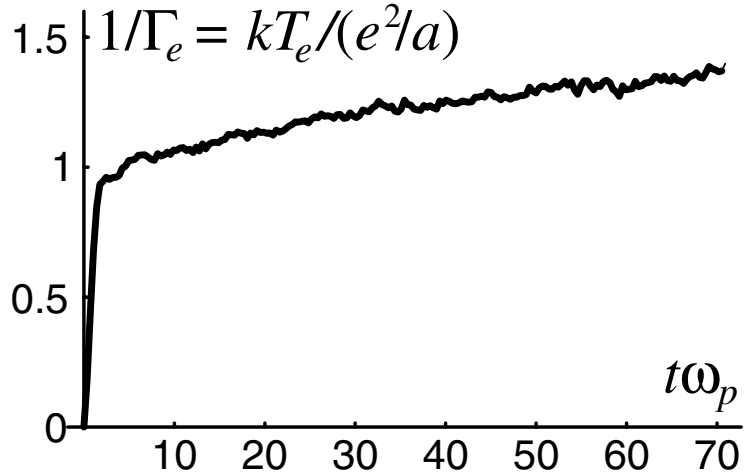}
\caption {Simulated time evolution of electron temperature scaled
by the electron density
  with $T_{\rm e}(0)=0$. Reused with permission from \cite{kon02prl}. Copyright 2002, American Physical Society.}
\label{eheat_kuz}
\end{figure}
While the temperature measurements of \cite{rfl04} were limited to
times larger than a few microseconds due to the finite duration of
the electric field pulse, detailed numerical studies of the
short-time plasma dynamics have been performed by several groups
\cite{mck02,kon02prl,kon02}.  Since an accurate molecular dynamics
treatment requires huge numerical efforts, in
\cite{mck02,kon02prl,kon02} this aspect of the plasma dynamics has
been studied for a smaller model system with a reduced ion-electron
mass ratio.  Fig.\ \ref{eheat_kuz} shows the calculated time
evolution of the scaled electron temperature $1/\Gamma_{\rm e}$, for
the extreme case of both zero initial ion temperature and zero
initial electron kinetic energy, i.e., for an ionizing laser
frequency that exactly matches the ionization potential of the cold
atoms.  Naively, one might expect very strong particle correlations
at these low temperatures.  However, the electron component is not
created in a correlated equilibrium state and heats up in the same
way as dicussed in Section \ref{section_DIH} for the ionic case.
Immediately after photoionization spatial electron-electron and
electron-ion correlations are practically negligible.  They begin to
develop on a timescale of the inverse electronic plasma frequency
$\omega_{\rm p,e}^{-1}=\sqrt{m_{\rm e}\varepsilon_{\rm
0}/e^2\rho_{\rm e}}$ lowering the electron potential energy, which
is released as heat to the electrons.  Due to the small electron
mass the correlation build-up is much faster than for the ions,
heating up the electrons within a few tens of nanoseconds.  Since
there are two contributions from electron-electron and from
electron-ion correlations, we may estimate the final equilibrium
excess correlation energy as two times the expression for $u_{\rm
corr}$ given in \eq{ueqanalyt}.  Energy conservation, $3\Gamma_{\rm
e}^{-1}/2=2u_{\rm corr}$, yields $\Gamma_{\rm e}\approx 1$, in
agreement with the temperature dynamics shown in \fig{eheat_kuz}. An
estimate of the released energy for the expansion velocity
measurement of \fig{expansionpaper} yields a plateau velocity of
$13$\,m/s.  The disorder induced heating alone is therefore not
sufficient to explain the observed enhanced plasma expansion.


\subsubsection{Rydberg atoms formed through three-body recombination}
\label{section_recombination}

\begin{figure}
  \centering
  \includegraphics[width=4in]{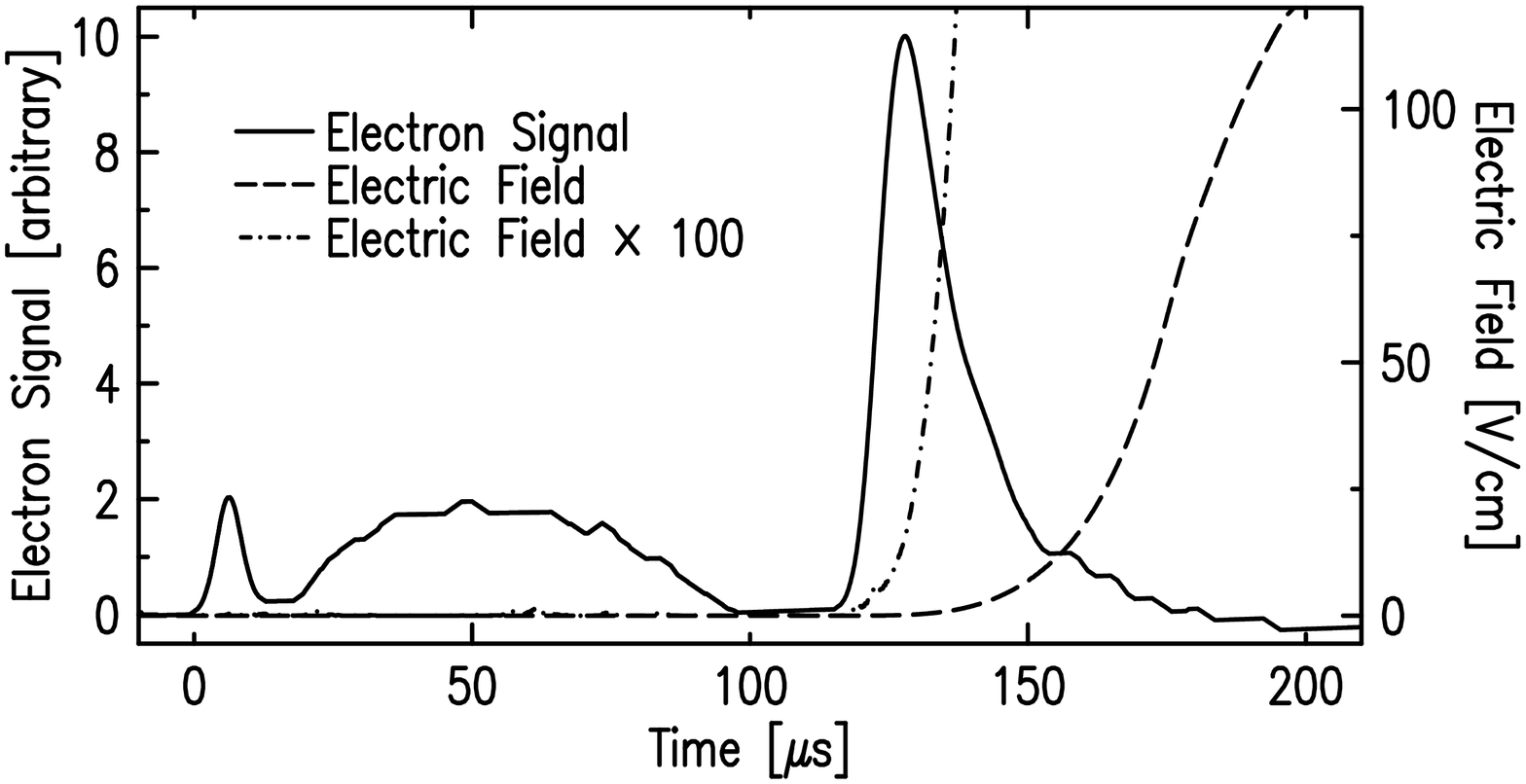}\\
\caption{Electron signal from a plasma created by photoionization of
$10^{5}$ atoms at $t=0$, with $E_{\mathrm{e}}/k_{B}=206\,$K. The
first and second features represent free electrons escaping from the
plasma. The third feature arises from ionization of Rydberg atoms. A
field of $5\,\rm{mV/cm}$ is present before the large field ramp
commences at about $120\,\mu$s.  Electrons from the first two
features are collected and detected with 10 times less efficiency
than the Rydberg electrons.  Reused with permission from
\cite{klk01}. Copyright 2001, American Physical
Society.}\label{fullsignal}
\end{figure}

In three-body recombination (TBR) \cite{sea59,mke69}, an electron
and an ion recombine to form a highly excited Rydberg atom, while
 a second electron participates in the collision to
conserve energy and momentum.  TBR is an important process in
ultracold neutral plasmas, which may be surprising because at
typical plasma temperatures of 1000 K or more, TBR is only important
in dense plasmas.  At low densities on the order of
$10^{10}$\,cm$^{-3}$, recombination typically proceeds by radiative
recombination (RR) and dielectronic recombination (DR)
\cite{sea59,gra92}.

TBR dominates in ultracold plasmas because the total three-body rate
varies with temperature as $T^{-9/2}$.  This divergent temperature
dependence has motivated investigations into whether TBR theory must
be modified in the ultracold regime \cite{hah97,hah00,mty95,tya00}.
In addition to the fundamental interest in TBR at low temperatures,
the process is important in the formation of cold antihydrogen
through positron-antiproton recombination \cite{aab02,gbo02}. In ion
storage rings, collisions between ions and electrons in electron
coolers can be in the 10-100\,K energy range, and anomalously large
recombination rates have been observed for highly charged ions
\cite{ionrings}. Some authors have suggested an enhanced
contribution from TBR to explain a portion of the
excess\cite{psc99}.

In addition to the fast initial rise of the temperature due to
disorder-induced heating, \fig{eheat_kuz} indicates further electron
heating on a slower timescale, which is due to the formation of
highly excited Rydberg atoms.
Figure \ref{fullsignal} shows experimental evidence of recombination
in an ultracold neutral plasma.  The origin of the time axis
corresponds to the photoionization pulse.  Immediately after that,
some electrons leave the sample and are directed to the detector by
a small DC field (5 mV/cm), producing the first peak in the signal.
The ions are essentially immobile on this timescale, and the
resulting excess positive charge in the plasma creates the Coulomb
potential well that traps the remaining electrons \cite{kkb99}. The
broad peak corresponds to the escape of free electrons as the
Coulomb potential well vanishes as the plasma expands. After the
free electrons have escaped, the electric field is increased to
$120\,$V/cm in 50-100\,$\mu$s. This field can ionize Rydberg atoms
bound by as much as $70\,$K, corresponding to a principal quantum
number of about $n=47$.  From the number of electrons reaching the
detector, the number of Rydberg atoms formed can be inferred, and
from the field at which the atoms ionize the distribution of Rydberg
atoms as a function of $n$ can be constructed.  The Rydberg atoms
survive for hundreds of microseconds with no significant change in
their distribution.  This implies that most of the atoms are in high
angular momentum states  which have long radiative lifetimes
\cite{gal94}.

\begin{figure}[t]
\centering
\includegraphics[width=2.5in]{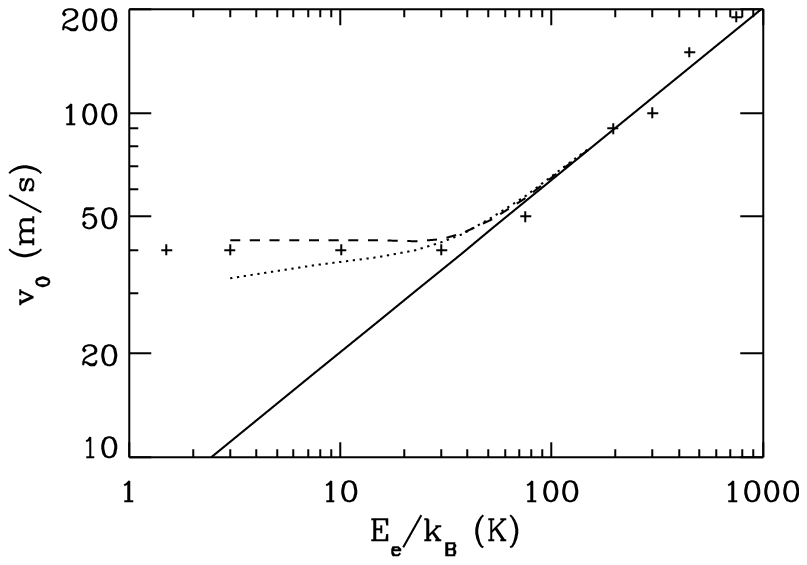}
\includegraphics[width=2.5in]{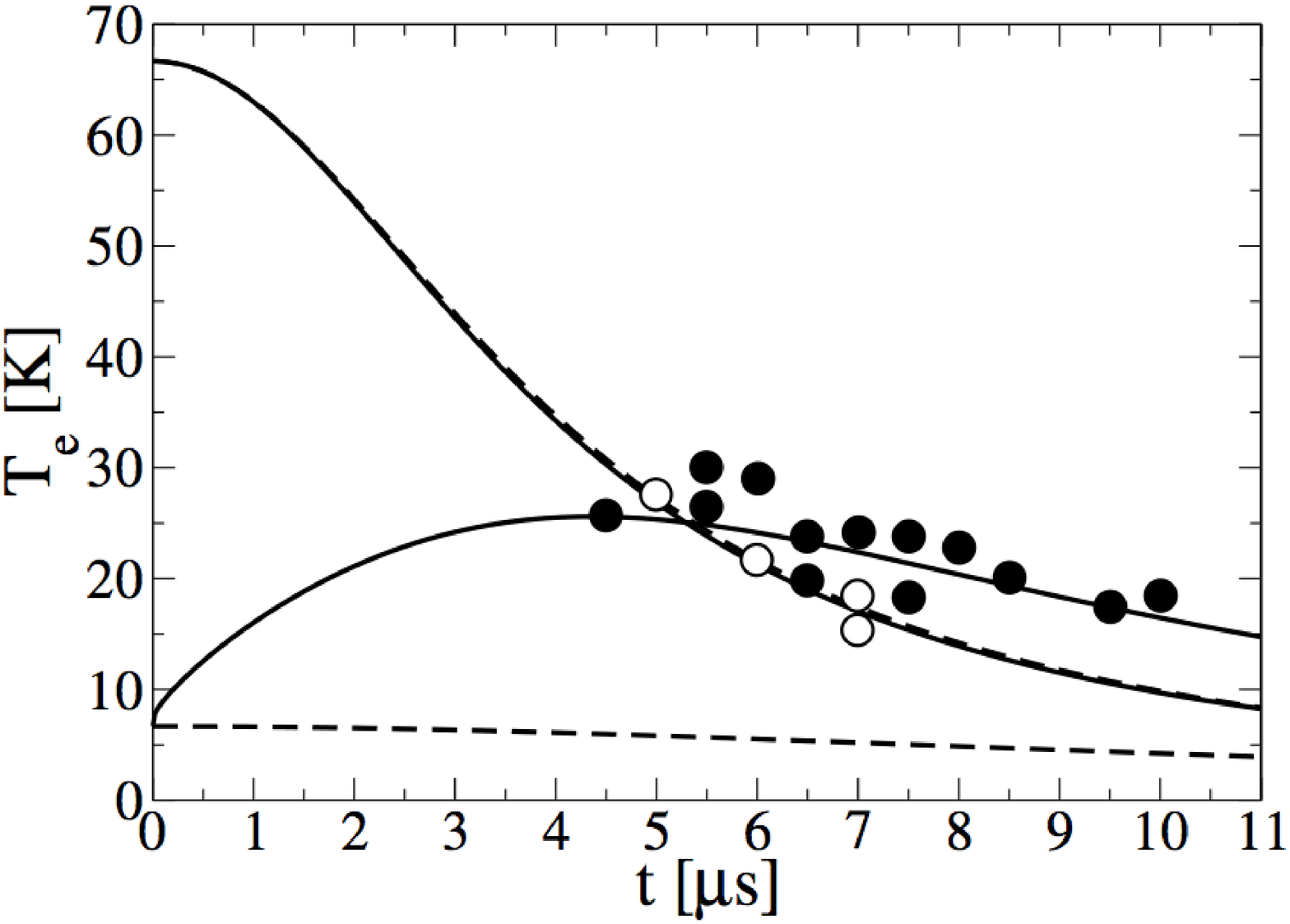}
\caption {(left): Comparison of PICMCC-calculations \cite{rha02} and
  measurements \cite{kkb00} of the plasma expansion velocity.
  Initial state parameters are the same as in fig.\ref{expansionpaper}. Reused with permission from
  \cite{rha02}. Copyright 2002, American Physical Society. (right): Calculated time evolution of the electron
  temperature \cite{ppa05conf} compared to experiment \cite{rfl04} for two different
  initial electron temperatures of $6.7$K (filled circles) and $66.7$ (open circles).
  The remaining parameters are the same as in fig.\ref{etemp_exp}. Reused with permission from
  \cite{ppa05conf}. Copyright 2005, Institute of Physics.}
\label{coll_fig}
\end{figure}

The classical three-body recombination theory, which is well confirmed
for high temperature plasmas, predicts a total recombination rate of
\begin{equation} \label{tbr_total}
K_{\rm tbr}^{\rm (tot)}\approx3.8\cdot 10^{-9}
T_{\rm e}^{-9/2}\rho_{\rm e}^{2}\:\: s^{-1}\;,
\end{equation}
where $T_{\rm e}$ is given in K and the density $\rho_{\rm e}$ is in
cm$^{-3}$.  For typical parameters of $T=9$K and $\rho_{\rm
e}=3\cdot10^9$cm$^{-1}$ this yields an extremely large recombination
rate of $K_{\rm tbr}^{\rm (tot)}=2.4\cdot 10^6$\ s$^{-1}$.  Hence,
one would expect the plasma to almost completely convert into a
Rydberg gas within a few microseconds.  Contrary to this, only a few
percent of recombination is observed after 12$\mu$s for the above
initial conditions \cite{klk01}.  This has led to a series of
theoretical investigations questioning the validity of the standard
recombination theory \cite{tya00} and proposing strong coupling
effects on the recombination rate \cite{hah97,hah02}.  Subsequent
theory and experiments \cite{klk01,kon02,rha02,rfl04} indicate the
opposite, namely that the plasma is driven out of the strongly
coupled regime, by subsequent electron-atom collisions involving the
formed Rydberg atoms, which tend to additionally heat the electron
gas. The corresponding increase in temperature is in fact more rapid
than from direct recombination and quickly quenches further
recombination due to the strong $T_{\rm e}^{-9/2}$ dependence of the
rate.  The simple hydrodynamic description with standard collision
rates \cite{mke69}, as described in Section \ref{section_coll},
yields a surprisingly good description of the plasma dynamics.  As
shown in \fig{coll_fig} the enhanced expansion velocity of Fig.\
\ref{expansionpaper} can indeed be reproduced upon inclusion of
electron-Rydberg-atom collisions, and the hydrodynamic model even
yields a quantitative description of the temperature measurement
\cite{rfl04}.

The measured distribution of principal quantum numbers for the
Rydberg atoms also revealed that moderately tightly bound levels
($n<50$) were populated in the plasma, and that the binding energy
of the Rydberg atoms closely matched the excess energy that appeared
in the expansion as shown in \fig{expansionpaper} \cite{klk01}.

 Although the discussion about the validity of the conventional
 recombination theory under the conditions of an ultracold plasma is still not
 settled, the presented measurements \cite{klk01,rfl04} together with
 supporting theoretical calculations provide strong arguments for the
 applicability of the classical recombination rates even in the low
-temperature regime.

%% file: CoulCoupPar-jm.tex
As stated many times before in this review, fundamental interest in
ultracold neutral plasmas to a large extent stems from the possibility
of creating strongly coupled plasmas.  For a system of charges
embedded in a uniform neutralizing background, formally called a
one-component plasma (OCP) \cite{don99}, local spatial correlations
characteristic of a strongly coupled fluid appear for $\Gamma \ge 2$.
As $\Gamma$ further increases and eventually exceeds a value of $174$
the system crystallizes.  While the OCP model was originally developed
to describe dense neutron stars, with ions embedded in a degenerate
electron background, a clean experimental realization of an OCP is
provided by laser cooled ion clouds confined in Penning
\cite{gil88,ibt98,don99} or Paul \cite{dbh98,kdr03} traps.

In a real two-component plasma (TCP) the situation can be very
different. Besides the possibility of forming bound states,
electrons will tend to shield the ion-ion interaction potential.
For weakly coupled electrons, within the Debye-H\"uckel theory, the
Coulomb interaction is replaced by an interaction potential $V_{\rm
ii}=\exp(-r/\lambda_{\rm D})$ of Yukawa-type.  This shielding can
drastically reduce the ion-ion interaction energy and the
correlations, as approximately expressed by an effective coupling
constant $\Gamma_{\rm i}^{\star} = \Gamma_{\rm i}^{-\kappa}$
\cite{kon02,fha94}, where $\kappa=a/\lambda_{\rm
D}=\sqrt{3\Gamma_{\rm e}}$.  While this is certainly a good
approximation for $\Gamma_{\rm e}\ll 1$ it breaks down as
$\kappa>1$.

Ultracold plasmas  provide the first precise experimental studies of
two-component plasmas with both components in or near the strongly
coupled regime.  As we have discussed, however, various heating
mechanisms currently preclude the exploration of plasmas deep in the
strongly coupled regime ($\Gamma \gg 1$).

\subsubsection{Electronic Coulomb Coupling Parameter}
\label{section_electroncoupl}
\begin{figure}[t]
\centering
\includegraphics[width=4.0in]{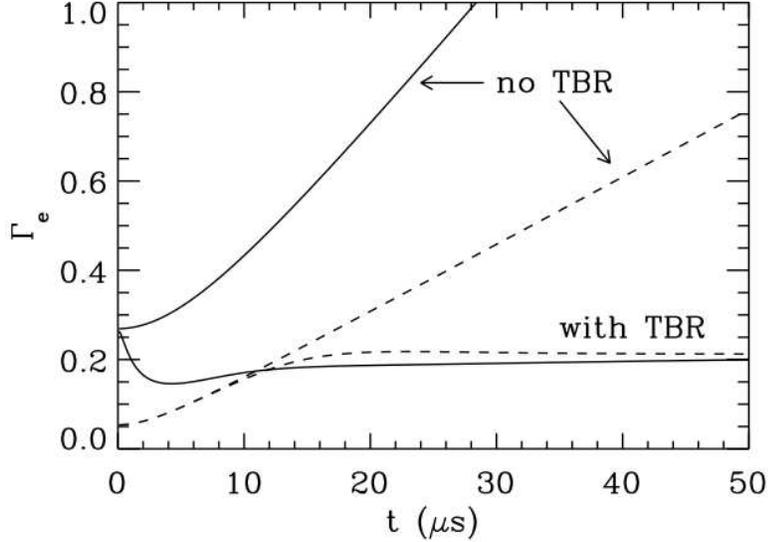}
\caption {Time evolution of the electron Coulomb coupling parameter
for an initial average density of $10^{15}$\,cm$^{-3}$ and two
different initial electron temperatures of $10$\,K (solid lines) and
$50$\,K (dashed lines). Two simulations, including and excluding
Rydberg atom formation are shown for each parameter set. Reused with
permission from \cite{rha02}. Copyright 2002, American Physical
Society.} \label{Ge_rob}
\end{figure}
 At the lowest possible initial electron
temperatures of about $0.1$K, limited by the linewidth of the
ionizing laser, and typical densities of $10^9$cm$^{-3}$ very large
Coulomb parameters of about $30$ might be expected.  As outlined in
the previous section there are however various heating mechanisms
quickly decreasing this value.  After an initial disorder induced
heating phase, the electrons are further heated by Rydberg atom
formation and subsequent electron-atom collisions.  Interestingly,
$\Gamma_{\rm e}$ approaches an almost universal value largely
independent of the initial plasma parameters.  This is demonstrated
in \fig{Ge_rob}, showing results of PICMCC simulations (see Section
\ref{sectionhydrodyn}) for two different initial temperatures
\cite{rha02}.  When three-body recombination is turned off,
$\Gamma_{\rm e}$ increases in time due to the adiabatic cooling of
the electron gas.  When the formation of Rydberg atoms is taken into
account this increase is strongly suppressed by the heating of the
electron gas.  While the initial evolution is qualitatively
different for the high and low temperature case, respectively, both
curves approach a value of $\Gamma_{\rm e}\approx0.2$ within about
$10\mu$s. This prediction has been confirmed by subsequent
temperature measurements, discussed in Section
\ref{sectionelectronheating}. Although such a PICMCC treatment can
not describe any type of particle correlations of the plasma and
hence neglect the disorder-induced electron heating, the measured
value of the plateau is consistent with the theoretical result of
$0.2$, within the experimental uncertainty. This suggests that
Rydberg atom formation indeed constitutes the dominant heating
mechanism and correlation effects do not play a significant role for
the electron dynamics.

\subsubsection{Ionic Coulomb Coupling Parameter}
\label{section_ioncoupl} This situation is different for the ionic
component, whose temperature is practically not affected by
inelastic collisions due to the much larger mass of the ions.
Heating can therefore only arise from correlation build-up. PICMCC
simulations (Fig.\ \ref{Ge_rob}) are unable to describe
$\Gamma_{\mathrm{i}}$, since an accurate treatment of ion-ion
correlations is required, as can be achieved within the hybrid-MD
approach introduced in Section \ref{hyb}.

In order to calculate $\Gamma_{\rm i}$ one needs to know the ion
temperature and density. Experimentally, $T_{\rm i,ave}$ can be
determined from the absorption spectra \cite{csl04} or ion
fluorescence \cite{cdd05}. The density can be determined from
absorption images, knowledge of the neutral atom parameters and the
ionization fraction, or  the timescale of the disorder-induced
heating. In order to estimate the effect of screening,  the electron
temperature must be known. Finally, the inhomogeneity of the density
distribution must be taken into account. One way to do this is to
calculate the density-averaged coupling parameter, $\Gamma_{\rm
i,avg}^*= \langle{\rm exp}[-\kappa(r)]e^2 / [4\pi \varepsilon_0 a(r)
k_B T_{\rm i,ave}]\rangle$. Measurements find for strontium
plasmas $\Gamma_{\rm i,avg}^{*}\approx 2$ after the disorder-induced
heating phase for a wide range of initial conditions \cite{csl04}.
With lower $T_{\rm e}$ and higher density, $\Gamma_{\rm i,ave}^{*}$
was only slightly higher. A similar result, $\Gamma\approx 4$, was
found for calcium plasmas \cite{cdd05}.

The surprisingly small variation in $\Gamma_{\rm i,avg}^*$ suggests
that disorder-induced heating is a natural feedback mechanism that
leads to equilibration into a state just barely in the strongly
coupled fluid regime  independent of the initial conditions. This
conclusion is supported by the qualitative discussion on
disorder-induced heating in Section \ref{sectionDIH}, where it was
argued that the final temperature should be of the order of the
Coulomb interaction energy between neighboring ions, which directly
translates into $\Gamma \approx 1$.

The question remains of how the Coulomb coupling parameter evolves
in time as the system expands.  The simple hydrodynamical model
(Eq.\ \ref{expansion}) predicts that the  ion temperature decrease
during the expansion should overcompensate the accompanying decrease
in density, so that the Coulomb coupling parameter should increase
$\propto \rho^{-1/3}$ without bounds as the system expands.
Experimentally, such a prediction is hard to check, since the
directed radial expansion velocity quickly overwhelms the thermal
velocity component.  Theoretically, on the other hand, it is
straightforward to separate  the thermal velocity components in the
hybrid-MD simulation \cite{ppr04archive}.

\begin{figure}[tb]
\centerline{\includegraphics[width=4in,clip=true]{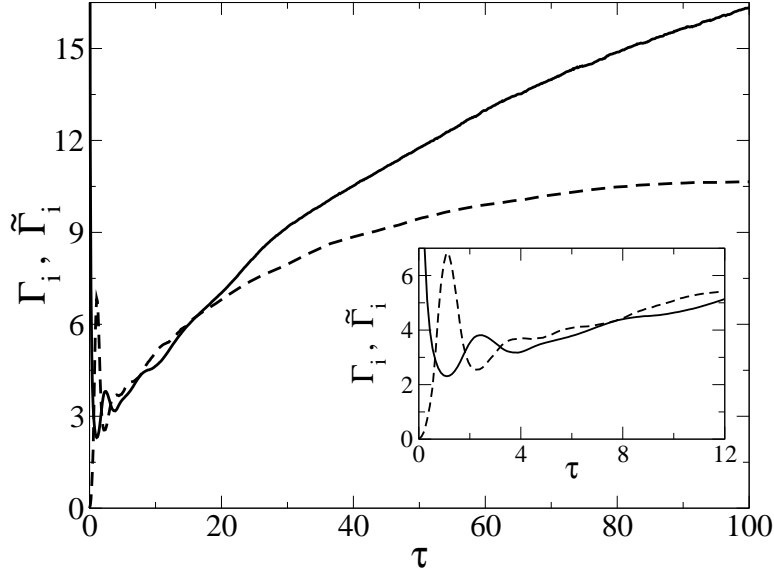}}
\caption{\label{f2pprprl05} Ionic Coulomb coupling parameter for a
plasma with $N_{\rm{i}}(0)=5\cdot10^4$,
$\bar{\rho}_{\rm{i}}(0)=1.1\cdot10^9$cm$^{-3}$ and
$T_{\rm{e}}(0)=50$K. The solid line shows the Coulomb coupling
parameter calculated from the average temperature and density, the
dashed line marks the coupling parameter extracted from pair
correlation functions (see text).  Inset: blow-up of the short-time
behavior.  Reused with permission from \cite{ppr05prl}. Copyright
2005, American Physical Society.}
\end{figure}

Figure \ref{f2pprprl05} shows that $\Gamma_{\rm i}$ does increase as
the system expands.
Hence, one might conclude that it is sufficient to simply wait long
enough to allow the system to cool down in order to reach the
crystallized state. However, it is questionable whether one should
speak of a ``crystal'' if the density drops below that of
interstellar space by the time it reaches a $\Gamma$ of 200. But
there is also a fundamental problem; the coupling parameter
$\Gamma_{\rm i}$ measures the amount of correlations in a plasma if
the system is in {\em in thermodynamic equilibrium}, but the plasma
is created far from equilibrium in a completely uncorrelated initial
state. Disorder-induced heating leads to local local thermal
equilibrium, but the plasma continuously expands, so a true static
equilibrium is never reached. One must compare the coupling
parameter formally calculated from temperature and density with the
spatial correlation properties of the system at all times.

In \cite{ppr05prl}, this question was studied by extracting an
alternative coupling parameter $\tilde{\Gamma}_{\rm i}$ from the
pair correlation functions $g_{\rm ii}(r)$ obtained with the
hybrid-MD approach (dashed line in \fig{f2pprprl05}).  In line with
the preceding discussion, $\Gamma_{\rm i}$ and $\tilde{\Gamma}_{\rm
i}$ initially have no similarity at all, reflecting the fact that
the plasma is very far away from any kind of equilibrium.  (In
addition, Fig.\ 3(a) of reference \cite{ppr05prl} shows that the
pair correlation function at early times is not well fit by that of
an equilibrium system, so that a parametrization of $g_{\rm ii}$ by
a single scalar quantity $\tilde{\Gamma}_{\rm i}$ does not
adequately describe the plasma's spatial correlation properties.
For $5<\omega_{\rm p,i}t<20$, the two curves for $\Gamma_{\rm i}$
and $\tilde{\Gamma}_{\rm i}$ approach each other and run in
parallel, suggesting that the plasma is close to a changing
quasi-steady state during this stage.

For $\omega_{\rm p,i}t>20$, however, the two curves diverge again, and
$\tilde{\Gamma}_{\rm i}$ settles towards a finite value of $\tilde{\Gamma}_{\rm i}
\approx 11$ deep in the strongly coupled regime, but still far away
from a value where the occurrence of long-range order might be
observed.  In other words, while the system expands and cools down
(reflected in the further increase of $\Gamma_{\rm i}$), the spatial
correlations in the system (parametrized by $\tilde{\Gamma}_{\rm i}$) do not
change anymore.  On a formal level, this freeze-out of
correlations is caused by another violation of the Bogoliubov
assumption \cite{bog46}, namely a cross-over of the timescale for
relaxation of correlations with the hydrodynamic timescale on which
macroscopic system parameters change. More intuitively, the
build-up of spatial correlations requires strong interactions between the
ions in order to cause a rearrangement of their spatial correlational structure.
As the system expands, the timescale on which such rearrangement
takes place becomes longer and longer as the inter-ionic forces
become weaker and weaker.  At some point, the system simply becomes
so dilute that the rearrangement does not occur anymore,
and the spatial structure of the plasma freezes out in a selfsimilar
expansion.  As a consequence, the plasma remains in a state governed
by strong short-range correlations, but no long-range order should
form.

%% file: future_strong_intro-jm.tex
As emphasized in various sections of this article, one of the main
motivations for studying ultracold plasmas from a plasma-physics
perspective is the fact that they are strongly coupled, and one
might hope to create crystalline plasmas. Coulomb crystallization is
actively studied in one-component systems, such as magnetically
confined ions \cite{jen05,dub05} and dusty plasmas
\cite{tmd94,abp04,bba06}. Equivalent studies in two-component
plasmas, on the other hand, are more scarce, as their experimental
realization poses several additional constraints \cite{bff05}.  In
this respect ultracold neutral plasmas have attracted considerable
attention as a promising opportunity to observe this phenomenon in
the laboratory.

However, as discussed above, three-body recombination strongly heats
the plasma electrons, while the disorder-induced heating rooted in
the (spatially) uncorrelated nature of the initial plasma state
(i.e., after creation of the plasma by photoionization) rapidly
heats electrons and ions.  This was realized soon after the first
experiment, and different suggestions have been put forward to
mitigate these effects.

For ions, these suggestions may be divided into two groups. On the
one hand, one may  avoid the disorder-induced heating. Since this
heating effect is directly related to the uncorrelated initial state
of the plasma, such approaches always aim at introducing some amount
of correlation in the spatial distribution of the atoms
\emph{before} the plasma is formed by photoionization. We will
discuss such approaches in Section \ref{OtherApproaches} below.  On
the other hand, one might  counteract the disorder-induced heating
by removing energy from the system during its evolution rather than
avoiding the heating from the outset.  In this case, one needs a way
to couple the system to its environment in order to dissipate part
of its internal energy, i.e., to cool it. Doppler laser cooling has
been suggested as a way to achieve this goal.  Since it is, in our
opinion, the most promising perspective for achieving plasma
crystallization at this point, we will discuss it in detail in
Section \ref{LaserCoolingIons}. Finally, a recent proposal to
increase the coupling of the plasma electrons by adding Rydberg
atoms to the plasma, the only suggestion so far aiming at the
\emph{electronic} component of the plasma, will be briefly discussed
in Section \ref{IceCubes}.

%% file: IonCoupling-jm.tex
\subsubsection{Laser-cooling the ions}
\label{LaserCoolingIons} \label{section_lcool} As stated above, the
idea behind laser cooling the plasma ions is to counteract
disorder-induced heating by removing energy from the system via an
external cooling mechanism. Such an approach is experimentally
feasible if alkaline earth elements are used for the plasma, as
their singly charged ions have a single outer electron which
provides a convenient cooling transition for currently available
laser systems. Hence, a ``standard'' Doppler cooling scheme may be
used to cool the ions, with an optical molasses created by six
counterpropagating lasers in a 3d configuration (cf.\ Fig.\
\ref{apparatus}).  The plasma ions can then be cooled during the
whole evolution of the system.

Experimentally, the scheme outlined above has not yet been realized.
However, there is little doubt that it should be
feasible\footnote{It should be noted, though, that the required
cooling rates are more easily achieved for light ions such as Be.}.
In fact, a first step in this direction has already been taken with
the Doppler imaging of the ion dynamics described in detail in
Section \ref{sectionopticalprobes}, since the same transition that
is used for the imaging can also be used for the cooling.
Theoretically, the above scenario has been studied in detail in
\cite{ppr04,ppr05jpb}.

Originally, the idea of Doppler laser cooling the ions was proposed
in \cite{kon02} and \cite{kag03}.  These studies also correctly
pointed out the potentially crucial role of elastic electron-ion
collisions in the plasma in such a scenario.  In a freely evolving
plasma, such collisions are usually negligible, due to the fact that
the timescale for electron-ion thermalization is much longer than
the timescale of the plasma expansion.  Hence, the heating of the
ions during the first microsecond of evolution due to these
collisions is limited to a few milli-Kelvin, which is negligible
compared to the additional disorder-induced heating which is orders
of magnitude stronger.  In the case of laser cooling, however, the
final ion temperature one is aiming for is also in the milli-Kelvin
range, i.e., of the same order as the energy transferred from
electrons to the ions via elastic collisions. Hence, these
collisions should be taken into account when discussing the
possibility of plasma crystallization.  In \cite{kon02,kag03}, their
influence on the achievable ionic temperature was estimated using
reasonable values for initial-state temperatures and densities.  It
was concluded that the ionic Coulomb coupling parameter would be
limited to a value $\Gamma_{\rm i} \approx 50$ due to these elastic
collisions, outside of the crystallized regime. However, these
estimates  assumed a constant heating rate corresponding to the
chosen initial conditions.  What was not taken into account is the
expansion of the plasma, which strongly reduces  the electron
temperature and density, which reduces the collisional heating. As a
consequence, these estimates are too pessimistic.

A detailed numerical investigation of the laser cooling process was
carried out in \cite{ppr04,ppr05jpb}, using both the hydrodynamic
approach and the hybrid-MD approach introduced in Section
\ref{sectiontheoreticalaspects}.  Insight into the modifications
induced by the laser cooling can be obtained from a hydrodynamical
description.  On the level of a kinetic description, the influence
of the cooling laser can be modeled via a Fokker-Planck term  in the
time evolution of the ionic distribution $f_{\rm i}$, \beq
\left(\frac{\partial f_{\rm i}}{\partial
t}\right)_c=\beta\left[\nabla_{\mathbf{v}} \left({\mathbf{v}}f_{\rm
i}\right)+\frac{k_BT_D}{m_{\rm i}}\Delta_{\mathbf{v}}f_{\rm
i}\right]\,, \eeq where $\beta$ is the cooling rate and $T_D$ the
Doppler temperature.  A derivation analogous to Section
\ref{sectionhydrodyn} leads to the macroscopic equations for the
evolution,
\begin{subequations}\label{lascool_macro}
\begin{eqnarray}
\label{lascool_macroa}
\frac{\partial}{\partial t}\sigma^2&=&2\gamma\sigma^2  \\
\label{lascool_macrob}
\frac{\partial}{\partial t}\gamma&=&\frac{\left(k_{{\rm B}}T_{{\rm e}}
+k_{{\rm B}}T_{{\rm i}}\right)}{m_{{\rm i}}\sigma^2}-\gamma\left(\gamma-\beta\right) \\
\label{lascool_macroc}
\frac{\partial}{\partial t} (k_{{\rm B}}T_{{\rm i}})&=&-2\gamma
k_{{\rm B}}T_{{\rm i}}-2\beta\left(T_{\rm i}-T_c\right) \\
\label{lascool_macrod}
\frac{\partial}{\partial t} (k_{{\rm B}}T_{{\rm e}})&=&-2\gamma
k_{{\rm B}}T_{{\rm e}}\;.
\end{eqnarray}
\end{subequations}
As readily observed from \eq{lascool_macro} the laser cooling
dramatically modifies the expansion, not only quantitatively, but
even qualitatively.  While the width $\sigma$ of the plasma cloud
increases linearly in time for the free expansion, it only varies as
$\propto t^{1/4}$ in the laser-cooled case.  As a consequence, the
plasma expands much slowly  than in the case of free expansion. This
raises the hope of achieving a long-range ordered state, since we
have seen in the previous section that it is the decrease in density
and the corresponding increase in relaxation timescale that leads to
a freeze-out of spatial correlation properties for a freely
expanding plasma.  In the laser-cooling scenario, the much slower
decrease in density should allow for much longer times during which
efficient rearrangement processes may occur, a prerequisite for the
build-up of long-range order. One may, at first glance,
think that this problem might be circumvented in any case by just
increasing the cooling rate $\beta$ sufficiently.  However, this is
not the case, since a too rapid cooling could freeze the plasma in a
metastable disordered state. The system needs a certain time for
spatial rearrangement to occur, which will not happen if the cooling
is too fast.

\begin{figure}[tb]
\includegraphics[width=2.6in,clip=true]{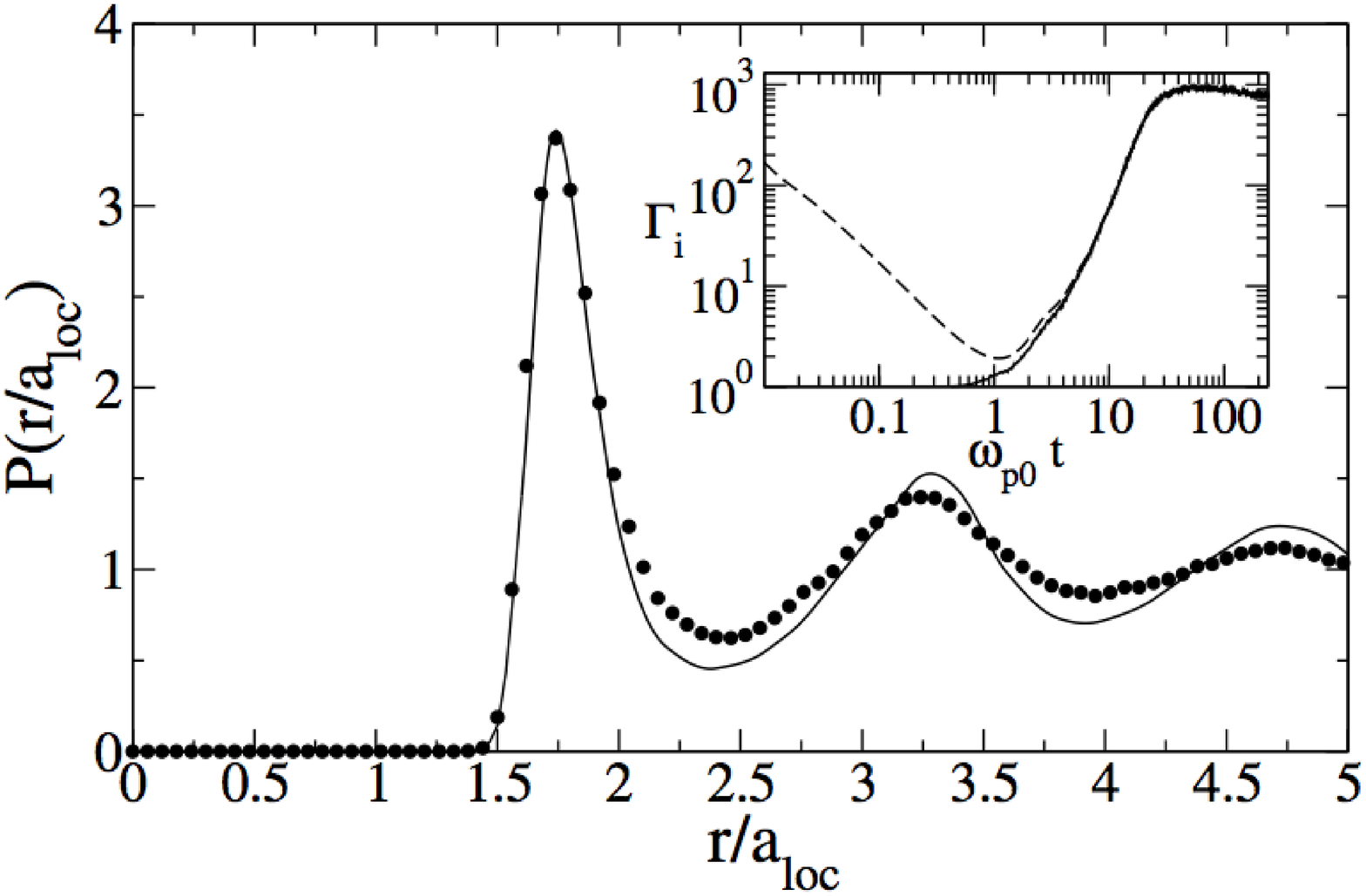}\hfill
\includegraphics[width=2.6in,clip=true]{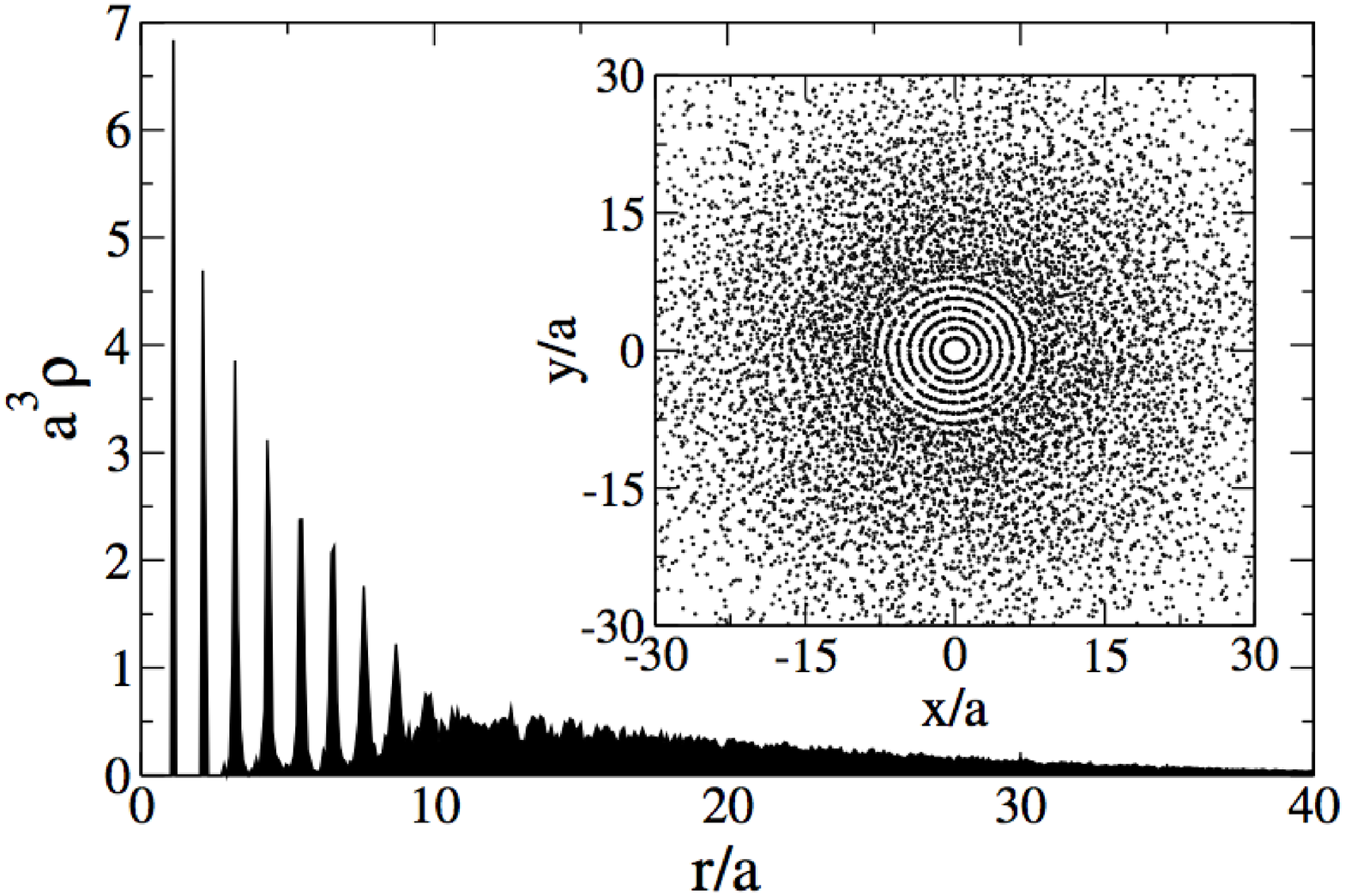}
\caption{\label{crystal} Left: Distribution of scaled ion distances
after an expansion time satisfying $\omega_{\rm p,i}t=240$ (dots).
The comparison with the correlation function of an OCP at
$\Gamma_{\rm i}=700$ indicates the development of strong order.  The
inset shows the evolution of the ionic Coulomb coupling parameter
determined from the correlation energy (solid line) and from the ion
temperature and density (dashed line). Right: Shell structure
formation during the plasma expansion demonstrated by the radial ion
density and a two dimensional cut through the plasma. Reused with
permission from \cite{ppr04}. Copyright 2004, American Physical
Society.}
\end{figure}

In order to obtain information about the build-up of spatial
correlations, one needs to go beyond a kinetic description of the
ion distribution function.  In \cite{ppr04} this was achieved within
the hybrid-MD approach of Section \ref{sectionhybrid}, where the
influence of the cooling lasers on the ions is described by a
Langevin term consisting of a dissipative cooling force and a
stochastic diffusive part.  In addition, elastic electron-ion
collisions should be taken into account in the description as
discussed above.  Such collisions can be implemented using a Monte
Carlo procedure described in detail in \cite{ppr05jpb}.  It is found
that the resulting structures formed during the system evolution may
 depend strongly on the initial-state parameters, such as number of
ions, electron temperature, and cooling rate.  Thus, the electronic
temperature must be chosen sufficiently high in order to keep
three-body recombination at a negligible level.  As discussed in
\cite{ppr04}, three-body recombination creates doubly-excited and
hence autoionizing states with an excess energy of several thousand
Kelvin above the continuum limit.  Hence, any autoionization process
results in a fast electron which would not be bound in the ionic
potential well, thereby destroying the plasma.

\begin{figure}[tb]
\includegraphics[width=1.8in,clip=true]{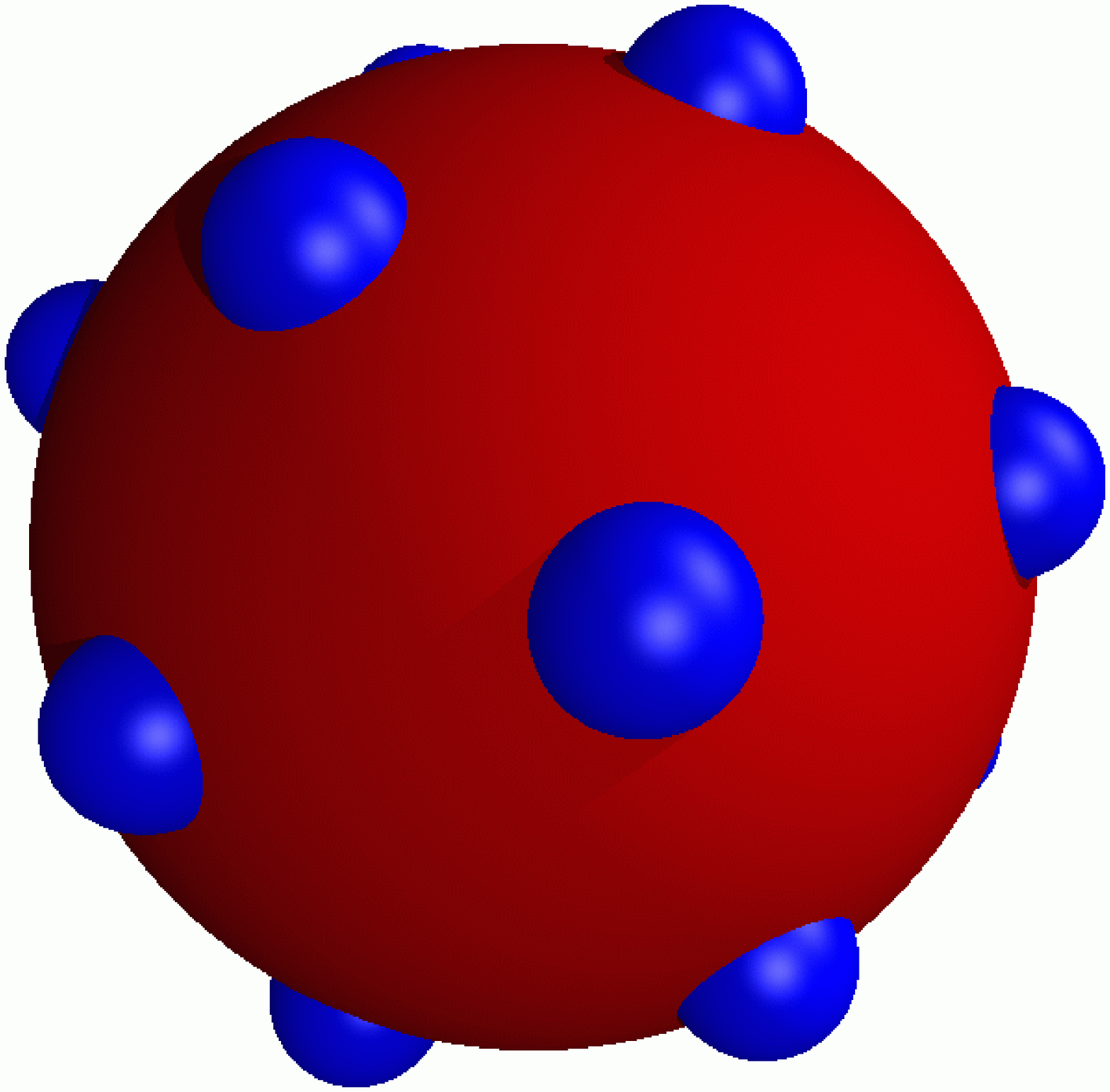}\hfill
\includegraphics[width=1.8in,clip=true]{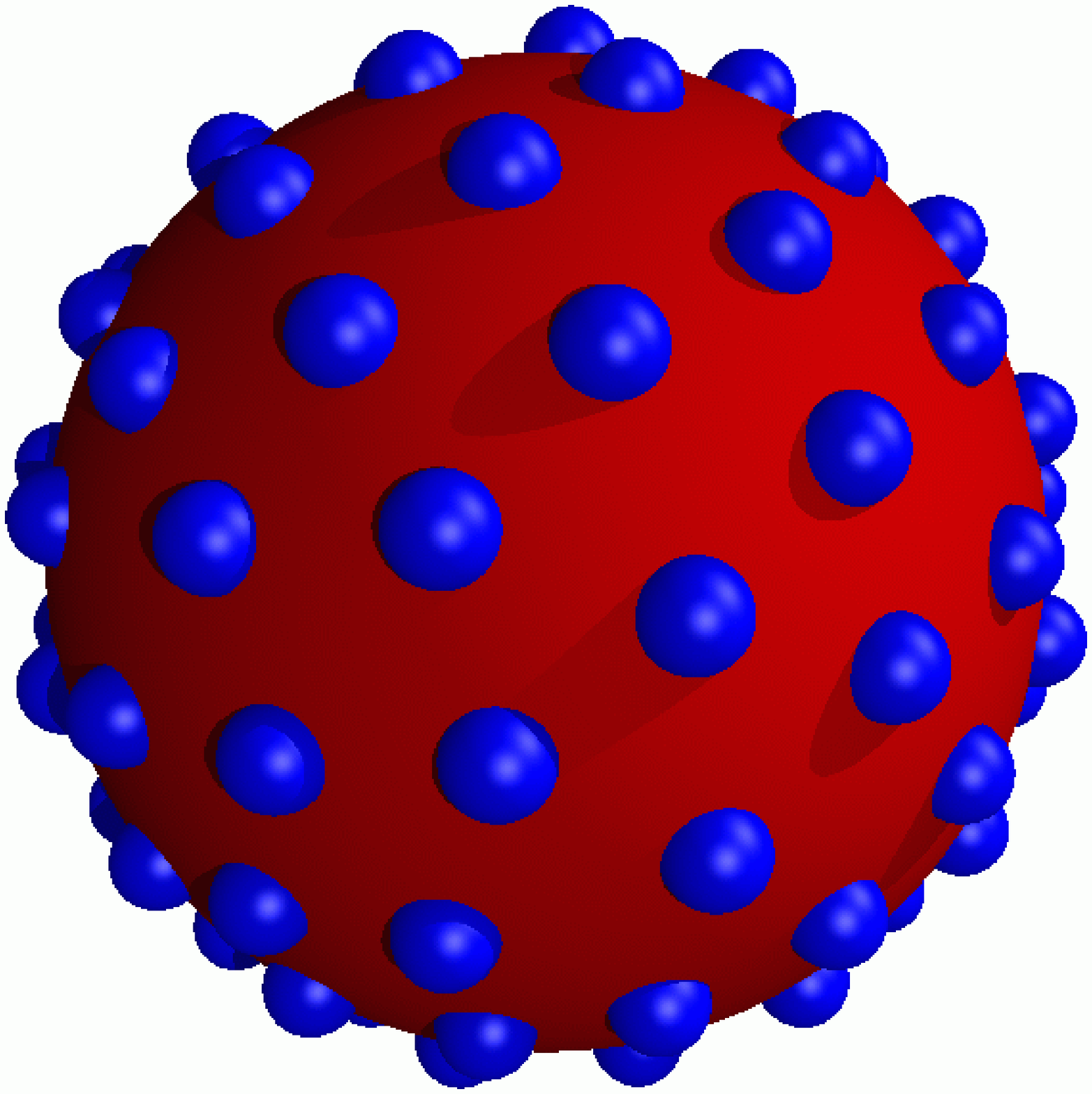}\hfill
\includegraphics[width=1.8in,clip=true]{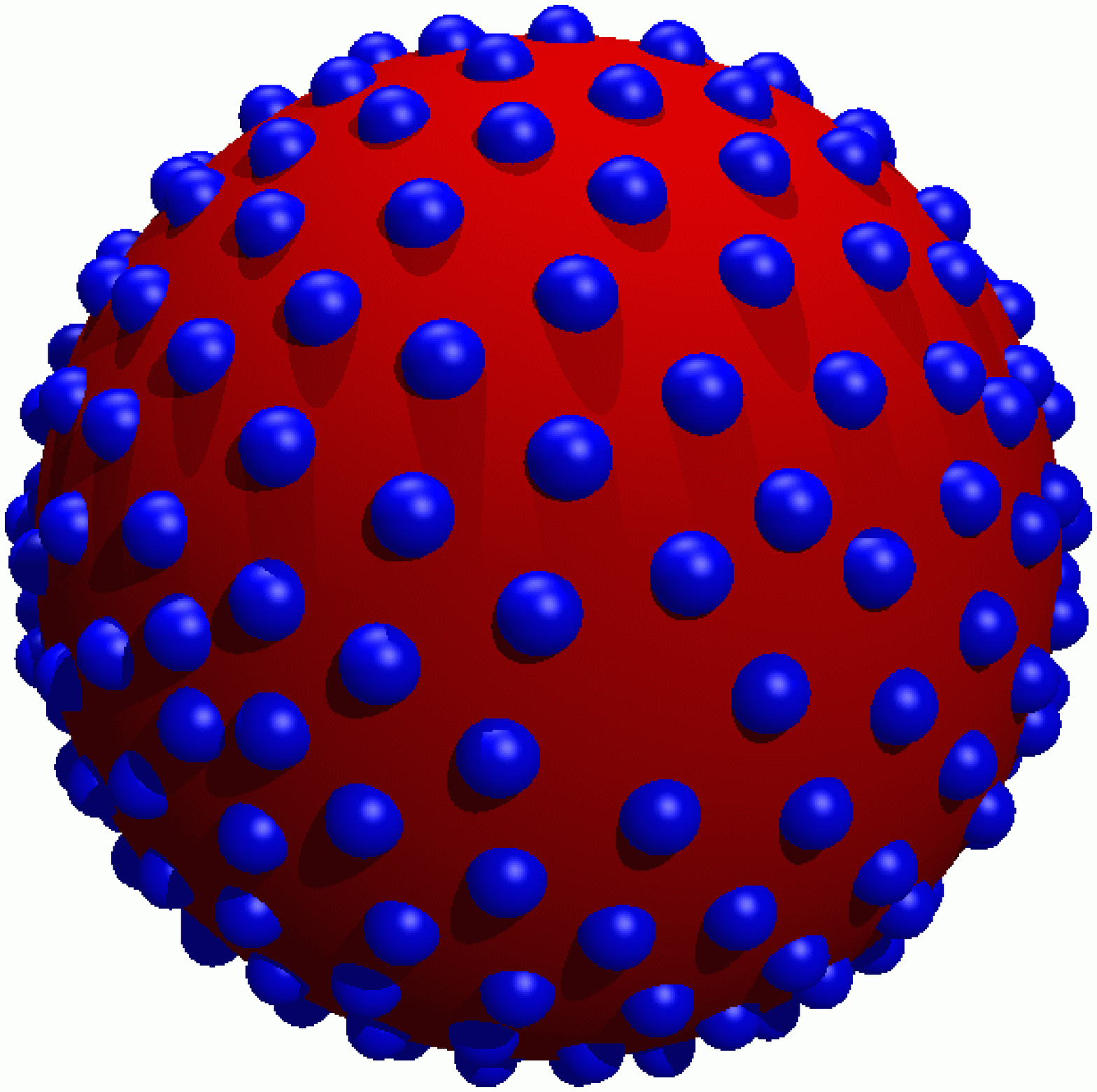}
\caption{\label{crystal2} Ionic intra-shell structure of the 1st,
3rd and 5th shell (from left to right). Reused with permission from
\cite{ppa05conf}, see also \cite{ppr04}.}
\end{figure}

There are also significant constraints on the cooling rate $\beta$.
As discussed above, if the cooling is too fast, the plasma will be
frozen in a disordered glass-like state.  On the other hand, if the
cooling is too slow, the density decreases too strongly for
effective inter-ionic interactions to take place.  As we have seen
in Section \ref{sectioncoulombcoupling}, reaching low temperatures
and correspondingly high coupling parameters does not yet ensure
spatial order in the plasma.  However, it was demonstrated in
\cite{ppr04,ppr05jpb} that, depending on particle number and cooling
rate, glass-like order or the formation of concentric shells may
occur (see Fig.\ \ref{crystal} and Fig.\ref{crystal2}).

While the precise dynamics of this structure formation is not yet
completely understood, it appears sufficiently robust that the
crystallization should be observable experimentally with current
state-of-the-art cooling techniques.  Such an experimental
confirmation, together with further detailed experimental and
theoretical studies, would allow significant physical insight into
the dynamics of strongly coupled two-component plasmas at the
liquid-crystal phase transition.  In particular, there are both
similarities and differences with laser-cooled ion clouds in ion
traps, where shell structures are also observed and intensively
studied \cite{gil88,bir92,bol03,hor02,dub88,sch03c,tot88}. It may
appear surprising that long-range order may occur in the present
case of an unconfined plasma where, in contrast to trapped ion
clouds, the crystallization process can only take place over a
finite amount of time. Pronounced differences also exist in the
dynamics of the crystallization process, which starts at the outer
edge of the ion cloud in the case of ions in a trap, but is observed
to proceed from the center of the plasma in the present scenario.
Moreover, the spherical structure of small ion crystals can be
attributed to the spherical symmetry of the trapping potential,
while in the present case it must be traced to the spherical
symmetry of the initial density distribution. This raises the
question of what is the expected symmetry resulting from an
ellipsoidal initial-state density \cite{cdd05}.

\subsubsection{Other Approaches}
\label{OtherApproaches}

The crystallized regime would be reached more directly by avoiding
the disorder-induced heating right from the start, rather than
counteracting it by external cooling.    Avoiding (or at least
weakening) it requires a certain amount of spatial order in the
initial state, i.e., one must generate a significant degree of
spatial correlation before the plasma is created by photoionization.
A number of  proposals exist for how such an ordered state may be
obtained, namely ({\em i}) using fermionic atoms cooled below the
Fermi temperature, so that the Fermi hole around each atom prevents
the occurrence of small interatomic distances \cite{mur01}; ({\em
ii}) an intermediate step of exciting atoms into high Rydberg
states, so that the interatomic spacing is at least twice the radius
of the corresponding Rydberg state \cite{gmu03}; and ({\em iii})
using optical lattices to arrange the atoms
\cite{gmu03,ppr04jphysb}. Calculations show that, indeed, these
proposals could achieve a significant reduction of the
disorder-induced heating, and the latter may even result in
correlation-induced cooling of the plasma \cite{ppr04jphysb}.
However, realizing the crystallized regime in this way turns to be
very challenging experimentally and the laser-cooling approach
discussed in the previous section seems to be more practical  with
current experimental capabilities.

%% file: icecubes-jm.tex
All the scenarios discussed in the previous Section
\ref{IonCoupling} aim at increasing the coupling of the \emph{ions}
in order to reach the crystallized regime.  On the other hand, it
would also be highly desirable to achieve strong coupling of the
plasma \emph{electrons}.  Many previously employed assumptions, such
as the dependence of three-body recombination rates on density and
temperature, or the Thomson cut-off for the existence of Rydberg
states are expected to fail in this regime. Moreover, this would
provide an intriguing realization of a classical two-component
system where both components are strongly coupled, with exciting
prospects for observing new and interesting phenomena.

\begin{figure}[bt]
\centerline{\includegraphics[width=4in,clip=true]{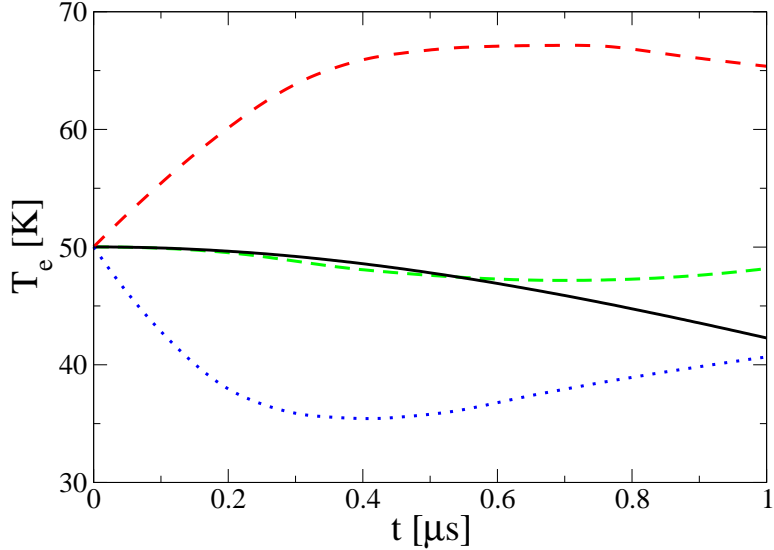}}
\caption{\label{icefig} Time evolution of $T_{\rm e}$ for a plasma
of $N_i = 80,000$ cesium ions with $T_{\rm e}(0) = 50$\,K and
$\sigma(0) = 155$\,$\mu$m, after addition of $N_{\rm R} = 80,000$
Rydberg atoms with principal quantum numbers $n=26,31,36$ (dashed
lines), compared to that of the pure plasma without additional
Rydberg atoms (solid line). Reused with permission from
\cite{vct05}. Copyright 2005, American Physical Society.}
\end{figure}

The only proposal so far targeting the electronic coupling parameter
was introduced in \cite{vct05}.  The basic idea consists of
deliberately adding Rydberg atoms to the plasma, which should then
be collisionally ionized by the free plasma electrons, thereby
reducing the kinetic energy of the electrons and cooling them.  Such
a scenario has been studied systematically in \cite{pczvpp06}, and
it was found that a moderate but still significant manipulation of
the electron temperature is possible in this way. Figure
\ref{icefig} shows the time evolution of $T_{\rm e}$ for a plasma
when Rydberg atoms with varying principal quantum number $n$ are
added at $t=0$.
As can be seen, adding Rydberg atoms with $n=36$ to the plasma leads to a
significant reduction of $T_{\rm e}$ compared to the pure plasma without additional
Rydberg atoms, in agreement with the original proposal \cite{vct05}. On the
other hand, this effect strongly depends on the principal quantum number
of the added Rydberg atoms. For lower excitation ($n=26$ in the present case),
even the reverse effect of a heating of the plasma electrons rather than a
cooling can be observed. This behavior, which might be surprising at first
glance, is connected with the existence of a bottleneck in the transition
rates for the inelastic electron-impact induced bound-bound and bound-free
transitions of the atoms. If the binding energy $E_{\rm b}$ of an atom is
larger than a critical (temperature-dependent) energy $E_{\rm c} \approx 4
k_{\rm B} T_{\rm e}$, the atom is predominantly driven to even lower states by
collisions with the plasma electrons, thereby heating them. If, on the
other hand, $E_{\rm b}$ is smaller than $E_{\rm c}$, collisions tend to
excite the atom to higher bound states or ionize it, and the plasma electrons
are cooled as originally envisioned.

Hence, by varying the principal quantum number of the added Rydberg
atoms, it is possible to control the rate and even the sign of the
change of the electron temperature. On the other hand, varying their
number influences the absolute magnitude of the temperature change
as well as the timescale over which cooling can be achieved.
Together, a remarkable amount of control over the time evolution of
the electronic temperature appears possible, in particular if one
considers the possibility of more sophisticated schemes of adding
the Rydberg atoms. Rather than adding all atoms simultaneously at
$t=0$, one may think of adding them continuously over a certain
amount of time, or in a number of individual pulses (or generally
speaking ``on demand''). One can also change the principal quantum
number of the atoms in time. (The latter might be necessary, e.g.,
to stay away from the bottleneck separating cooling from heating,
which changes as $T_{\rm e}$ changes in time.)

In addition to the temperature reduction discussed above, the
collisional ionization of Rydberg atoms also increases the density
of the plasma.  Thus, both effects act cooperatively, increasing the
Coulomb coupling parameter $\Gamma_{\rm e}$ as desired.  It was
found in \cite{pczvpp06} that, using relatively simple schemes for
the addition of the atoms, $\Gamma_{\rm e}$ reaches values of
$\Gamma_{\rm e} \approx 0.5$, as compared to $\Gamma_{\rm e}
\lesssim 0.2$ found in the experiments with pure
plasmas\footnote{Experimental results so far are consistent with
this theoretical prediction, but still inconclusive since current
schemes of extracting an electron temperature from the experiment
are not very precise.}.  In more advanced scenarios as outlined
above, reaching the strongly coupled state of $\Gamma_{\rm e}
> 1$ should most probably be realizable.  The coupling
strengths obtainable are  certainly far away from the regime where a
crystallization of the electrons might be observed.  Still,
pronounced liquid-like correlation effects may be anticipated, and
the first realization of a two-component system where both the
electronic and the ionic component are strongly coupled promises
exciting new developments.

%% file: future-jm.tex
In this article we have reviewed the development in research on
ultracold neutral plasmas since their first creation in 1999. We have described
numerous experimental tools to probe various aspects of the system behavior, ranging from optical imaging of individual ions to employing collective plasma excitations as probes
for macroscopic quantities. By combining such techniques, traditionally rooted in both atomic
and plasma physics, these experiments have studied a number of plasma physics
problems. They have revealed several surprises, which fosters continuing theoretical interest. We have outlined the
theoretical concepts, developed to describe and understand the
non-equilibrium dynamics of ultracold neutral plasmas. Our discussion has covered diverse approaches, such as simple reduced dynamical equations, kinetic or hydrodynamic descriptions and demanding microscopic approaches.

Based on the techniques introduced we have given a detailed description of the underlying
physical processes driving this quite complex dynamical behavior.
Within a hand-in-hand experimental and theoretical discussion a step-wise
developed physical picture of the plasma dynamics over various evolution stages
from its creation until its complete expansion into the surrounding vacuum has been given.
Thereby we have covered a wide spectrum of phenomena, ranging from low-temperature collisions and
Rydberg atom formation to the physics of collective plasma
excitations and effects of strong correlations between the plasma constituents.

This wealth of interesting physics revealed in the past years
demonstrates the rapid advancement in our understanding of these
fascinating systems, but also raises various open questions, which
we have pointed out in this article. Future studies in this
 young research field will certainly produce unexpected
results, but several trends are likely to continue for the field.

As discussed in Section \ref{sectionachievingstringcoupling}, the
pursuit of stronger coupling in both the electrons and ions will
remain an exciting and therefore intensively followed avenue of
research.  Experimental routes for increasing the ion coupling
constant, such as laser-cooling \cite{kon02,kag03} or photionizing a
pre-ordered sample \cite{mur01,ppr04jphysb}, may enable the
formation of Wigner Crystals \cite{ppr04}.  This will hopefully
allow exploration of the phase diagram of particles interacting
through a screened Yukawa potential \cite{hfd94,hfd96,hfd97} that
will compliment studies of dusty plasmas \cite{mtk99}.  Creation of
colder and denser electrons will show whether strongly coupled
electrons can be realized in ultracold plasmas or three-body
recombination precludes this possibility.

Exploring the regime of stronger electron coupling will necessarily
probe the complicated evolution of the electron temperature when
rapid thermalization, disorder-induced heating, threshold lowering,
and three-body recombination play significant roles.  Powerful
theoretical tools have been developed to address these phenomena
\cite{kon02prl,kon02,mck02,rha02,rha03,tya00,ppr04PRA}, but there is
still need for a comprehensive study of evolution in various regimes
of initial conditions and a disentangling of the many contributing
effects.  As emphasized in \cite{rha03}, the synthesis of ideas from
atomic physics and plasma physics is crucial for this research.
While techniques exist to  study electron temperature at later times
\cite{rfl04,fzr06}, at present there is no experimental diagnostic
to cleanly explore the electron equilibration phase because
electrons evolve so quickly.  The development of such diagnostic
would be a significant advance.

As the field grows, there is great possibility to increase the
connection to traditional areas of plasma physics.  The role of
strong-coupling and the equilibration and expansion dynamics
partly resemble the behavior of laser-produced plasmas in high
energy-density experiments. But as we have emphasized from the
start of this review, the slower time scales, excellent diagnostics,
and control over initial conditions in ultracold plasmas provide
unique opportunities to make significant contributions to these
other areas.

The recent success to produce cold Rydberg gases and plasmas in
strong magnetic fields of several Tesla \cite{cho05,cho05b} has
opened up a new regime, where the atomic scale and macroscopic
system dynamics is fundamentally different from the field-free
situation. Theory has just started to treat collisional
\cite{ro04,hvm06,psg06} and radiative \cite{gcr03R,toro06,psy06}
processes under such conditions, while the macroscopic dynamics of
magnetized, ultracold, $\mu$m-scale neutral plasmas still awaits a
deeper understanding -- all being essential questions for ongoing
efforts to produce antihydrogen atoms \cite{aab02,gbo02}.

New or improved experimental capabilities wait just over the horizon
and will certainly open many avenues of research.  As the resolution
for imaging techniques approaches $\sim 10$~$\mu$m, this will allow
studies of collective modes or other features of the ion density
distribution. The use of light scattering to directly observe
spatial correlations \cite{ibt98} and the application of magnetic
fields to  confine the plasma will also provide new possibilities.

In light of the rapid development reviewed in this article,
 the
study of ultracold neutral plasmas promises to remain an exciting
and growing field of research for many years to come.

%% file: quantities-jm.tex
\begin{itemize}
    \item[] CCP -- Coulomb Coupling Parameter
    \item[] LTE -- Local Thermal Equilibrium
    \item[] MD -- Molecular Dynamics
    \item[] OCP --  One Component Plasma
    \item[] TCP --  Two Component Plasma
    \item[] PIC -- Particle in Cell  codes
    \item[] PICMCC - Particle in Cell Monte Carlo Calculation
    \item[] TBR -- Three body recombination
    \item[] RR -- Radiative recombination
    \item[] DR -- Dielectronic recombination

\end{itemize}

\begin{itemize}
\item[*] in all expressions $\alpha$ stands either for ions
($\alpha=\rm i$) or electrons ($\alpha=\rm e$)
\item[] particle number \dotfill $N_{\alpha}$
\item[] plasma temperature \dotfill $T_{\alpha}$
\item[] plasma width \dotfill $\sigma$
\item[] plasma density \dotfill $\rho_{\alpha}$
\item[] peak density \dotfill $\rho_0 =N_{\alpha}/(2\pi \sigma^2)^{3/2}$
\item[] average density \dotfill $\bar{\rho} =N_{\alpha}/(4\pi \sigma^2)^{3/2}$
\item[] plasma frequency \dotfill $\omega_{\rm
p,\alpha}=[e^2\rho_{\rm \alpha}/(m_{\rm \alpha}\varepsilon_{\rm 0})]^{1/2}$
\item[]  Wigner-Seitz radius \dotfill $a_{\alpha}=[3/(4 \pi \rho_{\alpha})]^{1/3}$
\item[]  Coulomb coupling parameter \dotfill
$\Gamma_\mathrm{\alpha} =  e^2 / (4\pi \varepsilon_0 a_{\alpha} k_B
T)$,
\item[] Debye screening length \dotfill
$[\varepsilon_{0}k_{B}T_\mathrm{\alpha}/(e^{2}\rho_\mathrm{\alpha})]^{1/2}$
\item[] Br\"{u}ckner parameter \dotfill $r_s=a_{\alpha}/a_{0}$
\item[] (ratio of interparticle distance to Bohr radius)
\end{itemize}

%% file: app-absorptionspectrum-jm.tex
The Voigt profile for the ion absorption cross section reads
\begin{eqnarray} \label{absorptioncrossection}
   \hspace{-.25in}\alpha(\nu, T_\mathrm{i})&=&\int d s \frac{3^*
   \lambda^2}{2\pi}\frac{\gamma_0}{\gamma_\mathrm{eff}}\frac{1}{1+ 4\left(
   \frac{\nu-s}{\gamma_\mathrm{eff}/2\pi}\right)^2}
   \frac{1}{\sqrt{2\pi}\sigma_D(T_\mathrm{i})}
   \exp\left[-{(s-\nu_0)^2 \over 2\sigma_D(T_\mathrm{i})^2}\right],
 \end{eqnarray}
where the parameters are described below \eq{equationtieffspectrum}
in Section \ref{sectionopticalprobes}.

The expansion of the plasma described in Section
\ref{sectionexpansion} eventually becomes the dominant source of
Doppler broadening for the ion absorption spectrum in an ultracold
neutral plasma (Fig.\ \ref{spectrum}).  Fortunately, the expansion
is approximated well by a self-similar expansion of a Gaussian
density distribution, and the dynamics takes a particularly simple
form when there are no heat sources for the electrons and
ion-electron thermalization is negligible. (See Section
\ref{section_coll} and \cite{kkb00,rha03,ppr04}.)  These assumptions
are reasonable for plasmas of low density and high initial electron
energy $E_e$ , although the exact regime of validity is the subject
of current study.  In this regime, the dynamics is determined by the
parameters $T_{\mathrm e}(0)$ and $\sigma$, where $T_{\mathrm e}(0)$
is the electron temperature after any initial heating (Section
\ref{sectionelectronheating}).  To be more specific, we assume that
electron heating is only important at very early times before plasma
expansion is important, and subsequently the electron temperature
evolves only under the influence of the adiabatic expansion.  Hence,
$k_BT_{\mathrm e}(0)$ may be slightly higher than $\frac 23
E_\mathrm{e}$.
%

 The expansion velocity profile $\mathbf{u}(\mathbf{r},t)$
 (Eq.\ (\ref{expvelocity})) gives rise to an average Doppler shift of the
 resonant frequency that varies with position,
\begin{eqnarray}\label{doppler shift}
 \delta \nu(\mathbf{r})=u_z/\lambda&=&\frac{r k_B T_{\mathrm e}(0) t}{m_\mathrm{i} \sigma^2 \lambda}
 \cos\theta
\end{eqnarray}
where $r\cos\theta$ is  the displacement from the center of the
cloud along the  direction of laser propagation and $t$ is the time after photoionization. 
The position-dependent Doppler shift must be included in the
exponent of the Gaussian describing the  Doppler broadening in the
Voigt convolution of the absorption cross section
$\alpha(\nu,T_\mathrm{i})$ (Eq.\ (\ref{absorptioncrossection})). The
latter becomes explicitly position dependent and takes the form
\begin{eqnarray}
    \label{absorptionfull}
 &&\alpha= \frac{3^* \lambda^2}{
  {2\pi}}\frac{\gamma_0}{\gamma_\mathrm{eff}}F(\mathbf r);\\
 &&F(\mathbf r) = \int d s
  {1 \over 1+ 4\left( { \nu-s \over \gamma_\mathrm{eff}/2\pi}\right)^2} {1
  \over \sqrt{2\pi} \sigma_D(T_\mathrm{i}(\mathbf{r}))}
  \exp\left[-{(s-\nu_0-\delta \nu(\mathbf{r}))^2 \over
  2\sigma_D(T_\mathrm{i}(\mathbf{r}))^2}\right]\,,\nonumber
\end{eqnarray}
where we have also included position dependence if the local thermal
equilibrium temperature (Eq.\ (\ref{iontemp})).
This variation of the ion temperature with position is the major
complication to calculation of the full absorption $S(\nu) = \int
d^{3}r\rho_\mathrm{i}(\mathbf r)\alpha(\nu,T_\mathrm{i},\mathbf r)$.
But by replacing $T_{\mathrm{i}}(\textbf{r})$ with the average ion
temperature in the plasma, $T_\mathrm{i,ave}=\int d^3r
\hspace{.025in} T_{\mathrm{i}}(\textbf{r})$ \cite{kcg05}, and
utilizing the spherical symmetry of the plasma, the integral over
space can be evaluated, and the ion temperature can be replaced with
an effective ion temperature ($T_\mathrm{i,eff}$). The result is
\eq{equationtieffspectrum} in Section \ref{sectionopticalprobes}.
One can analytically show that for a time $t$ after photoionization

\begin{equation}\label{equationtieffrelatedtoexpansion}
    T_\mathrm{i,eff}=T_\mathrm{i,ave}+ T_\mathrm{e}(0)t^2/
    (\tau_\mathrm{exp}^{2}+t^2).
\end{equation}
The characteristic plasma expansion time is $\tau_\mathrm{exp}$ from
\eq{exptime}. This enables a clear and well-defined analysis of  the
data, such as shown in
 Fig.\ \ref{tempevolutiondata}. 
 In principle, it also
 separates the various contributions to the Doppler broadening of
 the spectrum, although the expansion energy typically becomes an order of magnitude greater than the
 thermal energy, making quantitative measurement of the thermal
 energy difficult at later times.
One can also consider comparison of
Eqs.\ (\ref{equationtieffspectrum})  and (\ref{equationtieffrelatedtoexpansion})
with
data as a check of the self-similar expansion model (Eq.\ (\ref{expansion})).

References \cite{kcg05} and \cite{lcg06}
 numerically justify approximate expressions relating $T_\mathrm{i,eff}$
 to $T_\mathrm{i,ave}$, $T_\mathrm{e}(0)$, time, and other physical parameters
when the ion temperature is not assumed to be constant.

%% file: quasineutrality.tex
The assumption of quasineutrality constitutes a central part for the
theoretical description and the analysis of ultracold plasma
experiments. The condition of approximately equal electron and ion
densities ($\rho_{\rm e}\approx\rho_{\rm i}$) implies a well defined
charge separation, but this fact often causes confusion and is
rarely discussed in the literature. We provide a more detailed
discussion of its physical meaning in this section.

We start our discussion with the set of Vlasov equations \beq
\label{vlasov_e_app} \frac{\partial f_{\rm e}}{\partial
t}+{\bf{v}}_{\rm e}\frac{\partial f_{\rm e}}{\partial {\bf{r}}_{{\rm
e}}} - \frac{e} {m_{\rm e}}\frac{\partial f_{\rm e}}{\partial
{\bf{v}}_{{\rm e}}} \frac{\partial \varphi}{\partial {\bf{r}}_{\rm
e}} =0 \; , \eeq \beq \label{vlasov_i_app} \frac{\partial f_{\rm
i}}{\partial t}+{\bf{v}}_{\rm i}\frac{\partial f_{\rm i}}{\partial
{\bf{r}}_{{\rm i}}} + \frac{e} {m_{\rm i}}\frac{\partial f_{\rm
i}}{\partial {\bf{v}}_{{\rm i}}} \frac{\partial \varphi}{\partial
{\bf{r}}_{\rm i}} =0 \; , \eeq for the phase space distribution
functions $f_{\alpha}({\bf r}_{\alpha},{\bf v}_{\alpha})$, where
 $m_{\alpha}$ is the electron ($\alpha={\rm e}$) and ion ($\alpha={\rm i}$) mass  and
$\varphi$ is the total mean-field potential determined by the
Poisson equation \beq \label{poisson_app} \Delta
\varphi=\frac{e}{\varepsilon_0}\left(\rho_{\rm e}-\rho_{\rm
i}\right)\;, \eeq with $\rho_{\rm alpha}=\int f_{\alpha}d{\bf
v}_{\alpha}$.  Most straightforwardly, the solution of these
equations is obtained by calculating the potential $\varphi$ from
Eq.\ (\ref{poisson_app}) at every instant of time to propagate the
phase space densities according to eqs.(\ref{vlasov_e_app}) and
(\ref{vlasov_i_app}). However, the Vlasov equations could be used
alternatively to calculate the electrostatic potential, while the
Poisson equation determines the total charge density in this case.
To do this we consider the hydrodynamic velocities \beq
\label{u_app} {\bf u}_{\alpha}({\bf r}_{\alpha})=N_{\alpha}^{-1}\int
{\bf v_{\alpha}}f_{\alpha}({\bf r}_{\alpha},{\bf v}_{\alpha})d{\bf
v}_{\alpha} \eeq whose time evolution is obtained by multiplying
eqs.(\ref{vlasov_e_app}) and (\ref{vlasov_i_app}) by ${\bf v}_{\rm
e}$ and ${\bf v}_{\rm i}$, respectively, and integrating over
velocities. Upon subtracting of the resulting equations and using
the continuity equations for the electron and ion densities we
obtain for the electrostatic potential \beq \label{potneutr_app}
e\frac{\partial\varphi}{\partial{\bf r}}=\frac{\rho_{\rm
e}\left(\frac{\partial {\bf u}_{\rm e}}{\partial t}+{\bf u}_{\rm
e}\frac{\partial {\bf u}_{\rm e}}{\partial {\bf r}}\right)-\rho_{\rm
i}\left(\frac{\partial {\bf u}_{\rm i}}{\partial t}+{\bf u}_{\rm
i}\frac{\partial {\bf u}_{\rm i}}{\partial {\bf
r}}\right)}{\rho_{\rm e}/m_{\rm e}+\rho_{\rm i}/m_{\rm
i}}+\frac{\frac{\partial}{\partial {\bf r}}\left(\frac{{\mathcal
P}_{\rm e}}{m_{\rm e}}-\frac{{\mathcal P}_{\rm i}}{m_{\rm
i}}\right)}{\rho_{\rm e}/m_{\rm e}+\rho_{\rm i}/m_{\rm i}} \eeq For
isotropic, thermal velocity distributions the pressure tensors
${\mathcal P}_{\alpha}$ become simple scalar fields given by
${\mathcal P}_{\alpha}=k_{\rm B}T_{\alpha}\rho_{\alpha}$. Since the
time evolution of the hydrodynamic velocities and the pressure
tensors follows from the self-consistent solution of the Vlasov
equations this expression is completely equivalent to the Poisson
equation (\ref{poisson_app}). While the first term in Eq.\
(\ref{potneutr_app}) describes local charge imbalances between the
two species, the second term arises solely from the thermal pressure
of the particles. The latter only vanishes for the unlikely special
case of ${\mathcal P}_{\rm i}=\frac{m_{\rm i}}{m_{\rm e}}{\mathcal
P}_{\rm e}$, and, hence, yields a finite contribution to the
electrostatic potential in general. On the other hand, the first can
be neglected within the quasineutral approximation, requiring almost
equal charge densities and local currents, i.e. $\rho_{\rm
i}\approx\rho_{\rm e}$ and ${\bf u}_{\rm i}={\bf u}_{\rm e}$. Thus
the electrostatic potential can be calculated from \beq
\label{pot2_app} e\frac{\partial\varphi}{\partial{\bf
r}}\approx\frac{\frac{\partial}{\partial {\bf
r}}\left(\frac{{\mathcal P}_{\rm e}}{m_{\rm e}}-\frac{{\mathcal
P}_{\rm i}}{m_{\rm i}}\right)}{\rho_{\rm e}/m_{\rm e}+\rho_{\rm
i}/m_{\rm i}}\approx k_{\rm B}T_{\rm e}\rho_{\rm
e}^{-1}\frac{\partial \rho_{\rm e}}{\partial {\bf r}}\approx k_{\rm
B}T_{\rm e}\rho_{\rm i}^{-1}\frac{\partial \rho_{\rm i}}{\partial
{\bf r}} \eeq which coincides with Eq.\ (\ref{quasineut}),
introduced in section \ref{colless_kin}. In the second step we have
employed an adiabatic approximation for the electrons, making use of
the relation $m_{\rm e}\ll m_{\rm i}$.

Note, that a direct application of the quasineutrality condition
$\rho_{\rm e}\approx\rho_{\rm i}$ in the Poisson equation
(\ref{poisson_app}) would have resulted in a vanishing charge
separation potential. The preceding discussion however shows that
this is not the case since the thermal pressures of the plasma
components generally yield a finite contribution of zero order in
the charge separation. Rather the potential is given by Eq.\
(\ref{pot2_app}), while the Poisson equation determines the
separation of charge densities $\delta=\rho_{\rm i}-\rho_{\rm e}$ of
the two species. However, we have $\delta/\rho_{\rm e}\ll 1$,
ensuring quasineutrality and corresponding to the condition for the
existence of a plasma state, introduced by Eq.\ (\ref{plasma_cond}).